\documentclass[twocolumn,astrosymb,trackchanges]{aastex701}

\usepackage{xspace}

\newcommand\kms{\ensuremath{{\rm km}~{\rm s}^{-1}}\xspace}
\newcommand\angstrom{\ensuremath{\mathring{\rm A}}\xspace}
\newcommand\ergs{\ensuremath{{\rm erg}~{\rm s}^{-1}}\xspace}
\newcommand\ergscm{\ensuremath{{\rm erg}~{\rm s}^{-1}~{\rm cm}^{-2}}\xspace}
\newcommand\msun{\ensuremath{M_\odot}\xspace}
\newcommand\msunyr{\ensuremath{M_\odot~{\rm yr}^{-1}}\xspace}

\newcommand\kmsMpc{\ensuremath{{\rm km}~{\rm s}^{-1}~{\rm Mpc}^{-1}}\xspace}

\newcommand\ovi{\ion{O}{6}\xspace}
\newcommand\oviwave{\ion{O}{6}~$\lambda\lambda$1032,1038\,\angstrom\xspace}

\newcommand\sfr{\ensuremath{{\rm SFR}}\xspace}
\newcommand\mdotovi{\ensuremath{\dot{M}_{\rm O\,VI}}\xspace}
\newcommand\mdotx{\ensuremath{\dot{M}_{\rm X,spec}}\xspace}
\newcommand\mdothcf{\ensuremath{\dot{M}_{\rm X,spec,hcf}}\xspace}
\newcommand\mdotcf{\ensuremath{\dot{M}_{\rm cf}}\xspace}
\newcommand\mdotmax{\ensuremath{\dot{M}_{\rm max}}\xspace}
\newcommand\mdotmaxap{\ensuremath{\dot{M}_{\rm max,ap}}\xspace}

\usepackage{threeparttable}
\usepackage{amsmath}
\usepackage{amssymb}
\usepackage{amsfonts}
\usepackage{mathbbol}
\usepackage{physics}
\usepackage{gensymb}
\usepackage{tabularx}
\usepackage{fontawesome}
\usepackage{hyperref}
\usepackage{xurl}

\let\tablenum\relax
\usepackage{siunitx}

\shorttitle{Reigniting the FUSE II}
\shortauthors{Reefe et al.}

\graphicspath{{./}{figures/}}

\begin{document}

\title{Reigniting the FUSE II: \ovi Emission in Massive Elliptical Galaxies}

\correspondingauthor{Michael Reefe}
\email{mreefe@mit.edu}

\author[0000-0003-4701-8497]{Michael Reefe}
\altaffiliation{National Science Foundation, Graduate Research Fellow}
\affiliation{Kavli Institute for Astrophysics \& Space Research, Massachusetts Institute of Technology, Cambridge, MA 02139, USA}
\email{mreefe@mit.edu}

\author[0000-0001-5226-8349]{Michael McDonald}
\affiliation{Kavli Institute for Astrophysics \& Space Research, Massachusetts Institute of Technology, Cambridge, MA 02139, USA}
\email{mcdonald@space.mit.edu}


\author[0000-0003-2630-9228]{Greg L. Bryan}
\affiliation{Department of Astronomy, Columbia University, New York, NY 10027, USA}
\email{gb2141@columbia.edu}

\author[0000-0002-2808-0853]{Megan Donahue}
\affiliation{Department of Physics and Astronomy, Michigan State University, East Lansing, MI 48824, USA}
\email{donahu42@msu.edu}

\author[0000-0002-9378-4072]{Andrew C. Fabian}
\affiliation{Institute of Astronomy, University of Cambridge, Madingley Road, Cambridge, CB3 0HA, UK}
\email{acf@ast.cam.ac.uk}

\author[0000-0003-2754-9258]{Massimo Gaspari}
\affiliation{Department of Physics, Informatics and Mathematics, University of Modena and Reggio Emilia, 41125 Modena, Italy}
\email{massimo.gaspari@unimore.it}


\author[0000-0002-3031-2326]{Eric Miller}
\affiliation{Kavli Institute for Astrophysics \& Space Research, Massachusetts Institute of Technology, Cambridge, MA 02139, USA}
\email{milleric@space.mit.edu}

\author[0000-0001-6638-4324]{Valeria Olivares}
\affiliation{Departamento de F\'isica, Universidad de Santiago de Chile, Av. Victor Jara 3659, Santiago 9170124, Chile}
\affiliation{Center for Interdisciplinary Research in Astrophysics and Space Exploration (CIRAS), Universidad de Santiago de Chile, Santiago
9170124, Chile}
\email{valeria.olivares@usach.cl}

\author[0000-0002-3514-0383]{G. Mark Voit}
\affiliation{Department of Physics and Astronomy, Michigan State University, East Lansing, MI 48824, USA}
\email{voit@msu.edu}


\begin{abstract}

This is the second paper in a series where we examine archival far-ultraviolet spectroscopy of 29 massive elliptical galaxies from the \textit{Far Ultraviolet Spectroscopic Explorer} (\textit{FUSE}) with modern methods.  In the first paper, we performed SED modeling with aperture-matched photometry and analyzed the young and old stellar populations in these galaxies. In this paper, we report extinction-corrected \oviwave fluxes and upper limits for each galaxy, which we use to infer the rate at which gas is cooling through $10^{5.5}$ K via multiphase condensation of the galaxy's CGM (\mdotovi) under the assumption of steady radiative cooling. By considering both Galactic and intrinsic extinction, we overwhelmingly find \mdotovi to be larger than values found previously in the literature by a median of $\sim 80\%$.  Additionally, we quantify correlations between \mdotovi and star formation rates, classical luminosity-based X-ray cooling rates, and X-ray spectroscopic cooling rates.  The results suggest that the efficiency of cooling from the $\gtrsim 10^7$ K circumgalactic medium evolves with host mass, being suppressed by a factor of $\sim 4$-$5$ more at the cluster scale than it is at the group and elliptical scales.  In contrast, under the steady cooling assumption, cooling from $10^{5.5}$ K to the molecular phase appears to be suppressed by an order of magnitude, independent of mass, requiring turbulent mixing layers, thermal conduction, multiphase recycling, a bottom-heavy initial mass function, or some other additional physics to explain.

\end{abstract}

\section{Introduction} \label{sec:intro}

In the right conditions (high enough density and/or low enough temperature), the hot X-ray-emitting Circumgalactic Medium (CGM) that surrounds all massive galaxies is expected to cool within less than a Hubble time, flowing into the central galaxy and providing a source of cold gas that can be used to form new stars \citep{1994ARA&A..32..277F}.  The idealized rate at which gas should cool and be deposited onto the central galaxy from such a cooling episode can be estimated based on the luminosity and temperature of the cooling gas.  However, when comparing this to the observed star formation rates (SFRs) in the central galaxy, they are 1--2 orders of magnitude smaller \citep{1987MNRAS.224...75J, 1989AJ.....98.2018M, 1999MNRAS.306..857C, 2008ApJ...681.1035O, 2012ApJS..199...23H, 2018ApJ...858...45M, 2022ApJ...940..140C}.  This discrepancy is known as the ``cooling flow problem.''  

Of the mechanisms which have been proposed as a resolution, the most promising is feedback from an active galactic nucleus, or AGN \citep[see reviews by][]{2007ARA&A..45..117M, 2012ARA&A..50..455F, 2012NJPh...14e5023M, 2020NatAs...4...10G, 2022PhR...973....1D}, which reheats and redistributes the gas. Essentially all systems which exhibit discrepant cooling rates and SFRs host a radio-loud AGN \citep{2009ApJ...704.1586S}, and the mechanical power output by their radio jets is consistent with the radiative losses of the cooling gas \citep{2006ApJ...652..216R, 2004ApJ...607..800B, 2012MNRAS.421.1360H}.  This provides a global energy balance, while still allowing local residual thermal instabilities to develop \citep{2015Natur.519..203V}.  This is backed up by work on hydrodynamical simulations, which can reproduce the observed behavior of these thermal instabilities across scales from clusters to groups to isolated ellipticals \citep{2011MNRAS.411..349G, 2011MNRAS.415.1549G, 2012MNRAS.424..190G, 2012MNRAS.419.3319M, 2014ApJ...789...54L, 2014ApJ...789..153L}.  

The ``classical'' cooling rate described above can then be seen as an upper limit, representing the idealized maximal cooling rate in the absence of any feedback, \mdotmax.  A complementary approach is to calculate the cooling rate of the gas at a particular temperature, $\dot{M}(T)$, under the assumption of a steady-state single-pass cooling flow, where gas flows through temperature $T$ en route from the hot phase to the cold phase.  This is a strong idealization, but one that drastically simplifies the calculation.  Observationally, $\dot{M}(T)$ is probed via UV and X-ray emission line spectroscopy, using lines whose emissivity peaks near $T \sim 10^{5.5}$ K and $10^{6.5}$ K, respectively.  These ``instantaneous'' measurements of the cooling rate lie between the luminosity-based X-ray measurements (\mdotmax) and the cold star-forming gas (probed by the \sfr), both of which are necessarily time-averaged on Myr--Gyr timescales.
Then, in the picture of ``reduced'' cooling limited by AGN feedback, the cooling rate should either decrease or remain flat as the temperature decreases.  At the same time, the observed masses of molecular gas reservoirs \citep{2001MNRAS.328..762E, 2015A&A...573A.111O, 2018A&A...618A.126O} imply that there cannot be a significant excess of cooling gas over the \sfr.  
If the steady-cooling picture is correct, then we expect: $\sfr \simeq \dot{M}(10^{5.5}\,{\rm K}) \simeq \dot{M}(10^{6.5}\,{\rm K}) < \mdotmax$.

However, making such emission line measurements raises more questions than answers.  The soft X-ray transitions of lines like \ion{O}{7}--\ion{O}{8} and \ion{Fe}{17}--\ion{Fe}{24} reveal that $\dot{M}(10^{6.5}\,{\rm K})$ are significantly smaller than $\mdotmax$, and sometimes even smaller than the $\sfr$, which should not be possible \citep{2001A&A...365L..99K, 2001A&A...365L..87T, 2001ApJ...560..194M, 2001A&A...365L.104P, 2003ApJ...590..207P, 2006PhR...427....1P}.  Whereas, simultaneously, lower ionization lines like \ovi and [\ion{Fe}{10}] produce $\dot{M}(10^{5.5}\,{\rm K})$ that are usually larger than both the \sfr and the $\dot{M}(10^{6.5}\,{\rm K})$ \citep{2001ApJ...560..187O, 2004A&A...421..503L, 2005ApJ...635.1031B, 2006ApJ...642..746B, 2014ApJ...791L..30M, 2015ApJ...811..111M, 2017ApJ...835..216D}.  There are a few potential explanations for these strange patterns:
\begin{enumerate}
    \item $\dot{M}(10^{5.5}\,{\rm K})$ are systematically overestimated, and $\dot{M}(10^{6.5}\,{\rm K})$ are more representative of the ``true'' cooling rate. Measured $\dot{M}(10^{5.5}\,{\rm K})$ may be enhanced relative to SFRs and $\dot{M}(10^{6.5}\,{\rm K})$ in a few ways, for example via thermal conduction \citep[AKA ``heating flows''][]{1993ApJ...405L..17C} or turbulent mixing layers \citep{1990MNRAS.244P..26B, 1993ApJ...407...83S}. 
    \item $\dot{M}(10^{6.5}\,{\rm K})$ are systematically underestimated, and $\dot{M}(10^{5.5}\,{\rm K})$ are closer to the ``true'' cooling rate.  Measured $\dot{M}(10^{6.5}\,{\rm K})$ may be reduced by absorption from cold gas which forms in situ as the hot atmosphere cools. This self-absorption model has recently been referred to as a ``hidden cooling flow'' \citep{2022MNRAS.515.3336F, 2023MNRAS.521.1794F, 2023MNRAS.524..716F, 2024MNRAS.535.2173F}.
    \item SFRs may be systematically overestimated.  In this case, situations where $\dot{M}(10^{6.5}\,{\rm K}) < \sfr$ are less problematic.  However, one of the above explanations is still required to explain why $\dot{M}(10^{5.5}\,{\rm K}) > \dot{M}(10^{6.5}\,{\rm K})$.
\end{enumerate}

It is worth noting that modern hydrodynamic simulations, under the framework of chaotic cold accretion, predict large stochasticity in time in all of the above quantities (in addition to turbulence, mixing, uplift, and recycling), far from a steady-state cooling flow as we have discussed.  In any individual system, then, there is no need for the instantaneous cooling rates and star formation rates to agree with each other, even if the long-term mass budget is closed \citep{2012ApJ...746...94G, 2026arXiv260527504B, 2026arXiv260527511C, 2026arXiv260527508P}.  On a population level, time variability is expected to increase the intrinsic scatter, but it may also contribute to systematic offsets because the tracers have different temporal kernels and nonlinear emissivity weightings, while recurrent condensation and reheating can cause the same material to cross the \ovi-emitting phase multiple times.

The central goal of this work is to address these proposed solutions with a set of updated measurements.  This is the second in a series of papers covering a comprehensive and homogeneous reanalysis of archival data on 29 massive elliptical galaxies from the \textit{Far Ultraviolet Spectroscopic Explorer} (\textit{FUSE}).  These data are well positioned to provide clarity on these problems, as they cover both the FUV-bright young stellar populations and the \oviwave doublet, which is a strong cooling line sensitive to the $10^{5.5}$ K regime.  

In \citet{2026arXiv260905675R} (hereafter Paper I) we went into detail about our methodology on data collection and reduction, providing the highest possible quality \textit{FUSE} spectra. We performed SED fitting of the \textit{FUSE} spectra with aperture-matched UV, optical, and IR photometry.  We modeled the old and young stellar populations, including dust absorption and re-emission and the UV upturn, from which we measured SFRs and stellar metallicities (among other parameters).  From this analysis, we observed a few key findings relating to the cooling flow problem, which we summarize here:
\begin{itemize}
    \item Compared to our modeled SFRs, canonical SFRs (especially of the more quiescent systems) have likely been systematically overestimated (or hit a noise floor). Scaling relations between SFR and the UV continuum \citep[e.g.][]{1998ARA&A..36..189K} often do not account for the UV upturn feature that is observed in massive quiescent galaxies \citep[though some studies have accounted for this, i.e.][]{2012ApJS..199...23H}.  H$\alpha$-based SFRs may similarly be inflated by H$\alpha$ emission from non-stellar sources (AGNs, radiative cooling, shock heating, etc.).
    \item The observed SFRs cannot be explained entirely by recycled gas from old stellar populations, as evidenced by their metallicities.  At least a portion of the star formation most likely comes from either cooling CGM gas or stripped cold gas from satellites.
\end{itemize}
These revelations provide hints at the aforementioned solutions to the cooling flow puzzle, but alone they are not sufficient to answer it.

Therefore, in this paper (Paper II), we now look to perform a reanalysis of the \oviwave emission in these systems.  All of the 29 systems we consider here have previously published measurements of \ovi \citep{2001ApJ...560..187O, 2004A&A...421..503L, 2005ApJ...635.1031B, 2006ApJ...642..746B}, but these measurements were made by different teams with different analysis techniques, and rely on many now-outdated assumptions.  Our goal is to provide a homogeneous sample of modern measurements, and treat statistical and systematic uncertainties with care. As explained in detail in the introduction to Paper I, our new \ovi measurements will use updated distances for each system, will account for both line-of-sight Galactic dust extinction \textit{and} intrinsic dust extinction within each host system, will use stellar population models tailored to each system (from Paper I) for continuum subtraction, and will use cooling simulations tailored to the metallicities and thermodynamics of each system to convert \ovi luminosities into cooling rates.

This paper is organized as follows. $\S$\ref{sec:sample} goes over the sample selection. $\S$\ref{sec:methods} reviews our methods for modeling the \ovi emission from \textit{FUSE}, simulating cooling gas in each system to convert the \ovi luminosities into cooling rates, and measuring classical cooling rates to compare with the \ovi cooling rates with \textit{Chandra}. $\S$\ref{sec:res} presents our results.  The implications of these results are discussed further in $\S$\ref{sec:discuss}, putting them into context with the cooling flow problem.  Finally, $\S$\ref{sec:conclusion} summarizes our findings and suggests the next steps needed to move forward.  Throughout this paper, we use a flat $\Lambda$CDM cosmology with $H_0=70$ \kmsMpc, $\Omega_m = 0.27$, and $\Omega_{\Lambda} = 0.73$.  Uncertainties are given as $1\sigma$ (68\% confidence) unless otherwise stated.

\section{Sample} \label{sec:sample}

Here, we briefly recount the basic properties and coverage of our sample.  For more details, we refer the reader to Table 1 in Paper I, which contains a full list of each system's name, coordinates, redshift, and distance.  Our sample consists of 29 out of the 30 massive elliptical galaxies which have archival far-ultraviolet spectroscopy from \textit{FUSE}.  One system, NGC 3585, had to be dropped from our sample, because all 16 ks of its \textit{FUSE} data are unusable due to a combination of telescope motions and high airglow contamination. Of the remaining 29 systems, 4 have multiple unique pointings: Abell 1795, Virgo (M 87), NGC 4636, and NGC 4649.  For Abell 1795 and NGC 4636, these pointings are reduced and analyzed separately, and the final \mdotovi measurements are combined.  For M 87 and NGC 4649, the pointings are co-spatial and only differ in position angle, so they are combined before the \ovi fitting stage.

These systems can be subdivided roughly into thirds based on their environment, with 8/29 being brightest cluster galaxies (BCGs), 12/29 being brightest group galaxies (BGGs), and 9/29 being non-central elliptical galaxies (ETGs), ranging in halo mass from $M_{500} \sim 10^{12}$--$10^{14}$ \msun.  For simplicity, we refer to BCGs by the names of their clusters.  All 29 systems are local, with 23/29 having redshifts $z < 0.01$, and the remaining 6 being only marginally further, with the furthest system being Abell 2597 at $z \sim 0.083$.  For such nearby systems, peculiar motions can have a profound impact on the measured redshift, and can deviate substantially from the Hubble flow (indeed, one of our targets, NGC 4406, actually shows a net blueshift).  Therefore, we use redshift-independent distances for all but the 5 furthest systems in our sample, obtained from NED\footnote{\url{https://ned.ipac.caltech.edu/}}.

\section{Methods} \label{sec:methods}

At the heart of our goal is calculating four measurements of cooling at different phases: \sfr, $\mdotovi \sim \dot{M}(10^{5.5}\,{\rm K})$, $\mdotx \sim \dot{M}(10^{6.5}\,{\rm K})$, and $\mdotmax$ for our sample.  Importantly, all of these quantities are model-dependent measures of a pure cooling-flow-equivalent mass deposition rate (except for the \sfr, which is instead model-dependent based on our stellar population and star formation history models from Paper I).  They do not represent actual measurements of a net mass deposition rate.  Nevertheless, there is precedent in the literature for working directly with the cooling rates, and we find that analyzing correlations in $\dot{M}$-space (as opposed to staying in luminosity-space) is conceptually more convenient for testing steady cooling models. In the following sections, we go over our methods for collecting each of these measurements.

\subsection{\sfr} \label{sec:sfr}

We have already calculated SFRs for our sample based on SED fitting of the UV spectroscopy alongside optical/IR photometry as described in Paper I.  Throughout this work, we use both the 100 Myr averaged and 10 Myr averaged SFRs from these models, applying on a case-by-case basis whichever one is a more appropriate match to the timescales probed by the other cooling rates we measure.  We refer the reader to Paper I for a detailed explanation of our methods for calculating these SFRs.

\subsection{\ovi cooling rates (\mdotovi)} \label{sec:mdot_ovi}

\subsubsection{\ovi models} \label{sec:ovi_models}

Our procedure for obtaining \oviwave fluxes for each system is as follows.  For each system (or each pointing, in the case of the systems with multiple pointings), we utilize our fully reduced \textit{FUSE} spectra from Paper I.  These spectra have been coadded across all exposures and detectors and resampled to a wavelength resolution of 0.2 \angstrom.  We subtract our median stellar continuum models from Paper I, leaving only a residual emission line spectrum, and focus our emission line modeling on a small window around the \ovi doublet, covering rest-frame 1025--1045 \angstrom.

We model the residual emission line spectrum with a pair of Gaussians for the \ovi doublet, locked in a $I(1032)/I(1038)=$ 2:1 intensity ratio \citep{2003ApJS..149..205M}, and with their kinematics tied. We model \ovi emission both from the host system and from foreground Galactic emission (as they may overlap in the low-$z$ systems).  The centroid of the Galactic lines may vary within $\pm 50$ \kms of their expected positions, and their full-widths at half-maximum (FWHMs) are restricted to $\leqslant 200$ \kms, allowing them to model the typical motions seen in the Milky Way halo.  The host lines' kinematics, meanwhile, have more freedom, with line-of-sight velocities allowed within $\pm 250$ \kms (relative to the systemic redshift) and FWHMs within $\leqslant 600$ \kms.  These lines are then attenuated by a few factors, which we go through one-by-one in the following paragraphs.

(1) Galactic H$_2$ absorption affects both sets of \ovi lines.  This is modeled following the same methodology as Paper I, taking H$_2$ templates from \citet{2003PASP..115..651M} and assuming a continuous temperature distribution, which sets the relative normalizations of each transition.  The H$_2$ column density, $\log(N_{\rm H_2,gal}/{\rm cm}^2)$, which sets the overall normalization of the templates, is fixed to the values found from our continuum models from Paper I.

(2) Absorption from H$_2$ intrinsic to the host system affects only the host's \ovi lines.  This is modeled in exactly the same way as the Galactic H$_2$ absorption (apart from the redshift), with normalizations $\log(N_{\rm H_2,int}/{\rm cm}^2)$ again taken from our continuum models from Paper I.

(3) Absorption from metal lines originating in the Galactic ISM may also affect the host's \ovi lines if they happen to overlap at the given redshift.  Therefore, we explicitly model a small set of these ISM lines that fall within a wavelength range that overlaps with the \ovi lines in each system.  The lines that we potentially include in the model are \ion{O}{1} $\lambda\lambda$988.6,988.7,988.8,1039.2; \ion{N}{3} $\lambda\lambda$989.8,991.5,991.6; \ion{C}{2} $\lambda$1036.3; \ion{Ar}{1} $\lambda$1048.2; \ion{N}{2} $\lambda$1084.0; \ion{N}{1} $\lambda\lambda$1134.2,1134.4,1135.0; \ion{Ca}{2} $\lambda\lambda$1135.5,1135.6; and \ion{Fe}{2} $\lambda\lambda$1142.4,1143.2,1144.9.  These lines' kinematics are restricted in the same way as the Galactic \ovi emission lines.

(4) Galactic dust extinction affects both sets of \ovi lines. $E(B-V)_{\rm gal}$ values are obtained from the \citet{2011ApJ...737..103S} dust maps, and correction factors are calculated using a \citet{1989ApJ...345..245C} extinction law with an $R_V=3.1$, appropriate for the Milky Way.  Since these corrections are well known, they are applied directly to the data, before any fitting is done.

(5) Extinction from dust intrinsic to the host system only affects the host's \ovi lines.  We obtain $E(B-V)_{\rm int}$ values from our stellar population models from Paper I.  However, since these reddening values are derived from the young and old stellar populations, $E(B-V)_{\rm int,yng}$ and $E(B-V)_{\rm int,old}$, it is not appropriate to apply them directly to CGM gas in the $10^{5.5}$ K phase.  In general, it is difficult to extrapolate an appropriate reddening to use, as little is known about the distribution and properties of dust in the $10^{5.5}$ K phase of massive elliptical galaxies.  Therefore, we make a few assumptions about the dust to make the analysis simpler.  First, we assume that $E(B-V)_{\rm int,old}$ is an appropriate proxy for the level of diffuse, volume-filling dust that exists throughout the host galaxy: $E(B-V)_{\rm int,old} \approx E(B-V)_{\rm int,diff}$.  Then, we assume that $E(B-V)_{\rm int,yng}$ can be converted into a value appropriate for the $10^4$ K phase existing primarily in \ion{H}{2} regions around young stars with a conversion factor $C_E$, such that $E(B-V)_{\rm int,yng}=C_E E(B-V)_{\rm int,neb}$.  Finally, we assume that the $10^4$ K gas phase has covering factor $C_f$ relative to the $10^{5.5}$ K \ovi-emitting phase, such that the total attenuation factor of the \ovi is given by:
\begin{equation}
\begin{split}
    E_{\rm int,O\,VI} = (1-C_f) & 10^{-0.4k'(\lambda)E(B-V)_{\rm int,old}} \\
    + C_f\,& 10^{-0.4k'(\lambda)E(B-V)_{\rm int,yng}/C_E}
\end{split}
\end{equation}
where $k'(\lambda)$ uses a \citet{2000ApJ...533..682C} extinction law with an $R_V=4.05$.  

It now remains to find appropriate values for $C_E$ and $C_f$.  For the former, we use a value of $C_E=0.17 \pm 0.05$, obtained in Paper I by comparing our $E(B-V)_{\rm int,yng}$ estimates to reddenings estimated from the Balmer decrement in the same systems, done by \citet{1999MNRAS.306..857C}.  We note that this is substantially lower than the typically used value of 0.44 from \citet{2000ApJ...533..682C}, which has been derived from actively star-forming and starbursting galaxies and is likely not appropriate to use for massive elliptical galaxies.  From paper I, we have interpreted the lower value of $C_E$ to be reflective of differences in the dust's geometry---ellipticals have dust distributions which are more concentrated around star-forming clumps and filaments, with little to no dust cospatial with the old stellar populations elsewhere in the galaxy.  Spiral galaxies, on the other hand, have dust distributed throughout the disk and bulge, cospatial with both young and old stellar populations, while old stars in globular clusters throughout the halo may have much less dust.  With that said, the precise value of $0.17$ is quite uncertain, as it relies on $E(B-V)$ values obtained from the H$\alpha$/H$\beta$ ratio from \citet{1999MNRAS.306..857C}, which can be contaminated in BCGs by collisional excitation and other effects that enhance H$\alpha$ relative to the other Balmer lines.

For $C_f$, we use the following very simplistic and idealized argument.  If we assume the $10^4$ K and $10^{5.5}$ K are situated in concentric spherical shells, in pressure equilibrium ($P_1=P_2$), then by the ideal gas law ($P=(N/V)kT$) we obtain the constraint $V_1/V_2=N_1T_1/N_2T_2$.  The volume of the inner sphere is $V_1 \propto R_1^3$, while the outer shell is $V_2 \propto R_2^3-R_1^3$.  Whereas the corresponding cross-sectional areas (filled) are $A_1 \propto R_1^2$ and $A_2 \propto R_2^2$.  The covering fraction is then the fraction of the area which is covered by the dusty $10^4$ K phase, $A_1/A_2$, divided in half, since half of the outer shell is in front of the inner sphere and does not get covered by it.  Therefore,
\begin{equation}
    C_f \approx \frac{1}{2}\frac{A_1}{A_2} = \frac{1}{2}\bigg(\frac{N_2T_2}{N_1T_1} + 1\bigg)^{-2/3}
\end{equation}
This result depends on our assumption of the relative masses of the two clouds.  For an individual, isolated, rapid cooling event, total mass must be conserved ($N_1 = N_2$), which gives $C_f \sim 0.05$.  On the other hand, for a steady-state cooling flow with a constant $\dot{M}$, the amount of gas in each phase depends on its residence time, $N_i \propto t_{{\rm res}}(T_i)$, and since the cooling function peaks at $\sim 10^{5.5}$ K, the residence time of the \ovi-emitting phase is short compared to the $10^4$ K phase, $t_{\rm res}(10^4\,{\rm K})/t_{\rm res}(10^{5.5}\,{\rm K}) \sim 70$, leading to a much larger covering fraction of $C_f \sim 0.4$.  We take these two extreme cases as brackets on the true covering fraction, and thus assume for the remainder of our analysis that $C_f \sim 0.225 \pm 0.175$.  However, since this correction requires by far the most assumptions and carries the largest systematic uncertainties (in $C_E$ and $C_f$), we provide \ovi fluxes both with and without this correction applied.

We use Python's \textsc{lmfit} package \citep{lmfit} to optimize our models.  This procedure is performed in two steps.  First, we perform a global optimization with the differential evolution method to find the best fit values.  Second, to robustly estimate uncertainties on our model parameters, we perform 1000 bootstrap iterations.  In each iteration, the data are resampled using their uncertainties, and a local Levenberg-Marquardt least squares optimization is performed, with initial values taken from the global best fit.  The 15.9$^{\rm th}$, 50$^{\rm th}$, and 84.1$^{\rm st}$ percentiles are extracted from the bootstrap iterations, which we use to calculate statistical uncertainties ($\sigma_{\rm stat}$) on the model parameters.

The statistical uncertainties on \ovi fluxes from bootstrapping, however, will underestimate the true uncertainties, because they do not account for any uncertainty in the attenuations from H$_2$ and dust, since these model parameters are fixed from our previous round of continuum fits from Paper I.  They also do not account for uncertainties in the absolute flux calibration or background subtraction.  Therefore, we categorize these as systematic uncertainties ($\sigma_{\rm syst}$), and explicitly calculate them, separately from the statistical uncertainties.  We achieve this by taking the relevant model parameter uncertainties (on the H$_2$ column densities, dust reddenings, and background normalizations) from Paper I, and propagating them analytically to calculate the relevant uncertainties on \ovi flux.  Additionally, to account for the large uncertainty in the assumptions used to derive $C_f$, we also include its systematic uncertainty of $\pm 0.175$, derived above to bracket the two limiting cases we considered. This is propagated in addition to the uncertainties on the other model parameters.  For the uncertainty on absolute flux calibration, we use a value of 10\%, following the typical values quoted in the \textit{FUSE} data reduction analysis software \citep{2007PASP..119..527D}.  These systematic uncertainties are reported separately from the statistical uncertainties, since they are relevant for the true measurement of \ovi flux, but not for determining the statistical significance of the detection.

We use a detection threshold of $2.5\sigma_{\rm stat}$, which corresponds to a false positive rate of 1 in 161, more than sufficient for our sample size of 29 systems. For the systems where \ovi is not detected above this threshold, we perform a stacking analysis.  To create the stack, we first take each residual spectrum and remove any detected metal absorption lines using our best fitting model.  Then we convert each spectrum to luminosity units (${\rm erg}\,{\rm s}^{-1}\,{\rm \angstrom}^{-1}$) and stack in the rest frame, weighting by the (statistical) errors.  In addition to the full stacked spectrum of all nondetections, we also create stacked spectra containing only the systems within cluster halos, group halos, and elliptical halos.  We fit these stacked spectra with a simplified model that does not include any attenuation. We also do not tie the intensity ratio of the \ovi lines to 2:1, since the unmodeled attenuation may alter this ratio.  Our systematic error calculations only include the 10\% absolute calibration error, and therefore may be underestimated due to the missing attenuation effects.  Otherwise, our methods for fitting the stacked spectra remain identical to the individual systems.

\subsubsection{\textsc{Cloudy} simulations} \label{sec:cloudy}

To convert \ovi luminosities to cooling rates, we perform simulations of a cooling parcel of gas with \textsc{Cloudy} \citep{1998PASP..110..761F, 2023RMxAA..59..327C}.  The gas parcel starts out at a temperature and density representative of the hot CGM (see below), and is illuminated by a standard background of cosmic rays, cosmic microwave background, and X-ray background.  The simulation then iterates over time, tracking the gas parcel as it cools (either isobarically or isochorically), and logging emission from a set of cooling lines, including the \ovi doublet.  The simulation stops when the parcel reaches $10^4$ K.

We do not consider additional illumination from a central AGN in these simulations, as our sample consists overwhelmingly of radiatively inefficient, low Eddington ratio AGN.  Similarly, we do not consider illumination from young stellar populations, as even the hottest O-type stars do not produce a sufficient photon flux above 114 eV, the ionization potential of \ovi.  On the other hand, more substantial effects on the \ovi emission may arise from turbulence, mixing, and/or shocks driven by AGN mechanical feedback or supernovae (SNe).  As \textsc{Cloudy} is primarily a photoionization and radiative transfer code, it cannot easily model such effects, so they are ignored for the moment, making our cooling rate estimates valid in the limit of a pure, passive cooling flow.  We later consider these additional effects as corrective factors on our cooling rate estimates in $\S$\ref{sec:discuss}.

Each system in our sample has its own tailored set of simulations.  The initial gas temperature, $kT_i$, is set by the X-ray temperatures which we have measured from \textit{Chandra} imaging spectroscopy in Paper I.  The initial gas hydrogen density, meanwhile, is set to a universal constant of $n_{{\rm H},i} = 0.1$ cm$^{-3}$.  This is denser than most observed CGM gas, but our choice is motivated by the fact that cooling \ovi emission should primarily originate from clumpy overdensities of gas, and a density of 0.1 cm$^{-3}$ sits near the highest central densities observed in the CGM \citep{2006ApJ...640..691V}. Moreover, in the isobaric simulations, where the gas density reacts to the lower temperatures, our choice of initial density has essentially no impact on the results, as the gas quickly condenses to $\gtrsim 0.1$ cm$^{-3}$ before the \ovi emission peaks, regardless of its starting density.  The isochoric simulations, meanwhile, show a larger sensitivity to the initial $n_{\rm H}$, since this sets the density which they are locked to throughout the entire simulation.  Therefore, choosing an arguably large initial $n_{\rm H}$ of 0.1 cm$^{-3}$ is necessary to make the isochoric simulations appropriately reflect the most likely physical conditions in the denser clumps of cooling gas.

The gas metallicity in each simulation is also set based on measured values from \textit{Chandra} spectra in Paper I, using the proto-solar abundance set of \citet{2009LanB...4B..712L}.  However, large intrinsic variation is seen in individual elemental abundances relative to Fe, especially in the $\alpha$ elements, of which O is a part.  Therefore, we pull as many available individual $\alpha$ element abundances as we can find from high-resolution X-ray spectroscopy measurements (e.g. \textit{XMM}-RGS, \textit{XMM}-EPIC, \textit{Suzaku}, and \textit{XRISM}) in the literature.  Of particular importance is the O abundance, which we are able to find for 21/29 systems.  For the remaining systems, we fall back to categorical sample-wide abundance measurements for galaxy clusters and galaxy groups from the CHEERS project \citep{2017AnA...603A..80M}.  Otherwise, for elements which have no categorical CHEERS measurement (Ne) and no individual system measurement, the last fall back is to use the default scaled proto-solar abundances from \citet{2009LanB...4B..712L}.  All initial temperatures and abundances are given in Table \ref{tab:cloudy_init}.  

For each system, we run a total of 6 simulations: 2 nominal simulations for isobaric and isochoric cooling, 2 where the metal abundances are $1\sigma$ below their nominal values, and 2 where the metal abundances are $1\sigma$ above their nominal values.  From each simulation, we obtain a value $\Gamma$ (erg/g) that can be used to convert a given \ovi luminosity into a pure cooling-flow-equivalent mass flow rate:
\begin{equation}
    \mdotovi = L_{\rm O\,VI}/\Gamma
\end{equation}
The additional simulations at $\pm 1\sigma$ allow us to numerically estimate a systematic uncertainty in $\Gamma$, and therefore in the translation from luminosity to cooling rate, based on the uncertainty in the abundances.  Any additional systematics based on the uncertainty in the starting temperature $kT_i$ are assumed to be negligible.

We also run a set of generic simulations, not tailored to any individual system, for usage with our analysis of the stacked spectra.  For these simulations, we use the CHEERS average metal abundances where available, and we use an average initial temperature from our individual systems.

\begin{deluxetable*}{llllllllhhll}
\def\arraystretch{1.15}
\tabletypesize{\footnotesize}
\tablecaption{Initial temperatures and metal abundances for \textsc{Cloudy} simulations of each system}
\label{tab:cloudy_init}
\tablehead{
    \colhead{Name} & 
    \colhead{$kT_i$ (keV)} & 
    \colhead{Fe/H} & 
    \colhead{O/Fe} & 
    \colhead{Ne/Fe} & 
    \colhead{Mg/Fe} & 
    \colhead{Si/Fe} &
    \colhead{S/Fe} & 
    \nocolhead{Ar/Fe} & 
    \nocolhead{Ca/Fe} & 
    \colhead{Ni/Fe} & 
    \colhead{Ref\tablenotemark{*}}
}
\decimals
\startdata
\hline
\multicolumn{12}{c}{Clusters} \\
\hline
Perseus & $1.44_{-0.00}^{+0.00}$ & $0.12_{-0.00}^{+0.00}$ & $1.13_{-0.26}^{+0.26}$ & $0.99_{-0.30}^{+0.30}$ & $0.91_{-0.22}^{+0.22}$ & $0.82_{-0.11}^{+0.11}$ & $0.89_{-0.09}^{+0.09}$ & $0.84_{-0.08}^{+0.08}$ & $0.93_{-0.09}^{+0.09}$ & $0.96_{-0.09}^{+0.09}$ & S19 \\ 
Fornax & $0.97_{-0.01}^{+0.01}$ & $0.47_{-0.03}^{+0.05}$ & $0.52_{-0.04}^{+0.03}$ & $0.87_{-0.08}^{+0.07}$ & $1.06_{-0.07}^{+0.05}$ & $0.98_{-0.06}^{+0.04}$ & $0.89_{-0.07}^{+0.06}$ &   $-$                  &   $-$                  & $2.57_{-0.16}^{+0.14}$ & J09 \\ 
Virgo P1 & $1.06_{-0.01}^{+0.00}$ & $0.34_{-0.01}^{+0.02}$ & $1.04_{-0.25}^{+0.25}$ & $1.05_{-0.18}^{+0.18}$ & $0.80_{-0.17}^{+0.17}$ & $0.82_{-0.10}^{+0.10}$ & $0.95_{-0.13}^{+0.13}$ & $1.69_{-0.32}^{+0.32}$ & $1.52_{-0.39}^{+0.39}$ & $0.81_{-0.33}^{+0.33}$ & S10 \\ 
Virgo P2 & $1.08_{-0.00}^{+0.00}$ & $0.29_{-0.01}^{+0.01}$ & $1.04_{-0.25}^{+0.25}$ & $1.05_{-0.18}^{+0.18}$ & $0.80_{-0.17}^{+0.17}$ & $0.82_{-0.10}^{+0.10}$ & $0.95_{-0.13}^{+0.13}$ & $1.69_{-0.32}^{+0.32}$ & $1.52_{-0.39}^{+0.39}$ & $0.81_{-0.33}^{+0.33}$ & S10 \\ 
Abell 1795 P1 & $2.83_{-0.05}^{+0.04}$ & $0.41_{-0.01}^{+0.01}$ & $0.77_{-0.08}^{+0.08}$ & $1.02_{-0.16}^{+0.16}$ & $0.93_{-0.32}^{+0.32}$ & $0.90_{-0.16}^{+0.16}$ & $0.42_{-0.17}^{+0.17}$ &   $-$                  &   $-$                  &   $-$                  & T04 \\ 
Abell 1795 P2 & $3.10_{-0.06}^{+0.04}$ & $0.38_{-0.01}^{+0.01}$ & $0.77_{-0.08}^{+0.08}$ & $1.02_{-0.16}^{+0.16}$ & $0.93_{-0.32}^{+0.32}$ & $0.90_{-0.16}^{+0.16}$ & $0.42_{-0.17}^{+0.17}$ &   $-$                  &   $-$                  &   $-$                  & T04 \\ 
Abell 2029  & $3.58_{-0.15}^{+0.12}$ & $0.68_{-0.06}^{+0.10}$ & $1.22_{-0.07}^{+0.08}$ & $0.85_{-0.09}^{+0.10}$ & $0.93_{-0.07}^{+0.07}$ & $1.04_{-0.04}^{+0.04}$ & $0.72_{-0.05}^{+0.05}$ & $0.50_{-0.29}^{+0.31}$ & $0.91_{-0.29}^{+0.32}$ & $1.34_{-0.08}^{+0.10}$ & S25 \\ 
Abell 2597  & $2.04_{-0.08}^{+0.08}$ & $0.21_{-0.01}^{+0.01}$ & $1.32_{-0.06}^{+0.05}$ & $0.83_{-0.06}^{+0.07}$ & $0.81_{-0.06}^{+0.07}$ & $0.78_{-0.06}^{+0.06}$ & $0.85_{-0.06}^{+0.07}$ & $0.86_{-0.06}^{+0.07}$ & $0.84_{-0.06}^{+0.07}$ & $0.80_{-0.06}^{+0.07}$ & M05 \\ 
Abell 3112  & $2.55_{-0.14}^{+0.09}$ & $0.86_{-0.11}^{+0.33}$ & $0.62_{-0.11}^{+0.11}$ & $0.34_{-0.25}^{+0.25}$ &   $-$                  & $0.90_{-0.16}^{+0.16}$ & $0.28_{-0.13}^{+0.13}$ &   $-$                  &   $-$                  &   $-$                  & T04 \\ 
AWM 7  & $2.09_{-0.12}^{+0.16}$ & $1.33_{-1.33}^{+1.33}$ & $0.70_{-0.21}^{+0.23}$ & $1.64_{-0.18}^{+0.20}$ & $1.15_{-0.15}^{+0.15}$ & $0.83_{-0.08}^{+0.09}$ & $0.86_{-0.10}^{+0.10}$ &   $-$                  &   $-$                  & $1.29_{-0.27}^{+0.28}$ & S08 \\ 
\hline
\multicolumn{12}{c}{Groups} \\
\hline
IC 1459  & $0.69_{-0.02}^{+0.02}$ & $0.06_{-0.01}^{+0.01}$ & $0.61_{-0.65}^{+1.52}$ &   $-$                  & $0.52_{-0.50}^{+1.66}$ & $0.86_{-0.85}^{+2.06}$ &   $-$                  &   $-$                  &   $-$                  &   $-$                  & A03 \\ 
NGC 1316  & $0.79_{-0.01}^{+0.01}$ & $0.14_{-0.01}^{+0.01}$ & $1.04_{-0.29}^{+0.51}$ & $1.17_{-0.37}^{+0.63}$ & $1.11_{-0.34}^{+0.62}$ & $0.62_{-0.23}^{+0.39}$ & $0.65_{-0.24}^{+0.41}$ &   $-$                  &   $-$                  &   $-$                  & K14 \\ 
NGC 1395  & $0.83_{-0.05}^{+0.05}$ & $0.09_{-0.03}^{+0.07}$ &   $-$                  &   $-$                  &   $-$                  &   $-$                  &   $-$                  &   $-$                  &   $-$                  &   $-$                  & S13 \\ 
NGC 1407  & $0.79_{-0.01}^{+0.01}$ & $0.17_{-0.03}^{+0.02}$ & $0.32_{-0.13}^{+0.11}$ &   $-$                  & $1.09_{-0.14}^{+0.14}$ & $1.18_{-0.16}^{+0.18}$ & $2.16_{-0.66}^{+0.66}$ &   $-$                  &   $-$                  & $3.44_{-0.82}^{+1.08}$ & H06 \\ 
NGC 3115  & $0.74_{-0.04}^{+0.02}$ & $0.04_{-0.02}^{+0.01}$ &   $-$                  &   $-$                  &   $-$                  &   $-$                  &   $-$                  &   $-$                  &   $-$                  &   $-$                  &  \\ 
NGC 3379  & $0.78_{-0.12}^{+0.09}$ & $0.01_{-0.00}^{+0.01}$ &   $-$                  &   $-$                  &   $-$                  &   $-$                  &   $-$                  &   $-$                  &   $-$                  &   $-$                  &  \\ 
NGC 3607  & $0.70_{-0.24}^{+0.20}$ & $0.12_{-0.11}^{+0.12}$ &   $-$                  &   $-$                  & $0.62_{-0.38}^{+0.38}$ &   $-$                  &   $-$                  &   $-$                  &   $-$                  &   $-$                  & H06 \\ 
NGC 3923  & $0.63_{-0.01}^{+0.01}$ & $0.15_{-0.01}^{+0.01}$ & $0.38_{-0.08}^{+0.11}$ &   $-$                  & $0.94_{-0.20}^{+0.32}$ & $0.85_{-0.21}^{+0.31}$ & $0.34_{-0.11}^{+0.62}$ &   $-$                  &   $-$                  &   $-$                  & J09 \\ 
NGC 4125  & $0.62_{-0.04}^{+0.03}$ & $0.07_{-0.01}^{+0.02}$ & $0.73_{-0.19}^{+0.31}$ & $0.88_{-0.22}^{+0.37}$ & $0.82_{-0.22}^{+0.37}$ & $0.73_{-0.35}^{+0.52}$ & $0.77_{-0.37}^{+0.55}$ &   $-$                  &   $-$                  &   $-$                  & K14 \\ 
NGC 4636 P1 & $0.66_{-0.00}^{+0.00}$ & $0.22_{-0.00}^{+0.01}$ & $0.42_{-0.02}^{+0.03}$ & $0.74_{-0.06}^{+0.05}$ & $0.77_{-0.05}^{+0.05}$ & $0.98_{-0.06}^{+0.07}$ & $0.85_{-0.13}^{+0.13}$ &   $-$                  &   $-$                  & $0.31_{-0.13}^{+0.29}$ & J09 \\ 
NGC 4636 P2 & $0.67_{-0.01}^{+0.01}$ & $0.32_{-0.03}^{+0.02}$ & $0.42_{-0.02}^{+0.03}$ & $0.74_{-0.06}^{+0.05}$ & $0.77_{-0.05}^{+0.05}$ & $0.98_{-0.06}^{+0.07}$ & $0.85_{-0.13}^{+0.13}$ &   $-$                  &   $-$                  & $0.31_{-0.13}^{+0.29}$ & J09 \\ 
NGC 4649 P1 & $0.88_{-0.00}^{+0.00}$ & $0.31_{-0.01}^{+0.01}$ & $0.39_{-0.04}^{+0.05}$ & $0.40_{-0.09}^{+0.04}$ & $0.98_{-0.07}^{+0.07}$ & $0.94_{-0.06}^{+0.07}$ & $0.99_{-0.11}^{+0.11}$ &   $-$                  &   $-$                  & $2.12_{-0.20}^{+0.14}$ & J09 \\ 
NGC 4649 P2 & $0.88_{-0.00}^{+0.00}$ & $0.31_{-0.01}^{+0.01}$ & $0.39_{-0.04}^{+0.05}$ & $0.40_{-0.09}^{+0.04}$ & $0.98_{-0.07}^{+0.07}$ & $0.94_{-0.06}^{+0.07}$ & $0.99_{-0.11}^{+0.11}$ &   $-$                  &   $-$                  & $2.12_{-0.20}^{+0.14}$ & J09 \\ 
NGC 5846  & $0.75_{-0.01}^{+0.01}$ & $0.22_{-0.01}^{+0.02}$ & $0.18_{-0.07}^{+0.07}$ & $0.56_{-0.22}^{+0.23}$ & $0.74_{-0.07}^{+0.07}$ & $0.76_{-0.07}^{+0.07}$ & $1.12_{-0.23}^{+0.23}$ &   $-$                  &   $-$                  & $2.00_{-0.72}^{+1.15}$ & H06 \\ 
\hline
\multicolumn{12}{c}{Ellipticals} \\
\hline
NGC 1404  & $0.74_{-0.00}^{+0.00}$ & $0.27_{-0.01}^{+0.01}$ & $0.85_{-0.22}^{+0.31}$ & $0.85_{-0.21}^{+0.32}$ & $0.69_{-0.16}^{+0.26}$ & $1.19_{-0.29}^{+0.47}$ & $1.26_{-0.31}^{+0.50}$ &   $-$                  &   $-$                  &   $-$                  & K14 \\ 
NGC 1549  & $0.38_{-0.04}^{+0.07}$ & $0.12_{-0.08}^{+1.04}$ &   $-$                  &   $-$                  &   $-$                  &   $-$                  &   $-$                  &   $-$                  &   $-$                  &   $-$                  &  \\ 
NGC 4374  & $0.69_{-0.00}^{+0.00}$ & $0.13_{-0.01}^{+0.01}$ & $0.66_{-0.55}^{+0.21}$ &   $-$                  & $0.76_{-0.67}^{+0.21}$ & $1.64_{-1.10}^{+0.30}$ &   $-$                  &   $-$                  &   $-$                  &   $-$                  & A03 \\ 
NGC 4406  & $0.82_{-0.01}^{+0.01}$ & $0.18_{-0.02}^{+0.03}$ & $0.62_{-0.09}^{+0.11}$ & $0.60_{-0.13}^{+0.16}$ & $0.77_{-0.09}^{+0.10}$ & $0.75_{-0.09}^{+0.11}$ &   $-$                  &   $-$                  &   $-$                  & $0.54_{-0.26}^{+0.35}$ & J09 \\ 
NGC 4472  & $0.84_{-0.00}^{+0.00}$ & $0.26_{-0.01}^{+0.01}$ & $0.47_{-0.03}^{+0.04}$ & $0.54_{-0.04}^{+0.06}$ & $0.82_{-0.04}^{+0.05}$ & $0.79_{-0.04}^{+0.04}$ & $0.82_{-0.07}^{+0.07}$ &   $-$                  &   $-$                  & $1.22_{-0.14}^{+0.08}$ & J09 \\ 
NGC 4494  & $0.88_{-0.20}^{+0.19}$ & $0.03_{-0.02}^{+0.05}$ &   $-$                  &   $-$                  &   $-$                  &   $-$                  &   $-$                  &   $-$                  &   $-$                  &   $-$                  & H06 \\ 
NGC 4552  & $0.75_{-0.01}^{+0.01}$ & $0.12_{-0.01}^{+0.01}$ & $0.71_{-0.14}^{+0.19}$ &   $-$                  & $1.12_{-0.17}^{+0.31}$ & $0.99_{-0.18}^{+0.30}$ &   $-$                  &   $-$                  &   $-$                  &   $-$                  & J09 \\ 
NGC 4621  & $0.63_{-0.63}^{+0.00}$ & $1.24_{-1.24}^{+1.24}$ &   $-$                  &   $-$                  &   $-$                  &   $-$                  &   $-$                  &   $-$                  &   $-$                  &   $-$                  &  \\ 
NGC 5102  & $0.50_{-0.10}^{+0.10}$ & $0.01_{-0.01}^{+0.01}$ &   $-$                  &   $-$                  &   $-$                  &   $-$                  &   $-$                  &   $-$                  &   $-$                  &   $-$                  &  \\ 
\hline
\multicolumn{12}{c}{Averages} \\
\hline
Avg Cluster & $2.74_{-1.05}^{+1.05}$          & $0.82_{-0.24}^{+0.24}$ & $0.99_{-0.29}^{+0.29}$ & $-$ & $0.61_{-0.20}^{+0.20}$ & $0.96_{-0.30}^{+0.30}$ & $1.05_{-0.31}^{+0.31}$ & $1.06_{-0.35}^{+0.35}$ & $1.28_{-0.39}^{+0.39}$ & $1.95_{-0.84}^{+0.84}$ & M17 \\ 
Avg Group & $0.58_{-0.17}^{+0.17}$          & $0.81_{-0.20}^{+0.20}$ & $0.47_{-0.12}^{+0.12}$ & $-$  & $0.59_{-0.23}^{+0.23}$ & $0.94_{-0.30}^{+0.30}$ & $0.95_{-0.36}^{+0.36}$ & $1.06_{-0.40}^{+0.40}$ & $1.01_{-0.34}^{+0.34}$ & $-$  & M17 \\ 
\hline
\enddata
\tablenotetext{}{All abundances have been converted to \citet{2009LanB...4B..712L} units.  A small number of systems also have measurements of Ar/Fe and Ca/Fe, which are not shown here.}
\tablenotetext{*}{Literature references for $\alpha$/Fe abundances. S19: \citet{2019MNRAS.483.1701S}, J09: \citet{2009ApJ...696.2252J}, S10: \citet{2010MNRAS.405...91S}, T04: \citet{2004AnA...420..135T}, S25: \citet{2025PASJ...77S.254S}, M05: \citet{2005MNRAS.358..585M}, S08: \citet{2008PASJ...60S.333S}, A03: \citet{2003PhDT.........5A}, K14: \citet{2014ApJ...783....8K}, S13: \citet{2013ApJ...766...61S}, H06: \citet{2006ApJ...639..136H}, M17: \citet{2017AnA...603A..80M}}
\end{deluxetable*}

\subsubsection{Isobaric vs. isochoric cooling} \label{sec:mdot_ovi_comb}

From our \textsc{Cloudy} simulations, we obtain \mdotovi either assuming pure isobaric or isochoric cooling.  In reality, cooling should proceed along some thermodynamic trajectory in the $PV$ plane intermediate between these two extremes, with the relative importance of each process determined by how rapidly the cooling proceeds.  Isochoric cooling dominates in the ``fast'' regime, when the cooling time $t_{\rm cool}$ is much shorter than the sound-crossing time $t_{\rm sc}$ of the cooling region.  Isobaric cooling, then, dominates when the reverse is true: $t_{\rm cool} \gg t_{\rm sc}$.  From this, we develop a simple weighting scheme that we use to combine the isobaric (``cp'') and isochoric (``cd'') cooling rates. The weights are given by:
\begin{equation}
    w_{\rm cp} = \frac{t_{\rm cool}}{t_{\rm cool}+t_{\rm sc}}~, ~~~~ w_{\rm cd} = \frac{t_{\rm sc}}{t_{\rm cool}+t_{\rm sc}}
\end{equation}
It can be easily validated that these weights provide the correct asymptotic behavior in the two extreme cases presented above, while the two weights are equal in the intermediate case where $t_{\rm cool}=t_{\rm sc}$. These then determine the combined cooling rate:
\begin{equation}
    \dot{M}_{\rm O\,VI,comb} = w_{\rm cp}\dot{M}_{\rm O\,VI,cp} + w_{\rm cd}\dot{M}_{\rm O\,VI,cd}
\end{equation}
To calculate the weights, $t_{\rm cool}$ is extracted directly from the \textsc{Cloudy} simulations, and $t_{\rm sc}$ is calculated from the sound speed of the hot gas, $c_s$:
\begin{equation}
    c_s^2 = \gamma_{\rm ad}\frac{kT}{\mu m_p}
\end{equation}
where $\gamma_{\rm ad}=5/3$ is the adiabatic index for a non-relativistic thermal plasma.  To convert to a sound-crossing time, we assume a cooling region size of $\sim 1$ kpc. This choice is motivated by observational constraints on the sizes of cooling filaments in groups and clusters.  \citet{2025NatAs...9..449O} have mapped X-ray-bright filaments in a sample of nearby clusters, finding them to be unresolved in the X-ray, implying widths of $\lesssim 3$ kpc in Perseus and $\lesssim 1$ kpc in Centaurus.  Meanwhile, the optical H$\alpha$ emitting filaments in Centaurus have widths of $\sim 60$ pc \citep{2016MNRAS.461..922F}, which corresponds to $\sim 0.6$ kpc when scaling from $10^4$ K to $10^7$ K (assuming pressure equilibrium).

\subsection{X-ray spectroscopic cooling rates (\mdotx, \mdothcf)} \label{sec:mdot_x}

We do not calculate new X-ray spectroscopic cooling rates from scratch for this work.  Instead, we obtain them from \citet{2019MNRAS.485.1757L} (using what they define as $\dot{M}_{\rm 1\,cie+1\,cf}$), which has significant overlap with our sample (18/29 systems).  The remaining 11 systems do not have adequate data to measure an X-ray spectroscopic cooling rate.  In parallel, we also obtain X-ray spectroscopic cooling rates that have been calculated using hidden cooling flow models from \citet{2022MNRAS.515.3336F, 2023MNRAS.521.1794F, 2023MNRAS.524..716F, 2024MNRAS.535.2173F, 2024MNRAS.535.2697I, 2025arXiv250814785F}, which we label \mdothcf.  These measurements cover 15/29 systems, with those covered being largely the same subset of our sample as the unabsorbed measurements.

\subsection{Classical cooling rates (\mdotmax, \mdotmaxap)} \label{sec:mdot_c}

We calculate a new set of classical/maximal cooling rates \mdotmax using archival data from the \textit{Chandra X-ray Observatory}.  To do so, we require deprojected density, temperature, and metallicity profiles for each system. \textit{Chandra}-ACIS imaging data are retrieved and reprocessed following standard pipelines with CIAO (v4.18.0) and CALDB (v4.12.3).  Then, we follow the techniques and methodology of \citet{2006ApJ...640..691V} to calculate the deprojected thermodynamic profiles.  The 3D density profile is modeled using \citet{2006ApJ...640..691V}'s equation (3), and the 3D temperature profile is modeled using their equation (6).  We assume $n_p=0.92n_e$ and $n_H=0.83n_e$. 

Then, using these profiles in combination with the cooling function $\Lambda(kT,Z)$ \citep{1993ApJS...88..253S}, we calculate a cooling time profile:
\begin{equation}
    t_{\rm cool}= \frac{3}{2}\frac{(n_e+n_p)kT}{n_en_H\Lambda(kT,Z)}~.
\end{equation}
The profile $t_{\rm cool}(r)$ is then used to define the cooling radius $r_{\rm cool}$, within which we integrate the cooling rate profile $\mdotmax(r)$ to measure the total classical cooling rate, $\mdotmax \equiv \mdotmax(r<r_{\rm cool})$. The cooling radius itself is defined by $t_{\rm cool}(r_{\rm cool}) \equiv 3\,{\rm Gyr}$.  We choose 3 Gyr following \citet{2018ApJ...858...45M}.

The cooling rate profile is calculated within spherical shells, which we enumerate with $i$.  Then, 
\begin{equation}
    \mdotmax(i) = \frac{L_X(i)}{h(i)}
\end{equation}
where $L_X(i)$ is the (deprojected) X-ray luminosity in shell $i$, and $h(i) \equiv (5/2)(kT(i)/\mu m_p)$ is the thermal enthalpy per unit particle-mass in shell $i$. 

In addition to \mdotmax, which truly represents an upper limit on the cooling rate within $< r_{\rm cool}$, we also calculate projected aperture-matched cooling rates, which we label \mdotmaxap.  This is done by extracting a single spectrum integrated over the full rectangular \textit{FUSE} aperture and fitting a 2-temperature \texttt{phabs(apec+apec)} model with \textsc{PyXspec} \citep{2021ascl.soft01014G}, following the methods used in Paper I.  The integrated cooling rates are then calculated using the cooler temperature \texttt{apec} component, using the same formula as above for \mdotmax.


\section{Results} \label{sec:res}

\subsection{\ovi fluxes} \label{sec:res_ovi}

\begin{deluxetable*}{lllllllllll}
\def\arraystretch{1.15}
\tabletypesize{\footnotesize}
\tablecaption{All \ovi fluxes and kinematics, along with relevant attenuation parameters.}
\label{tab:ovi}
\tablehead{
    \colhead{Name} & \colhead{$E(B-V)_{\rm gal}$} & \colhead{$E(B-V)_{\rm neb}$} & \colhead{$E(B-V)_{\rm diff}$} & \colhead{$\log({\rm H_{2,gal}}/{\rm cm}^2)$} & \colhead{$\log({\rm H_{2,int}}/{\rm cm}^2)$} & \colhead{$F_{\rm O\,VI,c1}/10^{-15}$\tablenotemark{$*$}} & \colhead{$F_{\rm O\,VI,c2}/10^{-15}$\tablenotemark{$\dagger$}} & \colhead{$v_{\rm los}$} & \colhead{FWHM} & \colhead{$D$\tablenotemark{$\ddagger$}} \\
    \colhead{} & \colhead{mag} & \colhead{mag} & \colhead{mag} & \colhead{} & \colhead{} & \colhead{$\ergscm$} & \colhead{$\ergscm$}& \colhead{\kms} & \colhead{\kms} & \colhead{Mpc}
}
\startdata
\hline
\multicolumn{10}{c}{Clusters} \\
\hline
Perseus & $0.14_{-0.001}^{+0.001}$ & $1.9_{-0.2}^{+0.1}$ & $0.03_{-0.02}^{+0.03}$ & $19_{-1}^{+1}$ & $15.8_{-0.6}^{+0.6}$ & $23_{-6-16}^{+11+17}$ & $45_{-11-21}^{+22+22}$ & $60_{-20}^{+20}$ & $120_{-10}^{+110}$ & $69 \pm 3$\\ 
Fornax & $0.011_{-0.0}^{+0.0}$ & $0.03_{-0.02}^{+0.04}$ & $0.003_{-0.002}^{+0.002}$ & $15.09_{-0.07}^{+0.1}$ & $15.2_{-0.1}^{+0.2}$ & $11_{-3-1}^{+3+2}$ & $12_{-4-3}^{+4+3}$ & $-250_{-10}^{+50}$ & $400_{-200}^{+200}$ & $17.7 \pm 0.4$\\ 
Virgo & $0.02_{-0.001}^{+0.001}$ & $0.1_{-0.03}^{+0.03}$ & $0.017_{-0.005}^{+0.004}$ & $17.0_{-0.3}^{+0.3}$ & $15.4_{-0.2}^{+0.2}$ & $17_{-4-3}^{+4+3}$ & $25_{-6-5}^{+6+5}$ & $-130_{-80}^{+70}$ & $300_{-100}^{+100}$ & $16.8 \pm 0.3$\\ 
Abell 1795 P1 & $0.011_{-0.0}^{+0.0}$ & $0.4_{-0.3}^{+0.4}$ & $0.05_{-0.03}^{+0.04}$ & $19_{-2}^{+2}$ & $20_{-2}^{+1}$ & $<9$ & $<10$ & $-$ & $-$ & $284$\\ 
Abell 1795 P2 & $0.011_{-0.0}^{+0.0}$ & $0.4_{-0.3}^{+0.2}$ & $0.04_{-0.02}^{+0.02}$ & $16.2_{-0.6}^{+0.6}$ & $16.0_{-0.4}^{+0.4}$ & $3_{-1-1}^{+1+1}$ & $6_{-2-2}^{+2+2}$ & $0_{-200}^{+200}$ & $600_{-400}^{+0}$ & $284$\\ 
Abell 2029 & $0.034_{-0.001}^{+0.001}$ & $0.01_{-0.01}^{+0.02}$ & $0.002_{-0.001}^{+0.003}$ & $19_{-2}^{+2}$ & $17_{-1}^{+2}$ & $<7$ & $<7$ & $-$ & $-$ & $355$\\ 
Abell 2597 & $0.026_{-0.001}^{+0.001}$ & $0.3_{-0.2}^{+0.2}$ & $0.02_{-0.01}^{+0.01}$ & $17_{-1}^{+2}$ & $17.1_{-0.9}^{+1.0}$ & $3_{-1-1}^{+1+1}$ & $5_{-2-2}^{+2+2}$ & $0_{-100}^{+100}$ & $500_{-300}^{+100}$ & $378$\\ 
Abell 3112 & $0.01_{-0.0}^{+0.0}$ & $0.17_{-0.07}^{+0.06}$ & $0.03_{-0.01}^{+0.01}$ & $18_{-2}^{+2}$ & $19_{-2}^{+1}$ & $<5$ & $<6$ & $-$ & $-$ & $345$\\ 
AWM 7 & $0.096_{-0.004}^{+0.004}$ & $0.03_{-0.02}^{+0.03}$ & $0.005_{-0.003}^{+0.005}$ & $18_{-2}^{+2}$ & $19_{-3}^{+2}$ & $<18$ & $<18$ & $-$ & $-$ & $76$\\ 
\hline
\multicolumn{10}{c}{Groups} \\
\hline
IC 1459 & $0.014_{-0.001}^{+0.001}$ & $0.05_{-0.03}^{+0.03}$ & $0.009_{-0.005}^{+0.006}$ & $15.6_{-0.4}^{+0.5}$ & $16.1_{-0.5}^{+0.5}$ & $13_{-4-2}^{+4+2}$ & $16_{-5-4}^{+5+4}$ & $-20_{-80}^{+80}$ & $400_{-100}^{+100}$ & $26 \pm 2$\\ 
NGC 1316 & $0.018_{-0.0}^{+0.0}$ & $0.4_{-0.3}^{+0.4}$ & $0.021_{-0.01}^{+0.007}$ & $17.2_{-0.9}^{+0.7}$ & $17.2_{-0.6}^{+0.5}$ & $14_{-2-3}^{+2+4}$ & $23_{-4-10}^{+4+8}$ & $-30_{-40}^{+40}$ & $360_{-90}^{+70}$ & $19.2 \pm 0.6$\\ 
NGC 1395 & $0.019_{-0.0}^{+0.0}$ & $0.06_{-0.04}^{+0.07}$ & $0.001_{-0.001}^{+0.002}$ & $16.2_{-0.2}^{+0.2}$ & $15.4_{-0.2}^{+0.3}$ & $<4$ & $<6$ & $-$ & $-$ & $23 \pm 1$\\ 
NGC 1407 & $0.058_{-0.001}^{+0.001}$ & $0.07_{-0.05}^{+0.08}$ & $0.003_{-0.002}^{+0.004}$ & $19.31_{-0.07}^{+0.08}$ & $16.1_{-0.4}^{+0.4}$ & $<8$ & $<9$ & $-$ & $-$ & $24 \pm 2$\\ 
NGC 3115 & $0.04_{-0.001}^{+0.001}$ & $0.2_{-0.1}^{+0.2}$ & $0.001_{-0.001}^{+0.002}$ & $19.1_{-0.06}^{+0.05}$ & $16.1_{-0.3}^{+0.3}$ & $6_{-2-2}^{+2+2}$ & $7_{-2-3}^{+3+3}$ & $80_{-60}^{+50}$ & $200_{-100}^{+100}$ & $10.2 \pm 0.5$\\ 
NGC 3379 & $0.021_{-0.001}^{+0.001}$ & $0.007_{-0.005}^{+0.013}$ & $0.001_{-0.001}^{+0.002}$ & $15.9_{-0.6}^{+0.8}$ & $16.2_{-0.6}^{+0.7}$ & $<16$ & $<17$ & $-$ & $-$ & $11.1 \pm 0.3$\\ 
NGC 3607 & $0.018_{-0.0}^{+0.0}$ & $0.2_{-0.1}^{+0.2}$ & $0.017_{-0.005}^{+0.004}$ & $15.5_{-0.3}^{+0.4}$ & $15.5_{-0.3}^{+0.3}$ & $3_{-1-1}^{+1+2}$ & $5_{-1-2}^{+1+2}$ & $70_{-30}^{+30}$ & $170_{-50}^{+60}$ & $21 \pm 1$\\ 
NGC 3923 & $0.07_{-0.002}^{+0.002}$ & $0.007_{-0.005}^{+0.008}$ & $0.001_{-0.001}^{+0.001}$ & $17_{-1}^{+1}$ & $15.7_{-0.4}^{+0.6}$ & $<21$ & $<21$ & $-$ & $-$ & $22 \pm 2$\\ 
NGC 4125 & $0.016_{-0.001}^{+0.001}$ & $0.03_{-0.02}^{+0.02}$ & $0.005_{-0.003}^{+0.003}$ & $16.0_{-0.7}^{+0.8}$ & $15.6_{-0.4}^{+0.6}$ & $4_{-1-1}^{+2+1}$ & $4_{-2-1}^{+2+1}$ & $-100_{-100}^{+100}$ & $400_{-200}^{+200}$ & $23 \pm 2$\\ 
NGC 4636 P1 & $0.024_{-0.001}^{+0.001}$ & $0.006_{-0.004}^{+0.007}$ & $0.001_{-0.001}^{+0.001}$ & $17.8_{-0.7}^{+0.4}$ & $15.6_{-0.4}^{+0.4}$ & $4_{-1-2}^{+2+2}$ & $4_{-1-2}^{+2+3}$ & $-100_{-20}^{+90}$ & $120_{-10}^{+60}$ & $16.3 \pm 0.6$\\ 
NGC 4636 P2 & $0.024_{-0.001}^{+0.001}$ & $0.03_{-0.02}^{+0.04}$ & $0.004_{-0.003}^{+0.007}$ & $16.0_{-0.6}^{+0.8}$ & $17_{-2}^{+1}$ & $<7$ & $<8$ & $-$ & $-$ & $16.3 \pm 0.6$\\ 
NGC 4649 & $0.023_{-0.001}^{+0.001}$ & $0.12_{-0.06}^{+0.09}$ & $0.003_{-0.001}^{+0.002}$ & $19.09_{-0.04}^{+0.04}$ & $15.3_{-0.1}^{+0.2}$ & $16_{-4-2}^{+4+2}$ & $20_{-5-5}^{+5+5}$ & $70_{-60}^{+40}$ & $300_{-100}^{+100}$ & $16.7 \pm 0.4$\\ 
NGC 5846 & $0.048_{-0.001}^{+0.001}$ & $0.011_{-0.007}^{+0.01}$ & $0.002_{-0.001}^{+0.002}$ & $16.1_{-0.6}^{+0.6}$ & $16.1_{-0.6}^{+0.7}$ & $5_{-2-3}^{+2+2}$ & $5_{-2-3}^{+2+3}$ & $20_{-60}^{+20}$ & $120_{-10}^{+80}$ & $29 \pm 2$\\ 
\hline
\multicolumn{10}{c}{Ellipticals} \\
\hline
NGC 1404 & $0.01_{-0.0}^{+0.0}$ & $0.011_{-0.007}^{+0.007}$ & $0.002_{-0.001}^{+0.001}$ & $15.6_{-0.4}^{+0.5}$ & $15.3_{-0.2}^{+0.3}$ & $<6$ & $<7$ & $-$ & $-$ & $19.3 \pm 0.4$\\ 
NGC 1549 & $0.01_{-0.0}^{+0.0}$ & $0.2_{-0.1}^{+0.2}$ & $0.001_{-0.001}^{+0.001}$ & $15.8_{-0.4}^{+0.4}$ & $15.8_{-0.4}^{+0.4}$ & $4_{-1-2}^{+2+1}$ & $5_{-2-2}^{+2+2}$ & $-70_{-60}^{+50}$ & $170_{-50}^{+60}$ & $16.6 \pm 1.0$\\ 
NGC 4374 & $0.034_{-0.002}^{+0.002}$ & $0.03_{-0.02}^{+0.02}$ & $0.004_{-0.003}^{+0.004}$ & $17.9_{-0.8}^{+0.5}$ & $15.8_{-0.6}^{+0.6}$ & $<18$ & $<20$ & $-$ & $-$ & $17.0 \pm 0.4$\\ 
NGC 4406 & $0.025_{-0.0}^{+0.0}$ & $0.01_{-0.01}^{+0.02}$ & $0.002_{-0.002}^{+0.003}$ & $18_{-2}^{+1}$ & $17_{-1}^{+1}$ & $5_{-2-3}^{+2+3}$ & $5_{-2-3}^{+2+4}$ & $-40_{-30}^{+20}$ & $120_{-10}^{+20}$ & $16.4 \pm 0.4$\\ 
NGC 4472 & $0.019_{-0.0}^{+0.0}$ & $0.1_{-0.06}^{+0.07}$ & $0.002_{-0.001}^{+0.002}$ & $16.2_{-0.6}^{+0.4}$ & $16.0_{-0.4}^{+0.5}$ & $6_{-2-3}^{+2+1}$ & $7_{-2-4}^{+3+3}$ & $200_{-30}^{+30}$ & $120_{-10}^{+30}$ & $16.1 \pm 0.3$\\ 
NGC 4494 & $0.018_{-0.001}^{+0.001}$ & $0.04_{-0.02}^{+0.03}$ & $0.006_{-0.004}^{+0.005}$ & $15.8_{-0.5}^{+0.6}$ & $15.8_{-0.5}^{+0.6}$ & $2_{-1-2}^{+1+1}$ & $3_{-1-2}^{+1+2}$ & $-40_{-40}^{+30}$ & $160_{-40}^{+70}$ & $13.4 \pm 0.5$\\ 
NGC 4552 & $0.035_{-0.002}^{+0.002}$ & $0.04_{-0.03}^{+0.05}$ & $0.002_{-0.002}^{+0.003}$ & $19.45_{-0.05}^{+0.05}$ & $15.6_{-0.3}^{+0.4}$ & $7_{-2-1}^{+2+1}$ & $8_{-2-2}^{+3+2}$ & $240_{-60}^{+10}$ & $240_{-100}^{+80}$ & $16.5 \pm 0.5$\\ 
NGC 4621 & $0.027_{-0.001}^{+0.001}$ & $0.4_{-0.2}^{+0.4}$ & $0.002_{-0.001}^{+0.002}$ & $19.4_{-0.3}^{+0.2}$ & $16.5_{-0.9}^{+1.0}$ & $<11$ & $<13$ & $-$ & $-$ & $16.0 \pm 0.4$\\ 
NGC 5102 & $0.047_{-0.0}^{+0.0}$ & $0.04_{-0.02}^{+0.03}$ & $0.01_{-0.006}^{+0.008}$ & $18.5_{-0.2}^{+0.2}$ & $19.0_{-0.2}^{+0.2}$ & $18_{-6-4}^{+7+4}$ & $22_{-7-7}^{+9+7}$ & $210_{-30}^{+40}$ & $120_{-10}^{+60}$ & $10 \pm 4$\\ 
\hline
\enddata
\tablenotetext{}{All reported \ovi fluxes are for the brighter $\lambda$1032\angstrom line.  The dimmer $\lambda$1038\angstrom line is enforced to be half of this flux in our model.  The first pair of uncertainties is the statistical uncertainty $\sigma_{\rm stat}$, while the second pair is the systematic uncertainty $\sigma_{\rm syst}$.}
\tablenotetext{}{All measurements are made within the \textit{FUSE} LWRS aperture, which has an angular size of $30'' \times 30''$.}
\tablenotetext{*}{The flux after correcting for galactic metal absorption, galactic H$_2$ absorption, intrinsic H$_2$ absorption, and galactic dust extinction.}
\tablenotetext{\dagger}{The flux after correcting for galactic metal absorption, galactic H$_2$ absorption, intrinsic H$_2$ absorption, galactic dust extinction, \textit{and} intrinsic dust extinction.}
\tablenotetext{\ddagger}{The ``best'' distance for each system.  For nearby systems, these are redshift-independent distances obtained from \href{https://ned.ipac.caltech.edu/}{NED}.  For further systems, we use luminosity distances calculated from our standard cosmology.}
\end{deluxetable*}

Beginning with our \ovi emission models, we present the recovered line fluxes and kinematics, along with all relevant attenuation parameters (pulled from Paper I), in Table \ref{tab:ovi}. We present two measurements of the \ovi flux, with both measuring the flux of the brighter $\lambda$1032$\angstrom$ line. $F_{\rm O\,VI,c1}$ is the flux after correcting for galactic metal absorption, galactic H$_2$ absorption, intrinsic H$_2$ absorption, and galactic dust extinction.  $F_{\rm O\,VI,c2}$ is the flux after correcting for all of the same attenuation effects as $F_{\rm O\,VI,c1}$, in addition to intrinsic dust extinction.  Regarding kinematics, the line-of-sight velocities $v_{\rm los}$ are measured relative to  systemic redshifts from NED\footnote{\url{https://ned.ipac.caltech.edu/}}, reported in Table 1 of Paper I.  The FWHMs have not taken the instrumental resolution into account, but \textit{FUSE} has a high spectral resolution ($\sim 15$ \kms) compared to the typical velocity widths seen in lines, so this effect is negligible.

\begin{figure}
    \centering
    \includegraphics[width=\linewidth]{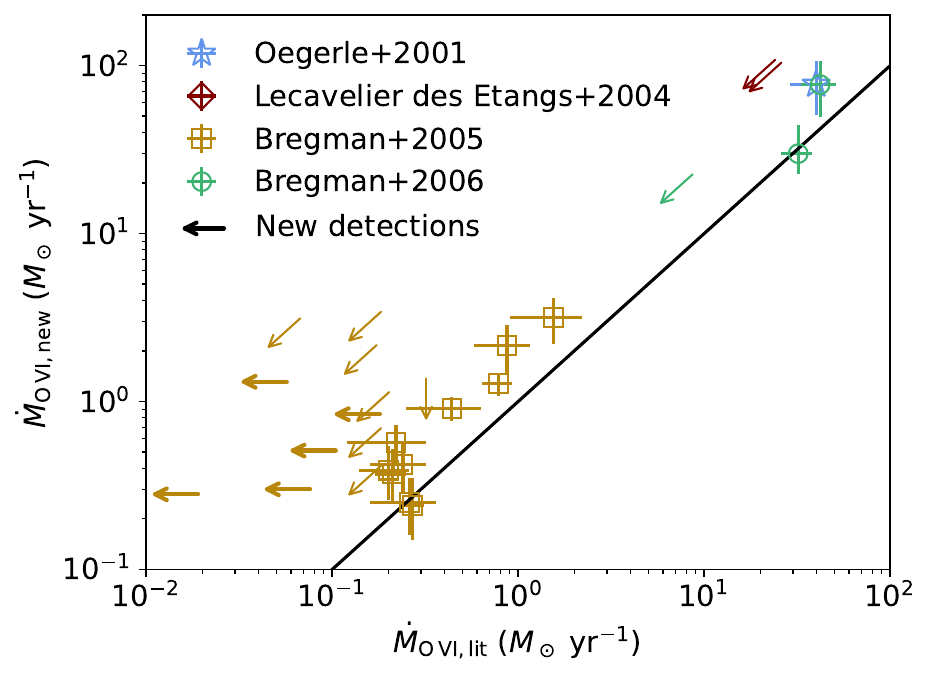}
    \caption{A comparison between values of the \ovi cooling rate pulled from the literature, $\dot{M}_{\rm O\,VI,lit}$, and our newly measured values $\dot{M}_{\rm O\,VI,new}$.  Previous studies are shown in different shapes and colors, defined in the legend.  The error bars show only the \textit{statistical} uncertainties $\sigma_{\rm stat}$, since previous studies did not explicitly provide systematic uncertainties.  Upper limits are annotated with arrows, with bold arrows indicating new detections in this study.  The solid line indicates where the two measurements are equal.  Our new measurements are almost unanimously higher than the literature values.}
    \label{fig:ovi_compare}
\end{figure}

We detect significant \ovi emission above our $2.5\sigma_{\rm stat}$ threshold in 19 out of our 29 systems (66\%), and for the remainder we report $3\sigma_{\rm stat}$ upper limits.  We compare these \ovi measurements to previous analyses of these data in the literature in Figure \ref{fig:ovi_compare} (converted to $\dot{M}$).  We share 13 detections in common with \citet{2001ApJ...560..187O, 2004A&A...421..503L, 2005ApJ...635.1031B, 2006ApJ...642..746B}, with the remaining 6 being newly detected.  Conversely, we do not find significant \ovi emission in 1 system which was previously reported as a detection (NGC 4374) in \citet{2005ApJ...635.1031B}.  It's likely that the updates to the \textit{FUSE} reduction pipeline, in combination with our updated methodology for fitting the stellar continuum and \ovi lines, are responsible for these differences.  

For the 13 systems where we share detections, our recovered line fluxes ($F_{\rm O\,VI,c1}$) are either in agreement within $\pm 1\sigma_{\rm syst}$ (8/13) or higher (5/13) than previously reported.  The marginally higher fluxes are likely a result of our careful treatment of both Galactic and intrinsic attenuation effects.  

The results of our stacking analysis on the non-detections are not presented in Table \ref{tab:ovi}, since the attenuation effects are no longer modeled, and the \ovi doublet's relative fluxes are no longer tied.  We find statistically significant \ovi emission in the stacked spectrum of all nondetections at the $3\sigma_{\rm stat}$ level, with luminosities of $6 \pm 2 \times 10^{37}~\ergs$ and $6 \pm 3 \times 10^{37}~\ergs$ for the $\lambda1032\angstrom$ and $\lambda1038\angstrom$ lines respectively, a line-of-sight velocity of $v_{\rm los} = 250_{-20}^{+0}$ \kms, and a FWHM of $270_{-110}^{+80}$ \kms.  

Interestingly, the ratio of the \ovi lines in the stacked spectrum deviates strongly from the expected 2:1 ratio that we observe and enforce in the individually detected systems.  The brighter $\lambda1032\angstrom$ line is strongly suppressed, resulting in a ratio closer to 1:1.  It is possible that this could be due to unmodeled H$_2$ absorption which preferentially affects the $\lambda1032\angstrom$ line.  The stacked systems do show evidence for both Galactic and intrinsic H$_2$ absorption, but we do not attempt to remove these effects before stacking due to numerical instabilities from dividing out factors close to 0.  This would also explain why the stacked \ovi line has a non-zero velocity shift (if its blue wing were absorbed more than its red wing).  Independently, resonant scattering of the $\lambda1032\angstrom$ line could also drive the ratio towards 1:1, and this phenomenon has been observed in the hot X-ray halos of galaxy clusters \citep{2011AstL...37..141Z}, creating apparent ``abundance drops'' in the core relative to the outskirts \citep{2025A&A...694A.149S}.

Looking at the stacked spectra split by halo, we find that \ovi emission is most significant in the groups, with a similar average luminosity and statistical significance as the whole stack, whereas the clusters and ellipticals both fall below our $2.5\sigma_{\rm stat}$ threshold.

\subsection{\ovi velocities}

For the 9 systems with measured FWHMs in \citet{2005ApJ...635.1031B}, our FWHM measurements are generally in good agreement, with 7/9 within $\pm 1\sigma$, and the remaining 2/9 narrower. We note that we have not corrected our FWHM measurements for instrumental broadening, because \textit{FUSE}'s instrumental resolution of $\sim 15$ \kms would lead to a negligible correction ($\lesssim 1$ \kms) on the velocities we measure.  No substantial literature comparison is available for the line-of-sight velocities, but we note that only a small number of systems (4) are significantly inconsistent with an offset of more than 1 binned pixel ($\sim 60~\kms$).

\begin{figure*}
    \centering
    \includegraphics[width=\columnwidth]{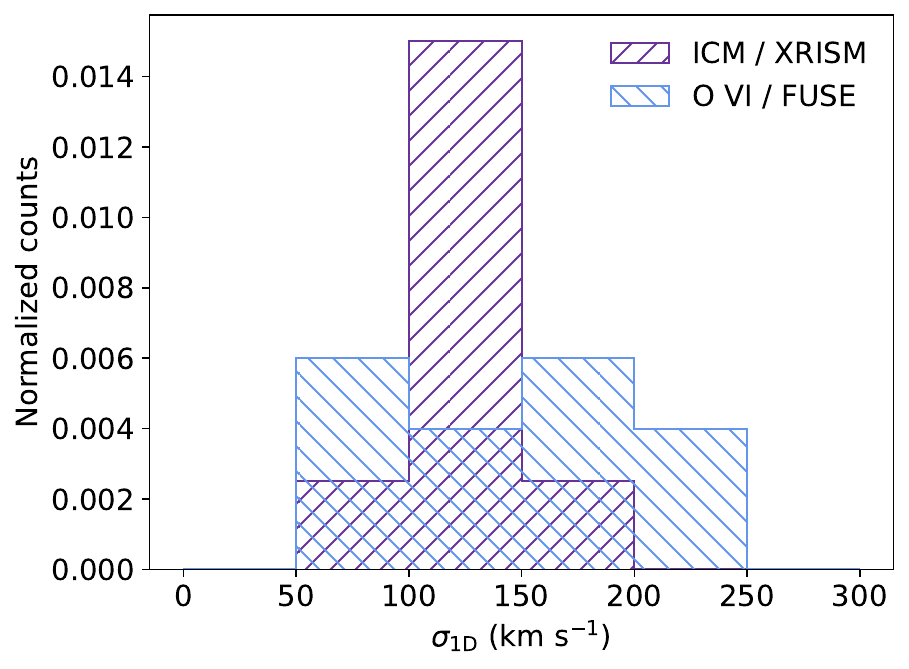}
    \includegraphics[width=\columnwidth]{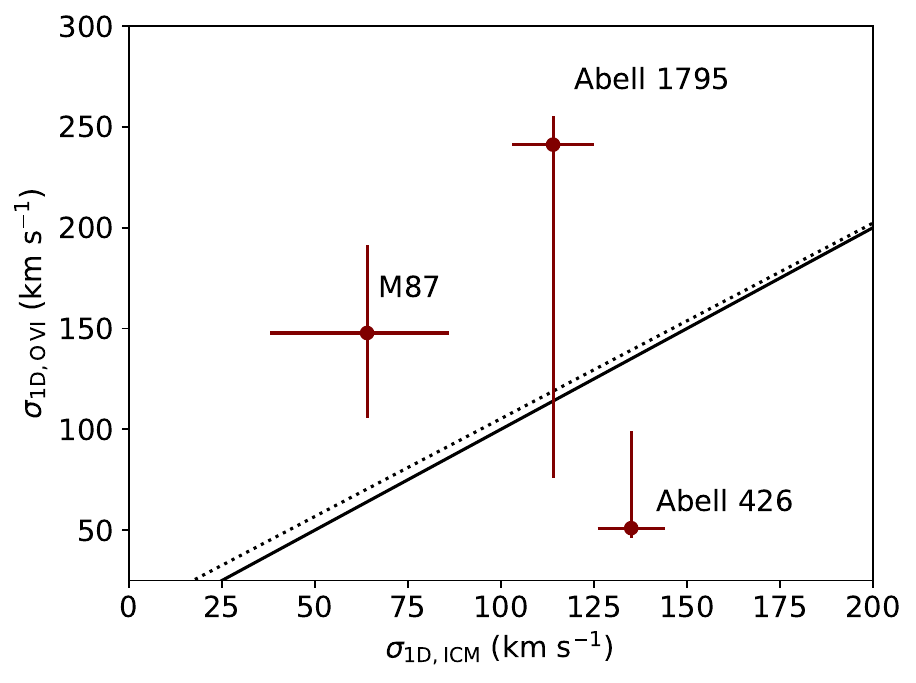}
    \caption{Left: Histograms of ICM velocity dispersions in cool core clusters from \textit{XRISM} (purple, forward-hatched) and \ovi velocity dispersions from \textit{FUSE} (blue, backward-hatched), binned to 50 \kms and normalized.  Right: The 3 systems which have both measurements available are plotted against each other. The solid line shows where the two velocity dispersions are equal, while the dotted line shows the best fit from \citet{2018ApJ...854..167G}, $\sigma_{\rm warm} = 0.97\sigma_{\rm hot}+8.3$ \kms.}
    \label{fig:vdisp_compare}
\end{figure*}

The systems, on average, show a modest level of dispersion in the $10^{5.5}~{\rm K}$ phase, with $\sigma_{\rm 1D} = {\rm FWHM}/2.355 \approx 110({\rm mean}) \pm 57({\rm stdev})~\kms$ (projected).  This is similar to the turbulent velocities in the hot ICM of cool core clusters, which have begun being measured within the past couple of years with \textit{XRISM}.  We compare our projected \ovi velocity dispersions against these ICM values in Figure \ref{fig:vdisp_compare}, using only our BCGs to ensure mass-matching is as close as possible.  The \textit{XRISM} values shown in the histogram (left panel) are for Abell 2029 \citep{2025PASJ...77S.242X}, Abell 1795 \citep{2026arXiv260608097S}, Centaurus \citep{2025Natur.638..365X}, Ophiuchus \citep{2025PASJ...77S.270F}, Perseus \citep{2026Natur.650..309T}, Hydra A \citep{2025arXiv250501494R}, PKS0745-191 \citep{2026PASJ...78..911T}, and M87 \citep{2026ApJ...998..210X}. The centroids are similar, but our \ovi measurements appear to have a broader distribution, which may be an artifact of a larger measurement error, rather than an intrinsic difference (our average error on $\sigma_{\rm 1D}$ is $\pm 41$ \kms, while the XRISM measurements average $\pm 15$ \kms).  There are 3 individual systems which have measurements of both velocities (right panel), which show a broad consistency, with a scatter around the 1:1 line.

There are a number of systematic effects that may cause our $\sigma_{\rm 1D, O\,VI}$ measurements to be biased low or high relative to $\sigma_{\rm 1D, ICM}$.  Perseus (Abell 426) is a prime example of the former effect---the brighter \ovi line in this system is heavily impacted by Galactic H$_2$ absorption, which alters its shape and width.  The kinematic measurements, then, rely heavily on the fainter \ovi line (see Appendix \ref{sec:appendix} for plots of each spectrum).  This effect results in the very asymmetric error bars that are observed.  Abell 1795, on the other hand, exemplifies the latter effect---the \ovi lines do not have a substantial contrast against the continuum in this system, resulting in line widths which may be broader than the true values.  Like Perseus, this is reflected in the highly asymmetric error bars.  

It is possible that the difference between $\sigma_{\rm 1D, O\,VI}$ and $\sigma_{\rm 1D, ICM}$ is physical, caused by differences in the \ovi's effective emitting depth, or by sampling different ensembles of \ovi-emitting clouds with varying kinematics \citep{2018ApJ...854..167G}.  However, the observed deviations are small enough, and the sample size is small enough, that it is difficult to pin down any one physical interpretation.  Based on the scatter, more and deeper measurements would be required to make any confident claims.  Therefore, we leave this to future work.

\subsection{Cooling rates} \label{sec:res_mdot}

\setlength{\tabcolsep}{4pt}
\begin{deluxetable*}{lllllllllll}
\def\arraystretch{1.15}
\tabletypesize{\footnotesize}
\tablecaption{All cooling rate measurements and upper limits.}
\label{tab:cooling_rates}
\tablehead{
    \colhead{Name} & \colhead{SFR$_{100}$\tablenotemark{*}} & \colhead{SFR$_{10}$\tablenotemark{*}} & \colhead{$\dot{M}_{\rm O\,VI,cp}$} & \colhead{$\dot{M}_{\rm O\,VI,cd}$} & \colhead{$\dot{M}_{\rm O\,VI,comb}$} & \colhead{$\dot{M}_{\rm O\,VI,snc}$\tablenotemark{$\dagger$}} & \colhead{\mdotx} & \colhead{$\dot{M}_{\rm X,spec,hcf}$} & \colhead{\mdotmaxap} & \colhead{\mdotmax}
}
\startdata
\hline
\multicolumn{11}{c}{Clusters} \\
\hline
Perseus & $34_{-8}^{+10}$ & $31_{-8}^{+9}$ & $30_{-7-16}^{+15+16}$ & $49_{-12-26}^{+24+26}$ & $30_{-7-15}^{+14+16}$ & $22_{-7-15}^{+14+15}$ & $18_{-6}^{+6}$ & $36_{-11}^{+11}$ & $24_{-6}^{+6}$ & $353_{-103}^{+71}$\\ 
Fornax & $0.21_{-0.03}^{+0.03}$ & $0.018_{-0.004}^{+0.005}$ & $0.8_{-0.2-0.2}^{+0.2+0.2}$ & $1.4_{-0.4-0.3}^{+0.4+0.4}$ & $0.8_{-0.2-0.2}^{+0.2+0.2}$ & $0.5_{-0.2-0.2}^{+0.2+0.2}$ & $0.04_{-0.01}^{+0.01}$ & $3_{-0}^{+2}$ & $0.29_{-0.07}^{+0.07}$ & $0.9_{-0.2}^{+0.4}$\\ 
Virgo & $0.04_{-0.02}^{+0.02}$ & $0.04_{-0.01}^{+0.02}$ & $0.9_{-0.1-0.2}^{+0.2+0.2}$ & $1.5_{-0.2-0.3}^{+0.3+0.3}$ & $0.9_{-0.1-0.2}^{+0.2+0.2}$ & $0.8_{-0.2-0.2}^{+0.2+0.2}$ & $0.62_{-0.03}^{+0.03}$ & $0.8_{-0.1}^{+0.1}$ & $0.9_{-0.2}^{+0.2}$ & $13_{-6}^{+3}$\\ 
Abell 1795 & $2.2_{-0.8}^{+1.1}$ & $2.0_{-0.7}^{+1.0}$ & $77_{-28-26}^{+29+27}$ & $127_{-46-44}^{+49+44}$ & $77_{-28-26}^{+29+27}$ & $75_{-28-26}^{+29+26}$ & $<22$ & $>22$ & $71_{-17}^{+17}$ & $177_{-55}^{+61}$\\ 
Abell 2029 & $<0.9$ & $<0.6$ & $<94$ & $<153$ & $<108$ & $<101$ & $<22$ & $-$ & $158_{-38}^{+38}$ & $235_{-97}^{+68}$\\ 
Abell 2597 & $2_{-1}^{+1}$ & $1_{-1}^{+1}$ & $77_{-27-25}^{+29+27}$ & $127_{-44-41}^{+47+45}$ & $77_{-27-25}^{+29+27}$ & $76_{-27-25}^{+29+27}$ & $11_{-4}^{+4}$ & $67_{-12}^{+84}$ & $185_{-44}^{+44}$ & $269_{-65}^{+67}$\\ 
Abell 3112 & $<2$ & $<0.9$ & $<98$ & $<161$ & $<112$ & $<107$ & $6_{-2}^{+2}$ & $99_{-51}^{+78}$ & $125_{-30}^{+30}$ & $152_{-37}^{+38}$\\ 
AWM 7 & $<0.2$ & $<0.2$ & $<20$ & $<33$ & $<23$ & $<21$ & $2.1_{-0.3}^{+0.3}$ & $-$ & $2.1_{-0.5}^{+0.5}$ & $4_{-1}^{+2}$\\ 
\hline
\multicolumn{11}{c}{Groups} \\
\hline
IC 1459 & $<0.1$ & $<0.1$ & $3_{-1-3}^{+1+5}$ & $5_{-2-5}^{+2+9}$ & $3_{-1-3}^{+1+5}$ & $3_{-1-3}^{+1+4}$ & $0.3_{-0.05}^{+0.06}$ & $3_{-1}^{+2}$ & $0.12_{-0.03}^{+0.03}$ & $0.16_{-0.04}^{+0.04}$\\ 
NGC 1316 & $<0.2$ & $<0.01$ & $1.3_{-0.2-0.6}^{+0.2+0.6}$ & $2.1_{-0.4-1.0}^{+0.3+1.1}$ & $1.3_{-0.2-0.6}^{+0.2+0.6}$ & $1.0_{-0.2-0.6}^{+0.2+0.6}$ & $0.26_{-0.04}^{+0.04}$ & $4_{-2}^{+2}$ & $0.14_{-0.03}^{+0.03}$ & $0.19_{-0.04}^{+0.04}$\\ 
NGC 1395 & $0.04_{-0.02}^{+0.06}$ & $0.008_{-0.006}^{+0.005}$ & $<1$ & $<2$ & $<1$ & $<0.3$ & $-$ & $-$ & $0.06_{-0.01}^{+0.01}$ & $0.15_{-0.06}^{+0.05}$\\ 
NGC 1407 & $0.04_{-0.01}^{+0.03}$ & $0.03_{-0.01}^{+0.01}$ & $<2$ & $<3$ & $<2$ & $<0.3$ & $-$ & $-$ & $0.19_{-0.05}^{+0.05}$ & $0.48_{-0.1}^{+0.1}$\\ 
NGC 3115 & $0.009_{-0.004}^{+0.007}$ & $0.007_{-0.003}^{+0.006}$ & $0.3_{-0.1-0.2}^{+0.1+0.2}$ & $0.5_{-0.1-0.3}^{+0.2+0.3}$ & $0.3_{-0.1-0.2}^{+0.1+0.2}$ & $<0.3$ & $-$ & $-$ & $0.007_{-0.002}^{+0.002}$ & $0.003_{-0.001}^{+0.001}$\\ 
NGC 3379 & $<0.04$ & $<0.007$ & $<3$ & $<5$ & $<3$ & $<1$ & $-$ & $-$ & $0.004_{-0.001}^{+0.001}$ & $0.034_{-0.007}^{+0.007}$\\ 
NGC 3607 & $0.03_{-0.01}^{+0.01}$ & $0.024_{-0.007}^{+0.012}$ & $0.6_{-0.2-0.6}^{+0.2+0.5}$ & $0.9_{-0.2-0.9}^{+0.3+0.8}$ & $0.6_{-0.1-0.6}^{+0.2+0.4}$ & $0.2_{-0.1-0.3}^{+0.2+0.3}$ & $-$ & $-$ & $0.03_{-0.007}^{+0.007}$ & $0.08_{-0.03}^{+0.03}$\\ 
NGC 3923 & $<0.03$ & $<0.02$ & $<3$ & $<5$ & $<4$ & $<2$ & $-$ & $-$ & $0.16_{-0.04}^{+0.04}$ & $0.23_{-0.05}^{+0.05}$\\ 
NGC 4125 & $<0.07$ & $<0.06$ & $0.5_{-0.2-0.2}^{+0.2+0.3}$ & $0.8_{-0.3-0.3}^{+0.4+0.5}$ & $0.5_{-0.2-0.2}^{+0.2+0.3}$ & $0.2_{-0.2-0.2}^{+0.2+0.2}$ & $-$ & $-$ & $0.06_{-0.01}^{+0.01}$ & $0.15_{-0.03}^{+0.06}$\\ 
NGC 4636 & $<0.02$ & $<0.009$ & $0.2_{-0.1-0.1}^{+0.1+0.2}$ & $0.4_{-0.2-0.2}^{+0.2+0.3}$ & $0.2_{-0.1-0.1}^{+0.1+0.2}$ & $0.2_{-0.1-0.1}^{+0.1+0.2}$ & $1.1_{-0.2}^{+0.2}$ & $0.9_{-0.1}^{+0.2}$ & $0.24_{-0.06}^{+0.06}$ & $1.2_{-0.2}^{+0.3}$\\ 
NGC 4649 & $0.18_{-0.05}^{+0.06}$ & $0.0_{-0.0}^{+0.005}$ & $1.3_{-0.2-0.3}^{+0.2+0.3}$ & $2.2_{-0.4-0.4}^{+0.4+0.4}$ & $1.3_{-0.2-0.3}^{+0.2+0.2}$ & $1.0_{-0.3-0.4}^{+0.3+0.3}$ & $<0.02$ & $3_{-1}^{+1}$ & $0.32_{-0.08}^{+0.08}$ & $0.8_{-0.2}^{+0.2}$\\ 
NGC 5846 & $<0.1$ & $<0.05$ & $2_{-1-2}^{+1+1}$ & $4_{-1-3}^{+1+2}$ & $2_{-1-1}^{+1+1}$ & $1_{-1-1}^{+1+1}$ & $0.43_{-0.06}^{+0.06}$ & $1.3_{-0.1}^{+0.2}$ & $0.5_{-0.1}^{+0.1}$ & $2.2_{-0.5}^{+0.5}$\\ 
\hline
\multicolumn{11}{c}{Ellipticals} \\
\hline
NGC 1404 & $<0.03$ & $<0.01$ & $<0.4$ & $<0.6$ & $<0.4$ & $<0.1$ & $0.6_{-0.3}^{+0.3}$ & $7.7_{-0.6}^{+0.5}$ & $0.32_{-0.08}^{+0.08}$ & $1.4_{-0.3}^{+0.3}$\\ 
NGC 1549 & $0.009_{-0.002}^{+0.005}$ & $0.007_{-0.002}^{+0.003}$ & $0.4_{-0.1-0.3}^{+0.2+1.4}$ & $0.6_{-0.2-0.5}^{+0.3+2.5}$ & $0.4_{-0.1-0.3}^{+0.2+1.4}$ & $0.2_{-0.1-0.2}^{+0.2+0.6}$ & $-$ & $-$ & $0.036_{-0.009}^{+0.009}$ & $0.06_{-0.01}^{+0.01}$\\ 
NGC 4374 & $<0.09$ & $<0.05$ & $<1$ & $<2$ & $<1$ & $<0.7$ & $0.2_{-0.1}^{+0.1}$ & $1.9_{-0.4}^{+0.1}$ & $0.14_{-0.03}^{+0.03}$ & $0.31_{-0.07}^{+0.06}$\\ 
NGC 4406 & $<0.04$ & $<0.02$ & $0.2_{-0.1-0.2}^{+0.1+0.2}$ & $0.4_{-0.2-0.3}^{+0.2+0.3}$ & $0.2_{-0.1-0.2}^{+0.1+0.2}$ & $0.1_{-0.1-0.2}^{+0.1+0.2}$ & $0.2_{-0.08}^{+0.08}$ & $-$ & $0.05_{-0.01}^{+0.01}$ & $0.09_{-0.04}^{+0.04}$\\ 
NGC 4472 & $0.07_{-0.03}^{+0.03}$ & $0.009_{-0.007}^{+0.006}$ & $0.4_{-0.1-0.2}^{+0.2+0.1}$ & $0.6_{-0.2-0.4}^{+0.2+0.2}$ & $0.4_{-0.1-0.2}^{+0.2+0.1}$ & $<0.3$ & $0.07_{-0.02}^{+0.02}$ & $1.1_{-0.0}^{+0.4}$ & $0.24_{-0.06}^{+0.06}$ & $1.0_{-0.2}^{+0.2}$\\ 
NGC 4494 & $<0.09$ & $<0.02$ & $0.3_{-0.1-0.3}^{+0.1+0.4}$ & $0.5_{-0.2-0.5}^{+0.2+0.7}$ & $0.3_{-0.1-0.3}^{+0.1+0.4}$ & $0.1_{-0.1-0.2}^{+0.1+0.2}$ & $-$ & $-$ & $0.014_{-0.003}^{+0.003}$ & $0.004_{-0.001}^{+0.001}$\\ 
NGC 4552 & $0.04_{-0.01}^{+0.01}$ & $0.008_{-0.002}^{+0.003}$ & $0.4_{-0.1-0.1}^{+0.2+0.2}$ & $0.7_{-0.2-0.2}^{+0.3+0.3}$ & $0.4_{-0.1-0.1}^{+0.2+0.2}$ & $0.2_{-0.1-0.1}^{+0.2+0.1}$ & $0_{-2}^{+2}$ & $0.65_{-0.07}^{+0.07}$ & $0.13_{-0.03}^{+0.03}$ & $0.19_{-0.05}^{+0.04}$\\ 
NGC 4621 & $0.0_{-0.0}^{+0.007}$ & $<0.02$ & $<0.7$ & $<1.1$ & $<0.7$ & $<0.2$ & $-$ & $-$ & $0.011_{-0.003}^{+0.003}$ & $0.15_{-0.03}^{+0.03}$\\ 
NGC 5102 & $0.002_{-0.002}^{+0.004}$ & $0.001_{-0.001}^{+0.003}$ & $2_{-1-3}^{+1+3}$ & $4_{-1-4}^{+2+5}$ & $2_{-1-3}^{+1+3}$ & $2_{-1-3}^{+1+3}$ & $-$ & $-$ & $0.003_{-0.001}^{+0.001}$ & $0.002_{-0.0}^{+0.0}$\\ 
\hline
\enddata
\tablenotetext{}{All measurements are given in units of \msunyr.}
\tablenotetext{}{For \mdotovi measurements, the first pair of uncertainties is the statistical uncertainty $\sigma_{\rm stat}$, while the second pair is the systematic uncertainty $\sigma_{\rm syst}$.}
\tablenotetext{*}{SFR$_{100}$ is averaged over the past 100 Myr, and SFR$_{10}$ is averaged over the past 10 Myr.}
\tablenotetext{\dagger}{$\dot{M}_{\rm O\,VI,snc}$ include a correction term for \ovi emission from supernova remnants.  See $\S$\ref{sec:ovi_snr_correction} for details.}
\end{deluxetable*}

All 6 of our cooling rate metrics are tabulated in Table \ref{tab:cooling_rates}, including the decomposition of the \ovi cooling rate into its purely isobaric ($\dot{M}_{\rm O\,VI,cp}$) and purely isochoric ($\dot{M}_{\rm O\,VI,cd}$) components.  We find $\Gamma_{\rm cp} = 8.9({\rm median}) \pm 4.2({\rm stdev}) \times 10^{12}\,{\rm erg}/{\rm g}$ for the isobaric models and $\Gamma_{\rm cd} = 5.4({\rm median}) \pm 2.6({\rm stdev}) \times 10^{12}\,{\rm erg}/{\rm g}$ for the isochoric models.  The systems with multiple pointings (Abell 1795 and NGC 4636) have also been combined into a single cooling rate measurement.  In general, we find that the combined \ovi cooling rates ($\mdotovi \equiv \dot{M}_{\rm O\,VI,comb}$) track the isobaric rates very closely.  This is a consequence of the cooling times generally being much longer than the sound-crossing times of a cooling cloud in the same system ($t_{\rm cool} \sim 100$ Myr and $t_{\rm sc} \sim 1$ Myr), meaning our weighting scheme ($\S$\ref{sec:mdot_ovi_comb}) will heavily favor the isobaric component in all but the most extreme cooling systems.  This result is somewhat dependent on the size of cooling cloud we assume (1 kpc), but even if we were to assume a $10\times$ larger size of 10 kpc, bringing up the typical $t_{\rm sc}$ to $\sim 10$ Myr, we would still weight 90\% towards the isobaric component, having a minimal effect on our results.  Cooling regions much larger than $\sim$10 kpc are physically challenging to imagine, given the typically observed sizes of clouds and filaments.

We also infer a population-wide level of average cooling from our stacked spectra and our generic \textsc{Cloudy} simulations.  From this, we find $\mdotovi = 0.11_{-0.04-0.04}^{+0.05+0.03}~\msunyr$ from the full stack of all non-detections.  For the stacks separated by halo size, we instead find $\mdotovi < 1.83~\msunyr$ (clusters), $\mdotovi = 0.18_{-0.06-0.06}^{+0.06+0.05}~\msunyr$ (groups), and $\mdotovi < 0.19~\msunyr$ (ellipticals).  These nondetections show a tentative mean signal, on the order of a few tenths of an \msunyr, but there is no monotonic population trend yet.

\section{Discussion: The Cooling Flow Problem} \label{sec:discuss}

With 6 different measurements of the cooling rate across different phases, we are now poised to look at the relationships between each of these measurements to come to a better understanding of how cooling proceeds in massive elliptical galaxies.  We remind the reader that all of these quantities are model-dependent measures of a pure cooling-flow-equivalent mass deposition rate and do not reflect an actual net mass deposition rate.

\subsection{Correlation modeling}

In the following subsections, we model correlations between pairs of cooling rates with a power law (linear in $\log_{10}x$, $\log_{10}y$) and find the best fitting parameters with \texttt{scipy}'s \texttt{minimize} function with a custom likelihood, implementing a generative model of the data which has arbitrary 2-dimensional uncertainties and some intrinsic scatter.  Following \citet{2010arXiv1008.4686H}, the most natural choice is to parametrize the line $y = mx + b \pm \sigma$ in terms of its angle $\theta = \arctan m$, perpendicular intercept $b_\perp = b\cos\theta$, and perpendicular scatter $\sigma_\perp = \sigma\cos\theta$, which has the advantage of treating all slopes equally (setting a flat prior in $\theta$, as opposed to $m$).  The likelihood is then generated by projecting $x$ and $y$ and their associated errors into an orthogonal displacement and variance from the line.

In addition, we incorporate both detections and lower/upper limits into the likelihood.  Given a detection with orthogonal distance $d_i$, model value $m_i$, and orthogonal uncertainty $\sigma_i$, let PDF$(d_i,m_i,\sigma_i)$ be the probability density function of a normal distribution, which defines this point's contribution to the total likelihood.  Then, in a similar fashion, given an upper limit $u_i$ and lower limit $l_i$ (both orthogonal to the line), let CDF$(u_i,m_i,\sigma_i)$ be the cumulative distribution function and let SF$(l_i,m_i,\sigma_i)$ be the survival function.  The total likelihood is then the sum over all detections and lower/upper limits:
\begin{equation}
\begin{split}
    \ln \mathcal{L} = &\sum_i \,\ln [{\rm PDF}(d_i, m_i, \sigma_i)] \\ 
                     +&\sum_j \,\ln [{\rm CDF}(u_j, m_j, \sigma_j)] \\
                     +&\sum_k \,\ln [{\rm SF}(l_k, m_k, \sigma_k)]
\end{split}
\end{equation}
Fits are performed with 300 bootstrapping iterations to estimate uncertainties on the model parameters.

\subsection{\ovi emission from supernova remnants} \label{sec:ovi_snr_correction}

Before examining relationships between the cooling rates in detail, we make note here of an additional correction that we apply to our \mdotovi values.  It is likely that a portion of the observed luminosity in the \ovi doublet may arise from supernova remnants (SNRs).  Therefore, to correct for this, we make an estimate of the \ovi luminosity $L_{\rm O\,VI}$ from SNRs, and subtract it from our observed $L_{\rm O\,VI}$ before converting them into cooling rates.

The total $L_{\rm O\,VI,SNe}$ from SNRs can be broken down into the typical rate of SNe, $\mathcal{R}$ (${\rm s}^{-1}$); the typical \ovi luminosity per event, $\mathcal{L}_{\rm O\,VI}$ (\ergs); and the typical time spent in the radiative \ovi-emitting phase per event, $t_{\rm rad}$ (${\rm s}$):
\begin{equation}
    L_{\rm O\,VI,SNe} = \mathcal{R}\mathcal{L_{\rm O\,VI}}t_{\rm rad}
\end{equation}
The rate $\mathcal{R}$ can further be broken down into rates for type Ia SNe ($r_{\rm Ia}$), which scales with total stellar mass, and core-collapse SNe ($r_{\rm cc}$), which scales with star formation rate.  Therefore:
\begin{equation}
    \mathcal{R} = r_{\rm Ia}(M_*/10^{10}\,M_\odot) + r_{\rm cc}(\sfr/\msunyr)~.
\end{equation}

\citet{2005A&A...433..807M} have estimated the specific rate of type Ia SNe, $r_{\rm Ia} = 0.044\,{\rm century}^{-1}$, while \citet{1990ApJS...74..833H} have estimated that of core-collapse SNe, $r_{\rm cc} = 0.77\,{\rm century}^{-1}$.  \citet{2006ApJS..165..480B} conducted a survey of supernova remnants in the Magellanic Clouds with \textit{FUSE}, from which we derive an average \ovi luminosity per event of $\mathcal{L}_{\rm O\,VI} = 2.4 \times 10^{35}\,\ergs$.  Finally, we assume a radiative timescale of $t_{\rm rad} = 10^5\,{\rm yr}$.  

\begin{figure}
    \centering
    \includegraphics[width=\columnwidth]{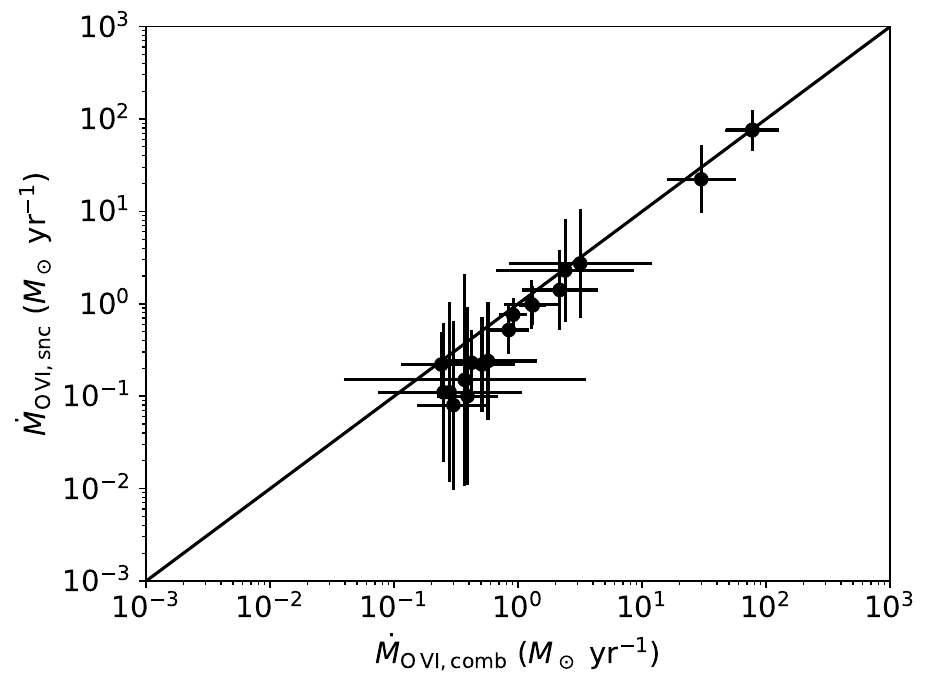}
    \caption{A comparison between \ovi-based cooling rates which have been corrected for \ovi contributions from SNRs ($\dot{M}_{\rm O\,VI,snc}$) against those which have not ($\dot{M}_{\rm O\,VI,comb}$).  The solid line shows 1:1 agreement.  The correction is small for cooling rates above $\sim 1$ \msunyr, and grows increasingly important at lower cooling rates.}
    \label{fig:mdot_ovi_comb_vs_snc}
\end{figure}

Putting these together yields $L_{\rm O\,VI,SNe}$ ranging from $\sim 10^{37}$--$10^{39}$ \ergs, depending on each system's stellar mass and \sfr.  In comparison to our observed \ovi luminosities of $\sim 10^{38}$--$10^{41}$ \ergs, this typically results in small corrections for most systems, but can become a dominant source of systematics in the systems with the faintest \ovi.  For those systems whose corrected rates become consistent with 0 within $\pm 1\sigma$, we consider them as upper limits for the remainder of the analysis.  This is true for four systems: NGC 3115, NGC 1549, NGC 4472, and NGC 4494.

The relationship between the uncorrected cooling rates ($\dot{M}_{\rm O\,VI,comb}$) and the corrected ones ($\dot{M}_{\rm O\,VI,snc}$) is shown in Figure \ref{fig:mdot_ovi_comb_vs_snc}.  The corrected cooling rates are also tabulated alongside the uncorrected rates in Table \ref{tab:cooling_rates}. 

\subsection{\mdotovi vs. \mdotmax and \mdotmaxap}

\begin{figure*}
    \centering
    \includegraphics[width=0.7\linewidth]{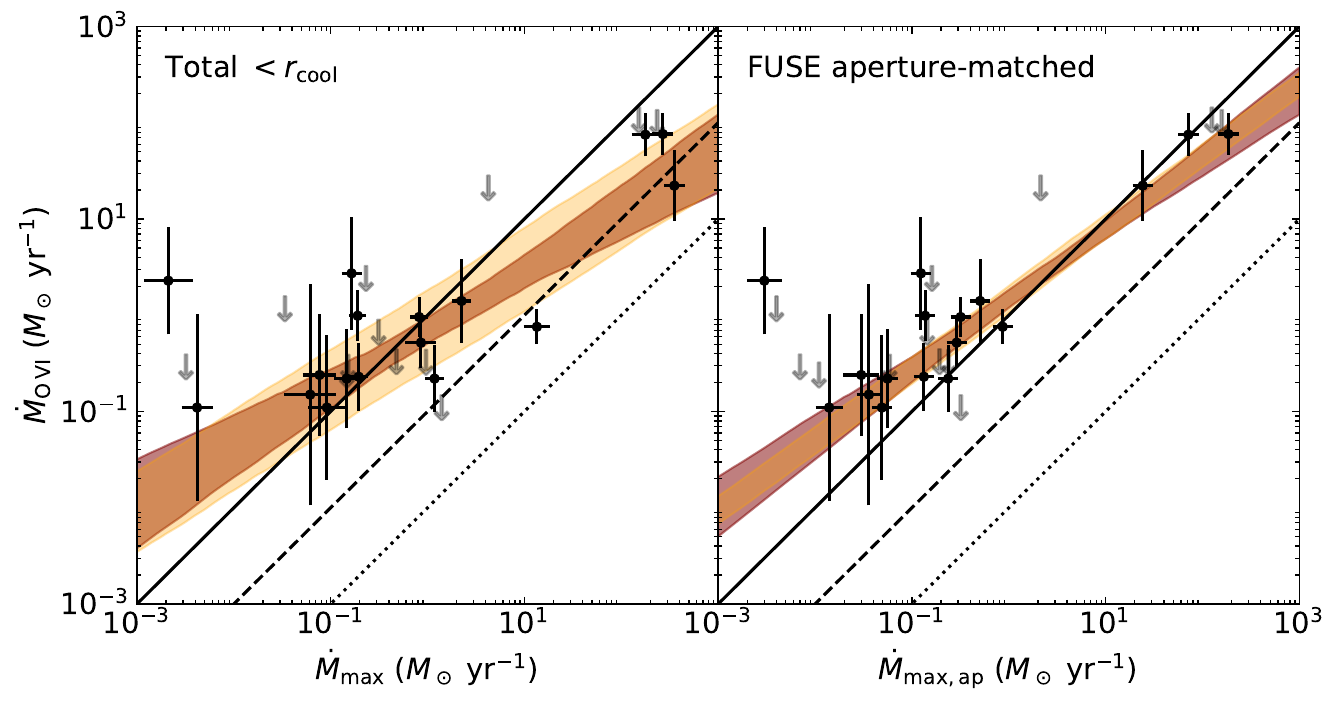}
    \caption{The \ovi cooling rate \mdotovi is plotted against the maximal cooling rate within $< r_{\rm cool}$ (\mdotmax; left panel) and within the \textit{FUSE} aperture (\mdotmaxap; right panel).  Error bars show the total statistical and systematic error, $\sigma_{\rm meas}^2 = \sigma_{\rm stat}^2+\sigma_{\rm syst}^2$.  Upper limits are shown with gray arrows.  The solid line shows where $y=x$, the dashed line shows where $y$ is 10\% of $x$, and the dotted line shows where $y$ is 1\% of $x$.  The red shaded region shows $\pm 1\sigma_{\rm meas}$ (measurement uncertainty from bootstrapping) from the line of best fit, while the yellow shaded region shows $\pm 1\sigma_{\rm intr}$ from the line of best fit, where $\sigma_{\rm intr}$ is the modeled intrinsic scatter of the data.  The best fits are $\log\mdotovi = (0.64_{-0.18}^{+0.12})\log\mdotmax - 0.16^{+0.14}_{-0.13} \pm 0.46_{-0.21}^{+0.10}$ and $\log\mdotovi = (0.75_{-0.12}^{+0.06})\log\mdotmaxap + 0.17_{-0.09}^{+0.10} \pm 0.18_{-0.18}^{+0.16}$.  Both relations have slopes $< 1$, indicative of a mass dependence in the cooling efficiency.}
    \label{fig:movi_mmax}
\end{figure*}

In Figure \ref{fig:movi_mmax}, we show \mdotovi against \mdotmax (left panel) and \mdotmaxap (right panel).  The two panels of this Figure can be thought of as demonstrating two potential extreme cases of cooling, based on the two different X-ray luminosity-based cooling rates on the abscissae.  In the first case, \mdotovi should be highly correlated with \mdotmax if there is a steady inwards flow of cooling gas, such that all of the X-ray-emitting gas within $< r_{\rm cool}$ is expected to end up within the \textit{FUSE} aperture before cooling through \ovi.  In the second case, \mdotovi should instead be highly correlated with \mdotmaxap if no ordered flow exists (if, for example, cooling is highly localized and clumpy/filamentary), and only the X-ray-emitting gas within the \textit{FUSE} aperture ends up contributing to the \ovi emission within the same aperture.

Directly comparing the two relations, we notice that the $\mdotovi/\mdotmaxap$ relation exhibits a noticeably lower intrinsic scatter than $\mdotovi/\mdotmax$, by roughly $0.3$ dex ($0.18$ dex compared to $0.46$ dex), indicating that the localized/clumpy cooling framework does a better job at reproducing \mdotovi on a system-by-system basis than the ordered flow model. At the same time, it is interesting to note that the median of $\mdotovi/\mdotmaxap$ lies $\sim 0.4$ dex above 100\%, whereas the median of $\mdotovi/\mdotmax$ is consistent with $\leqslant 100$\%.  Taken at face value, this implies that the aperture-matched \mdotmaxap is missing some of the X-ray luminosity which ultimately ends up contributing to \mdotovi, which the full integrated \mdotmax captures.  

Note that, since \mdotovi probes cooling on much shorter timescales than \mdotmax or \mdotmaxap, and cooling is expected to be stochastic \citep[i.e.][]{2012ApJ...746...94G, 2026arXiv260527504B, 2026arXiv260527511C, 2026arXiv260527508P}, it is not surprising that some systems exhibit ratios $\mdotovi/\mdotmax > 1$ and $\mdotovi/\mdotmaxap > 1$.  However, while this effect should be a driver behind the intrinsic scatter in $\mdotovi/\mdotmax$ and $\mdotovi/\mdotmaxap$, it should not preferentially shift it in any one direction, as is observed in $\mdotovi/\mdotmaxap$.  Instead, returning to our arguments laid out in the introduction, this behavior could be explained if our \mdotovi measurements are simply overestimated, due to, for example, thermal conduction \citep{1993ApJ...405L..17C} and/or turbulent mixing layers \citep{1990MNRAS.244P..26B, 1993ApJ...407...83S}, causing a large portion of the observed \mdotovi to be emitted from gas which has been recycled through heating and cooling many times. 

Differences in metal abundances from system to system, or inaccurate metal abundance measurements, may also be a driver of intrinsic scatter in these relations.  However, we find that our analysis of these population-level trends is insensitive to this variance.  Repeating our measurements and analysis using the average metal abundances from \citet{2017AnA...603A..80M} for \textit{all} systems yields trends that are within $\pm 1\sigma$ of our results, and thus do not change our conclusions.  This applies not only to the \mdotmax correlations we consider here, but also to all of the correlations that we discuss in the next few sections.

Regardless of which \mdotmax or \mdotmaxap is more appropriate for comparison with \mdotovi, a trend which is observable in both cases is a decrease in cooling efficiency with mass, where we interpret \mdotmax as a proxy for mass / halo size \citep[$\mdotmax \propto M^{1.8}$, as found by ][]{2018ApJ...858...45M}.  Cooling appears to be inefficient for the more massive systems ($\mdotmax \gtrsim 10~\msunyr$), being suppressed by a factor of a few (relative to \mdotmaxap) up to an order of magnitude (relative to \mdotmax), whereas efficiency increases as the mass scale decreases into the regimes of groups and ellipticals, approaching 100\% within the scatter.  In the case of \mdotmax, some of this trend may be explained by aperture effects, with the \textit{FUSE} aperture preferentially tracing larger physical scales in the more distant systems. The slope of the correlation is not as shallow when measured relative to \mdotmaxap ($0.75_{-0.12}^{+0.06}$ compared to $0.64_{-0.18}^{+0.12}$).  However, both of these cases exhibit slopes that are less than 1, at a confidence of $3\sigma$ (relative to \mdotmax) and $4.2\sigma$ (relative to \mdotmaxap), indicating that a mass-dependent cooling efficiency survives after removing any potential aperture systematics.

One possible interpretation of the sub-linear trends is a mass dependence in the spatial coupling of feedback. Such a dependence arises naturally if there exists some universal minimum entropy threshold for intracluster gas, an idea which has been invoked to explain the shape of the observed X-ray luminosity--temperature relation \citep{2001Natur.414..425V}.  Such a threshold is driven by non-gravitational effects, including radiative cooling, SNe, and AGN feedback, and it causes an observed excess in the entropy of intragroup gas, compared to that which would be extrapolated based on self-similar scaling arguments from more massive clusters, by a factor of 2--3 \citep{2003ApJ...593..272V}\footnote{Note that the entropy floor we are referring to here is specifically a floor in entropy measured on scales of $1/10^{\rm th}$ of the virial radius $R_{\rm vir}$, $K(0.1R_{\rm vir})$, as a function of halo temperature (see Figure 1 in \citet{2001Natur.414..425V} or Figures 12--13 in \citet{2003ApJ...593..272V}).   This is a distinct effect from the apparent floor in the innermost bins of radial entropy profiles (in other words the central entropy $K_0$, measured on much smaller scales than $0.1R_{\rm vir}$), measured for example in \citet{2009ApJS..182...12C} and similar works.  It has been recently argued that the latter effect may in fact not be physical, but rather purely an artifact of low spatial resolution and/or low counts \citep{2014MNRAS.438.2341P, 2018ApJ...862...39B}.  Meanwhile, the former effect which we have invoked here is robust, although studies with more data \citep[i.e.][]{2003MNRAS.343..331P} have shown that the excess entropy may not be a constant floor, but a proportionality $K(0.1R_{\rm vir}) \propto T^{0.65}$, lying above the self-similar prediction $K(0.1R_{\rm vir}) \propto T$.}.  This results in groups having relatively lower pressure in their cores.  Mechanical AGN feedback, in the form of jets, then encounters less integrated pressure as it passes through the core of a group than a cluster, which may lead to energy being deposited on larger scales in groups (outside of the core) than in clusters (inside of the core).  We emphasize that this concerns where feedback energy is deposited, not whether feedback operates: cavities, shocks, uplift, turbulence, and mixing are seen to be ubiquitous in groups and ellipticals in both simulations \citep{2011MNRAS.415.1549G, 2012MNRAS.424..190G} and observations \citep{2022A&A...666A..94O}.  In the cluster case, the denser core allows AGN feedback to couple efficiently to the gas and maintain a global thermodynamic equilibrium, while local thermal instabilities may still develop if the conditions for precipitation are met \citep{2012ApJ...746...94G, 2015Natur.519..203V}, allowing cooling to propagate at the $\sim 10\%$ level.

This interpretation also aligns with the ``black hole valve'' mechanism proposed by \citet{2020ApJ...899...70V}, which considers feedback from both AGN and type Ia SNe.  Uniquely from our discussion thus far, this mechanism considers how SNe Ia can drive outflows of cool gas.  The model is separated by a few distinct physical regimes which predict different qualitative behavior.  In systems with relatively high CGM pressure, the SNe-driven outflows are largely confined, leading to a mostly closed feedback loop that prevents significant star formation.  Whereas, in systems with lower CGM pressure, there exists a stagnation radius, above which gas can be ejected from the system by a combination of AGN uplift and SNe Ia-driven outflows, and below which cooling and condensation are concentrated onto the black hole.

The same mass dependence may also manifest in the temporal behavior of feedback.  Lower-mass systems with shallower potentials may produce individual outbursts with stronger fractional effects, leading to more burstiness than more massive systems.  Thus, the dependence of $\mdotovi/\mdotmax$ on mass could be shaped not only by differences in the feedback's radial distribution, but also its duty cycle or interface area of multiphase condensation.  The same processes that redistribute gas can also produce \ovi directly, in turbulent mixing layers and at shock interfaces. We therefore interpret $\mdotovi/\mdotmax$ as an apparent \ovi-equivalent cooling ratio, rather than as direct evidence for unimpeded cooling.

\subsection{\mdotovi vs. \mdotx and \mdothcf}

\begin{figure*}
    \centering
    \includegraphics[width=0.7\linewidth]{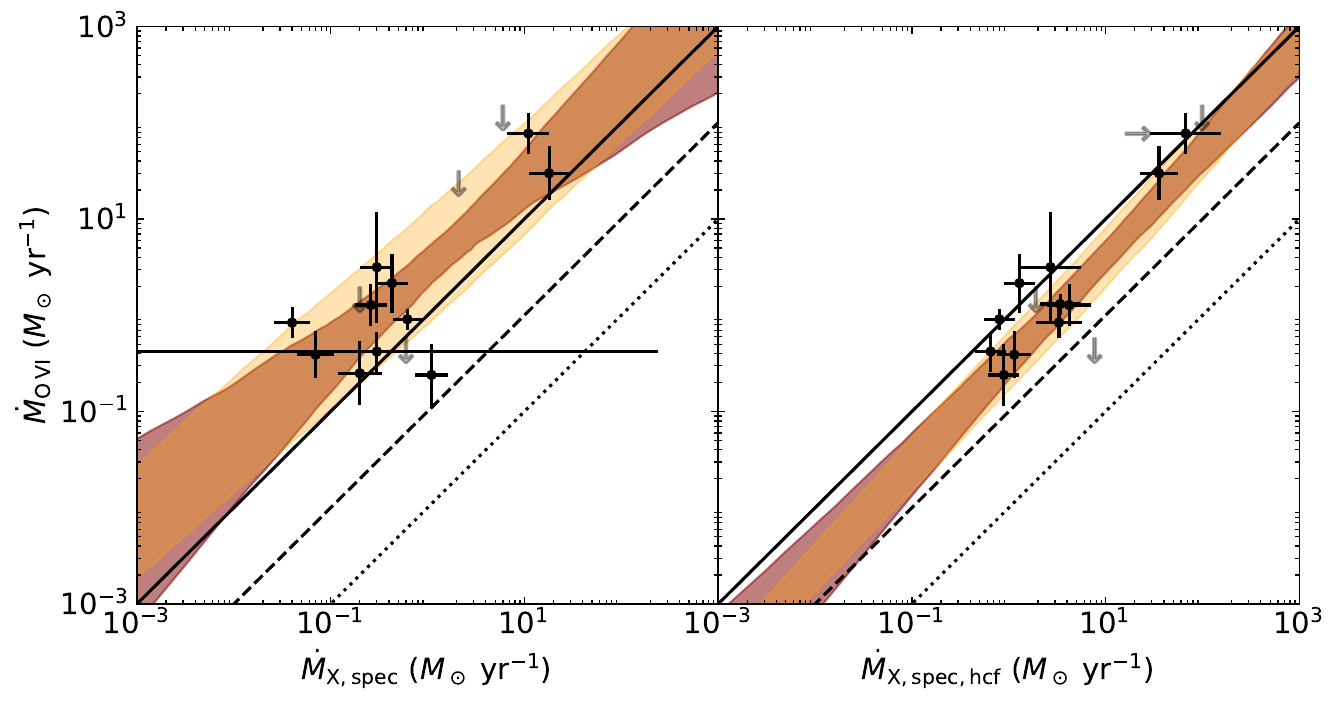}
    \caption{As Figure \ref{fig:movi_mmax}, but showing the \ovi cooling rate \mdotovi as a function of the X-ray spectroscopic cooling rates \mdotx (unabsorbed models) and \mdothcf (absorbed models).  Note that for this figure we use \mdotovi values which have not been corrected for SNR emission ($\dot{M}_{\rm O\,VI,comb}$ in Table \ref{tab:cooling_rates}).  The best fits are $\log\mdotovi = (0.91_{-0.32}^{+0.30})\log\mdotx + 0.53_{-0.25}^{+0.18} \pm 0.68_{-0.31}^{+0.12}$ and $\log\mdotovi = (1.09_{-0.16}^{+0.14})\log\mdothcf - 0.45_{-0.22}^{+0.21} \pm 0.37_{-0.37}^{+0.15}$.  The latter correlation reduces the scatter by $\sim 0.3$ dex, and shifts the X-ray cooling rates from underpredicting to overpredicting the \ovi.  These trends could be explained either with mixing layers, or with partially absorbed cooling (or both).}
    \label{fig:movi_mxspec}
\end{figure*}

We show \mdotovi against \mdotx and \mdothcf in Figure \ref{fig:movi_mxspec}.  Note that, uniquely for this figure, we use \mdotovi values which have \textit{not} been corrected for SNR emission.  Because the X-ray spectroscopic cooling rates would also require a similar correction, we opt to show the correlation between the ``raw'' values, rather than applying uncertain correction factors to both axes (which would both scale with stellar mass and \sfr, potentially introducing spurious correlations).  Given that these correction factors are mostly small ($\lesssim 20\%$; Figure \ref{fig:mdot_ovi_comb_vs_snc}), plotting the uncorrected values in this case has a limited impact on any inferred correlations, changing by $< 1\sigma$ compared to using corrected values.

Regarding the unabsorbed models (left panel), \mdotx is nearly always smaller than \mdotovi, which is inconsistent with a simple steady cooling flow.  This is a restatement of the classic puzzle of the missing soft X-rays \citep{2006PhR...427....1P}, now manifesting in the UV emission, as opposed to \mdotmax.  In the mixing layer scenario, this is naturally explained by the gas collisionally cooling through the soft X-ray temperatures where \mdotx is most sensitive, until it reaches the geometric mean temperature of the hot and cold phases \citep{1990MNRAS.244P..26B}. This heavily reduces the radiative contribution to the cooling above $\sim 10^{5.5}\,{\rm K}$.  Meanwhile, in the hidden cooling flow scenario, this low \mdotx is a consequence of absorption from cold in situ gas.  

Observing the correlation between \mdotovi and \mdothcf (right panel), as expected, \mdothcf are much larger than \mdotx, and are more consistent with our \mdotovi measurements, which have also been corrected for absorption.  The slope is close to unity and the intrinsic scatter is on the same order or smaller than the measurement uncertainty ($\lesssim 0.3$ dex), but \mdothcf now overpredicts \mdotovi by $\sim 0.5$ dex. Notably, this near-linear trend arises in part because the correction factor between absorbed and unabsorbed models appears to evolve with mass---with a typical increase in \mdothcf of $\sim 9\times$ in the groups and $\sim 6\times$ in the clusters.  Yet, the cooling rates in clusters remain suppressed relative to \mdotmax, as do \mdotovi.

The reduced scatter between \mdotovi and \mdothcf, compared to \mdotx, is consistent with an important role for absorbed (``hidden'') cooling models \citep{2022MNRAS.515.3336F, 2023MNRAS.521.1794F, 2023MNRAS.524..716F, 2024MNRAS.535.2173F}.  However, the $0.5$ dex offset indicates that these specific models, which treat the absorbent gas as fully interleaved with the emitting gas (in other words, a covering fraction of $\sim 1$), overpredict \mdotovi by a factor of $\sim 3$.  Reality, therefore, may lie somewhere in between the two extrema of completely unabsorbed (\mdotx) and maximally absorbed (\mdothcf), with a much smaller covering fraction, and with mixing layers potentially also playing a role in shifting down the relative magnitude of \mdotovi.  A lower covering fraction might also aid in explaining the lack of observed re-radiated emission from the (presumably) absorbed X-rays \citep{1995ApJ...452..164V}.

\subsection{\sfr vs. \mdotmax, \mdotx, and \mdotovi} \label{sec:sfr_mdot}

\begin{figure*}
    \centering
    \includegraphics[width=\linewidth]{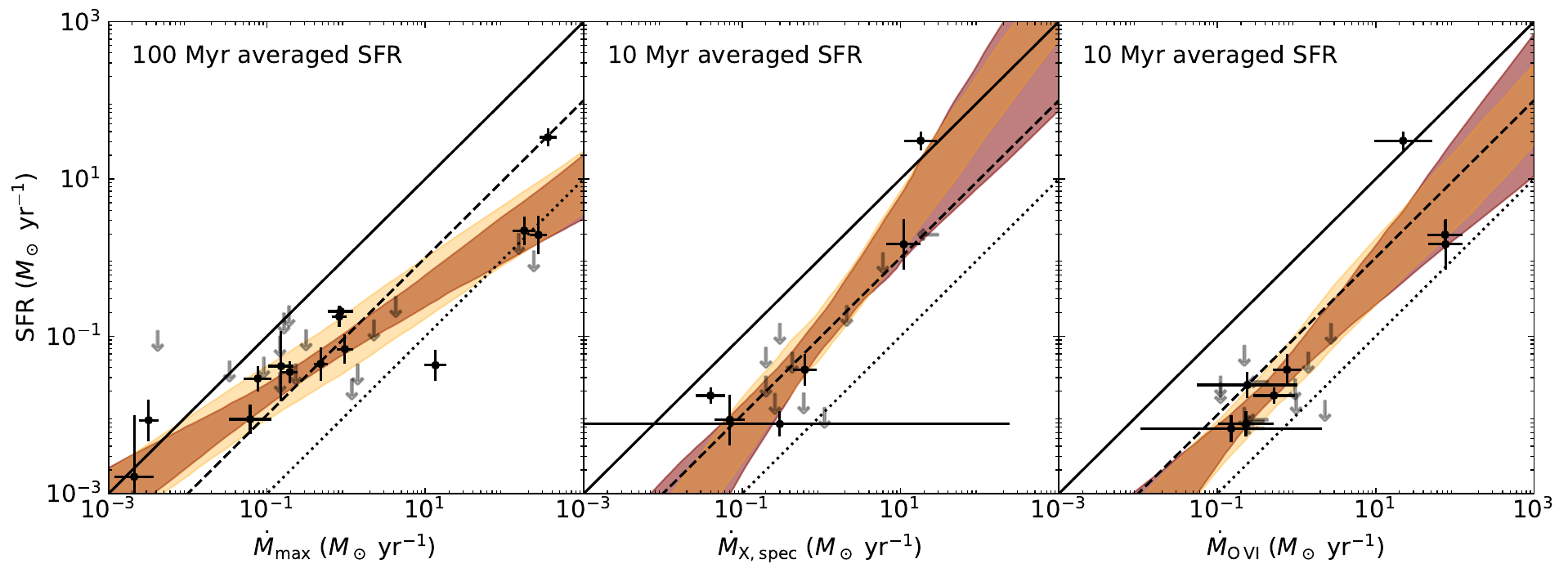}
    \caption{As Figure \ref{fig:movi_mmax}, but showing \sfr as a function of the maximal cooling rate \mdotmax, the X-ray spectroscopic cooling rate \mdotx, and the \ovi cooling rate \mdotovi.  The leftmost panel uses 100 Myr averaged SFRs, while the right two panels use 10 Myr averaged SFRs to more closely align with the timescales of the line emission.  The best fits are $\log{\rm SFR_{100}} = (0.68_{-0.14}^{+0.11})\log\mdotmax - 1.09_{-0.12}^{+0.14} \pm 0.43_{-0.21}^{+0.06}$, $\log{\rm SFR_{10}} = (1.39_{-0.43}^{+0.33})\log\mdotx - 0.90_{-0.28}^{+0.17} \pm 0.55_{-0.55}^{+0.04}$, and $\log{\rm SFR_{10}} = (1.10_{-0.29}^{+0.28})\log\mdotovi - 1.28_{-0.20}^{+0.16} \pm 0.48_{-0.48}^{+0.18}$.  \mdotovi--\sfr is consistent with a 1:1 relation, but with SFRs shifted an order of magnitude smaller than \mdotovi, requiring a mechanism besides pure cooling flows to explain.}
    \label{fig:sfr_movi}
\end{figure*}

We now turn to correlations with the \sfr (Figure \ref{fig:sfr_movi}).  In this Figure, we attempt to match the timescale that the SFR is measured over as best as possible to the timescale of the corresponding cooling rate it is being compared to.  The leftmost panel, which shows \mdotmax, uses 100 Myr averaged SFRs, whereas the right two panels, which show spectroscopic cooling rates, use the 10 Myr averaged SFRs. We observe a tight correlation around $\sfr \sim 0.05\mdotovi$, which as a consequence means that the \sfr--\mdotmax relation tracks closely with the earlier observed \mdotovi--\mdotmax relation (but shifted down $\sim 1.2$ dex).  This also implies that $\sfr \sim 0.05\mdotovi \sim 0.016\mdothcf$, due to the strong correlation between \mdotovi and \mdothcf.  In contrast, there is a marginally larger scatter between \sfr and \mdotx, which could be due to unmodeled absorption effects, or simply stochasticity in time.

The 10 Myr-averaged \sfr is systematically lower than \mdotovi, the \ovi cooling-flow-equivalent rate.  The interpretation of this offset and its significance depends upon one's interpretation of \mdotovi and the \sfr themselves.
Since \ovi may include interface emission and gas undergoing repeated thermal cycling, while star formation follows molecular-gas accumulation with a finite delay, this ratio should not be interpreted directly as a mass-conversion efficiency. The offset may instead reflect nonlinear \ovi weighting, multiphase recycling (including turbulent mixing layers), molecular-gas storage, reheating or outflows, and delayed or non-standard star formation.
In the following paragraphs, we discuss a few possible explanations/interpretations in further detail.

\begin{figure*}
    \centering
    \includegraphics[width=\linewidth]{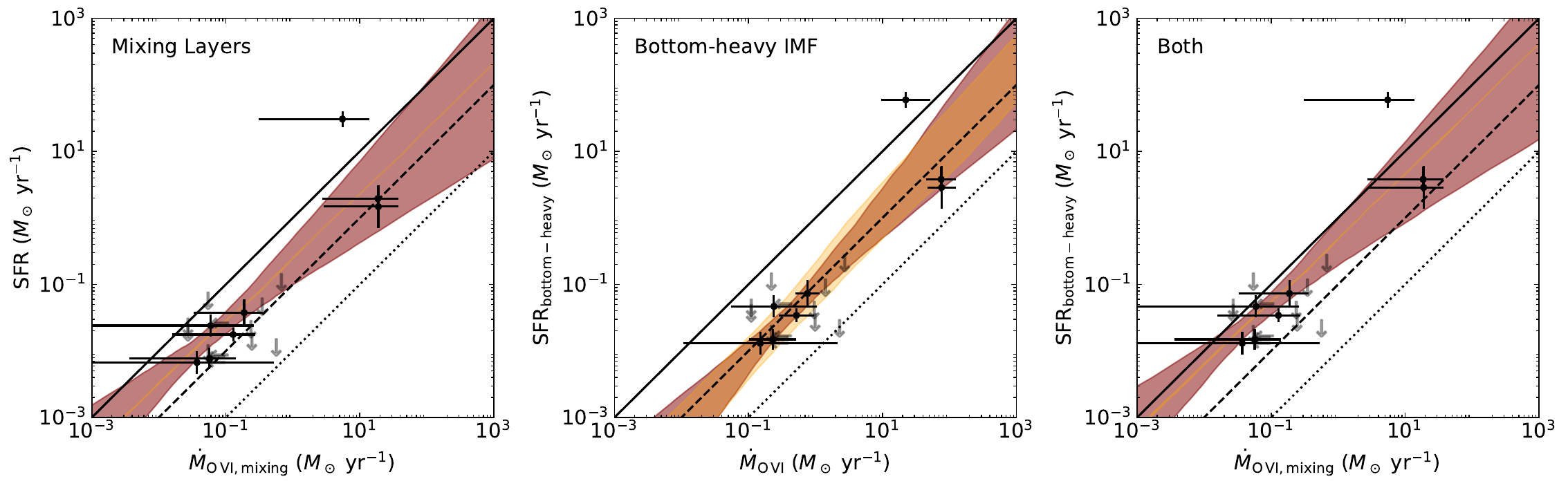}
    \caption{\sfr is shown as a function of \mdotovi, in 3 different physical cases, which are all shown in reference to the fiducial case of a Salpeter IMF and a pure cooling flow (shown in the right panel of Figure \ref{fig:sfr_movi}). In the left panel, a mixing layer model from \citet{2023ApJ...950...91C} with a hot phase temperature of $3 \times 10^7~{\rm K}$ and a metallicity of $0.316Z_\odot$ is used, which lowers the \mdotovi. In the middle panel, a bottom-heavy IMF with a slope of $-3$ below $1~\msun$ is used, which boosts the \sfr. In the right panel, both the bottom-heavy IMF and the mixing layer model are used, which brings the \mdotovi almost entirely within agreement with the \sfr (at the cost of making \mdotx measurements disagree with \mdotovi).}
    \label{fig:sfr_movi_models}
\end{figure*}

\textbf{(1) Mixing layers.}  Cold gas may be reheated by mixing with a quantity of hot gas and settling to an intermediate temperature \citep{1990MNRAS.244P..26B}.  In this mixing interface, gas may radiatively cool and turbulently reheat many times before eventually settling into the cold phase. This conceptually solves the issues raised by the steady one-pass cooling flow interpretation---star formation appears ``inefficient'', but in reality this is because the \ovi emission is enhanced by the gas recycling and does not reflect the \textit{net} cooling rate; at the same time, we do not see an accumulation of massive cold gas reservoirs, because the true \textit{net} cooling rate is (presumably) $\sim$equal to the SFR.

When gas gets recycled in a turbulent mixing layer, the reheating of cooled gas is not ``free''---by the conservation of energy, it must come from another energy source which gets depleted.  In other words, enhanced \ovi emission cannot simply be manufactured from nothing.  Rather, the observed ``boost'' in \ovi emission per unit mass may be produced by either (1) a redistribution of the cooling budget towards \ovi, and away from the hotter temperatures which are ``skipped'' via thermal conduction; and/or (2) a dissipation of the relative kinetic (shear) energy between the hot and cold phases into new thermal energy that is then radiated.

Hydrodynamical simulations of mixing layers have revealed further insights on these complex structures.  \citet{2010ApJ...719..523K} found that mixing happens on faster timescales than ionization or recombination, making non-equilibrium effects important in predicting ion densities within the mixing layer.  \citet{2020ApJ...894L..24F} argue that mixing layers have a fractal structure, and they have developed a model to describe the properties of such a configuration, including net mass flow rates.  Recently, \citet{2023ApJ...950...91C} have taken this model and developed a simple analytic framework that reproduces the behavior of the detailed 3D simulations and can be used to make predictions for emissivities of many emission lines, including \oviwave.  Their framework uses a steady-state one-dimensional mean-field description, assuming a planar interface with a monotonic temperature profile and constant mass flux.  Therefore, it does not capture the fractal nature seen in the 3D simulations, nor the literal cycling of fluid elements through the \ovi-emitting phase.  Instead, it characterizes turbulence using an effective conductivity and viscosity, controlled by a handful of scaling parameters which are tuned to match the 3D simulations.

\citet{2023ApJ...950...91C} provide accessible code notebooks\footnote{\url{https://github.com/ziruichen11/1.5D_mixing}} that can be used to reproduce their results, which we utilize to calculate predicted $\Gamma_{\rm O\,VI}$ values in a steady-state turbulent mixing layer.  We choose a hot phase temperature of $T_{\rm hot} = 3 \times 10^7~{\rm K}$, roughly equivalent to our average cluster ICM temperature of 2.74 keV (Table \ref{tab:cloudy_init}), while the cold phase temperature is fixed at $T_{\rm cold} = 3000~{\rm K}$ for these models.  Energetically, this places the mixed phase at $\sqrt{T_{\rm hot}T_{\rm cold}} \approx 300,000~{\rm K}$, right near the peak of \ovi emissivity.  We choose the closest grid point to a metallicity of $\sim 0.3Z_\odot$ (which is $0.316Z_\odot$), generally representative of the ICM.  The scaling parameters for the conductivity, viscosity, and shear velocity are left at their fiducial values \citep[see][for details]{2023ApJ...950...91C}. 

Then, the conversion from emission to net mass flow rate is simple to calculate:
\begin{equation}
    \Gamma_{\rm O\,VI,ML} = \frac{F_{\rm O\,VI,ML}}{\dot{m}_{\rm ML}}~,
\end{equation}
where $F_{\rm O\,VI,ML}$ is the \ovi flux (\ergscm) and $\dot{m}_{\rm ML}$ is the mass flux (g s$^{-1}$ cm$^{-2}$) in the mixing layer (ML).  We then compare this model to generic \textsc{Cloudy} isobaric/isochoric cooling models at the same metallicity $0.3Z_\odot$.  We find that the mixing layer's emission per unit mass is enhanced by a factor of $\Gamma_{\rm O\,VI,ML}/\Gamma_{\rm O\,VI,CF} \approx 4$ over the isobaric model, and $\approx 7$ over the isochoric model.  This has the effect of \textit{lowering} the inferred \mdotovi by the same factor.  Given that we have found most systems follow the isobaric path from our simplistic weighting scheme in $\S$\ref{sec:mdot_ovi_comb}, we use $4$ as our fiducial mixing layer enhancement factor.

We note, however, that the precise values obtained have a dependence on the simulation parameters for the conductivity, viscosity, and shear velocity.  Allowing these parameters to vary within plausible physical ranges results in a large systematic uncertainty, bringing the plausible range to $\Gamma_{\rm O\,VI,ML} \sim 1$--$7$. Additionally, the line emissivities in this model are calculated assuming ionization equilibrium, whereas (as mentioned already) mixing can drive non-equilibrium effects that may further alter the predicted $\Gamma_{\rm O\,VI,ML}$ values.  With these caveats in mind, we caution the reader that these models are shown as indicative estimates, rather than precise predictions. 

In the left panel of Figure \ref{fig:sfr_movi_models}, the \mdotovi are scaled down by a factor of $\sim 4$, matching our fiducial model, with uncertainties added in quadrature to reflect the uncertainty on this scale factor.  We observe that these models can reasonably account for a large portion of the discrepancy between the \mdotovi and \sfr measurements.  However, this would require the model parameters for conductivity, viscosity, and shear velocity to conspire in a way that is favorable to boosting the \ovi emission.  

We also note that the increase in \ovi emission comes at the expense of a decrease in emission from the higher ions of \ion{O}{7} and \ion{O}{8} from which \mdotx and \mdothcf are measured.  The \citet{2023ApJ...950...91C} models predict essentially all of the cooling above $\sim 10^6$ K to be conductive ($\Gamma_{\rm O\,VII/O\,VIII,ML}/\Gamma_{\rm O\,VII/O\,VIII,CF} \lesssim 0.02$), which is inconsistent with what we observe in Figure \ref{fig:movi_mxspec} in both \mdotx and \mdothcf.  This suggests that the simple analytic 1D turbulent mixing layer approximation cannot explain all of our observations, however this does not necessarily rule out turbulent mixing layers in general. In fact, the X-ray emission predicted by the 1D models is known to be inaccurate, with more recent work by \citet{2026arXiv260607741C} arguing that this is due to topological/geometric effects that are missing from the simple 1D structure. The volume-filling X-ray phase is not always in direct contact with cold clumps, which will have an effect on any spatially integrated emission measurements on scales larger than individual clumps.  Therefore, future mixing layer models may produce testable line ratios that are in better agreement with what we observe in \ovi, \ion{O}{7}, and \ion{O}{8}.

\textbf{(2) Non-Salpeter initial mass functions.}  Another possibility is that star formation proceeds in a non-standard way that results in an initial mass function (IMF) that is not the standard \citet{1955ApJ...121..161S} slope of $M^{-2.35}$ (which is what we have assumed thus far in this analysis).  Measurements of stellar mass and \sfr are dependent on the assumed IMF, which could result in \sfr measurements being systematically biased if the wrong IMF is used.  This can conceptually solve the stated problems if the SFRs have been underestimated.

Looking to alternate IMFs, the most widely used are \citet{2002Sci...295...82K} and \citet{2003PASP..115..763C}.  Both of these are so-called ``bottom-light'' IMFs, meaning the mass distribution below $1~\msun$ is suppressed compared to Salpeter.  As a consequence, they predict SFRs that are \textit{smaller} than Salpeter roughly by factors of 0.67 and 0.63, respectively \citep{2014ARAA..52..415M}.  Therefore, these cannot be used as explanations for our problem.

Instead, we require ``bottom-heavy'' IMFs, with the mass distribution $< 1~\msun$ being enhanced rather than suppressed.  These would predict \textit{larger} SFRs than Salpeter. Evidence for such IMFs has been observed in the stellar populations of massive elliptical galaxies by \citet{2010Natur.468..940V, 2012ApJ...760...70V, 2018MNRAS.474.4169O, 2022ApJ...932..103G}, with slopes below $1~\msun$ reaching super-Salpeter values of as low as $M^{-3}$. For a simple broken power law IMF with a slope of $-3$ below $1~\msun$ and $-2.35$ (Salpeter) above $1~\msun$, we estimate a correction factor of $\sim 1.94$ on the \sfr.   This is shown in the middle panel of Figure \ref{fig:sfr_movi_models}, where the SFRs have been scaled up by this factor. Therefore, similarly to the mixing layer scenario, we find evidence for a plausible correction that can account for \textit{some} of the discrepancy between \mdotovi and \sfr, but a super-Salpeter IMF alone cannot account for the full difference.  

It is worth mentioning that the IMF of the old stellar populations in these galaxies need not dictate the IMF of the currently forming stars.  While directly probing the IMF of the young stellar populations in massive early type galaxies is more difficult, there are conceptual arguments that might be made in favor of the bottom-heavy scenario.  \citet{2024MNRAS.531..267F} propose a model in which the IMF mode evolves over time, starting from Kroupa-like in times of rapid gravitational collapse, and transitioning to bottom-heavy after stellar and AGN feedback bring about a phase of subsonic cooling ($t_{\rm cool} > t_{\rm ff}$).  This cycle may repeat many times.  They argue that the bottom-heavy mode of star formation is driven by the high gas pressures (and high stellar velocities and densities) in the cores of elliptical galaxies, which drives down the Jeans mass $M_J \propto P^{-1/2}$ \citep{1994MNRAS.266..399F} to as low as $M_J \sim 0.022~\msun$ at $z=0$, or up to $M_J \sim 0.8~\msun$ at $z=5$.  Since $M_J$ sets the scales at which molecular clouds fragment, the actual masses of individual stellar objects may end up much lower, leading to a significant fraction forming below the H-burning limit of $0.08~\msun$.  These low-mass stars and brown dwarfs may also fall non-radiatively into the central black hole, having a tidal disruption radius smaller than the event horizon.  Therefore, they may effectively be considered baryonic dark matter.


\textbf{(3) Magnetic fields.}  We have argued that turbulent mixing layers and non-standard IMFs alone cannot fully account for the difference in the observed \mdotovi and \sfr values.  However, if both of these scenarios act together, the combined correction factor reaches well within the observed factor of $\sim 15$ (see the right panel of Figure \ref{fig:sfr_movi_models}).  Additionally, we have only considered a small subset of the possible parameter space of mixing layer models, and it is possible that more refined 3D simulations may be able to fully explain the observed differences without the need for non-standard IMFs.
Moreover, despite being less massive than initially expected, molecular gas reservoirs are common in these systems, which may be evidence of either an intrinsic imbalance in \mdotovi and \sfr over time (also accounting for other sinks for the molecular gas, like black hole accretion or reheating), or a long depletion timescale for the molecular gas.

Why does this molecular gas build up over time, rather than immediately collapsing into stars (i.e. what drives these long depletion timescales)? This could be related to a number of factors, one of which being magnetic fields.  It has been proposed that magnetic pressure plays an important role in supporting cold gas filaments in clusters against gravitational collapse \citep{2025NatAs...9..449O}, thus halting any new star formation.  This would require magnetic field strengths on the order of $B \sim 20$--$60~{\rm \mu G}$, which are physically plausible.  

We emphasize that magnetic fields alone cannot solve the observed discrepancy between \mdotovi and \sfr.  Without any other corrective factors from mixing layers or IMFs, $\mdotovi-\sfr$ is too large compared to observed molecular gas reservoirs.  Magnetic fields are therefore proposed only as a partial solution that may be invoked on top of the previous solutions.



\section{Conclusion} \label{sec:conclusion}

In this second paper in the \textit{Reigniting the FUSE} series, we have used archival far-ultraviolet spectroscopy from \textit{FUSE} of 29 massive elliptical galaxies to calculate updated measurements of cooling rates through the $\sim 10^{5.5}$ K phase from the \oviwave doublet (\mdotovi) with modern methodology.  Due to a careful treatment of both Galactic and intrinsic extinction, we recover \mdotovi that are on average $\sim 60\%$ larger than previously reported. In tandem, we have used archival \textit{Chandra}-ACIS imaging spectroscopy to calculate corresponding maximal cooling rates from the $\sim 10^7$ K X-ray halo (\mdotmax and \mdotcf).  

We have used these measurements together with star formation rates (\sfr) from the first paper in this series, and X-ray spectroscopic cooling rates gathered from the literature (\mdotx and \mdothcf), to perform an in-depth analysis of cooling gas in these galaxies.  Our findings are as follows:
\begin{itemize}
    \item The $\mdotovi/\mdotmax$ and $\mdotovi/\mdotmaxap$ ratios suggest that there is a mass dependence in cooling efficiency, decreasing from low-mass groups to massive clusters.  This may reflect differences in the physical scales of AGN feedback, acting primarily on larger radii in groups and smaller radii in clusters, or similar differences in other factors, such as duty cycle, as a function of mass.
    \item When the X-ray spectroscopic cooling rate is not corrected for absorption (\mdotx), we find that \mdotx underpredicts the cooling rate relative to \mdotovi. However, when it is assumed that the X-ray spectroscopic cooling rates are self-absorbing, as in the ``hidden cooling flow'' model, \mdothcf instead overpredicts the cooling rates relative to \mdotovi, while the scatter is reduced.  These results could be explained if cooling gas is partially absorbed, with some covering fraction $C_f \ll 1$, and/or if the \ovi is enhanced relative to the X-ray lines in turbulent mixing layers.  While the simplistic 1D mixing layer models we use predict inconsistent X-ray emission, much of the observed X-rays may originate from outside the mixing layers, since the hot phase is volume-filling, an effect which is being explored in more modern 3D mixing layer models \citep{2026arXiv260607741C} and may be rectified in the near future.
    \item The $\sfr/\mdotovi$ relation is tightly constrained around $\sfr \sim 0.05\mdotovi$ at all masses with low scatter, indicating a strong physical link between \sfr and \mdotovi.  The observed ratio can be explained by a combination of turbulent mixing layers (boosting the \ovi emission per unit mass) and/or a modestly bottom-heavy IMF (increasing the SFRs), both of which are well motivated in the literature.
    \item More generally, the mismatch among the \ovi-equivalent rate, X-ray cooling diagnostics, and \sfr may reflect a nonlinear, time-dependent CCA/feedback cycle in which condensation, turbulent interfaces, molecular-gas storage, AGN reheating, and star formation are phase-lagged, rather than a single steady conversion efficiency.
\end{itemize}

More data from systems spanning a higher dynamic range in cooling rates are required to confirm many of these findings and disentangle the complex web of possibilities that could explain the cooling rates.  In particular, upcoming and planned missions with far-ultraviolet spectroscopic capabilities (\textit{UVEX}, \textit{Aspera}, \textit{HWO}) will be critical for expanding the library of systems with \oviwave measurements.

All of the data presented in this analysis, including all values in Tables \ref{tab:ovi} and \ref{tab:cooling_rates}, are publicly available for download on Zenodo (DOI: \href{https://dx.doi.org/10.5281/zenodo.19120699}{10.5281/zenodo.19120699}) in machine-readable formats, along with the reduced \textit{FUSE} spectra, aperture-matched photometry, aperture-matched \textit{Chandra} spectra, and data from Paper I's stellar population analysis.

\begin{acknowledgements}


MR acknowledges support from the National Science Foundation Graduate Research Fellowship under Grant No. 2141064.  


MR and MM received funding for this project from the NASA Astrophysics Data Analysis Program under grant 80NSSC25K7560.


MG acknowledges support from the ERC Consolidator Grant BlackHoleWeather (101086804).

VO acknowledges support from the DICYT ESO-Chile Comite Mixto PS 1757, Fondecyt Regular 1251702, and CIRAS-AI Project, code FIUF137139-USACH.


This research is based on observations made with the \textit{FUSE} mission, obtained from the MAST data archive at the Space Telescope Science Institute, which is operated by the Association of Universities for Research in Astronomy, Inc., under NASA contract NAS 5–26555.


The scientific results reported in this article are based in part on data obtained from the Chandra Data Archive.  

\end{acknowledgements}

\appendix

\section{\ovi Spectra} \label{sec:appendix}

We present plots of each system's spectral fits to the \ovi line in Figures \ref{fig:ovi_fits_1}--\ref{fig:ovi_fits_6}.

\begin{figure*}[ht!]
    \centering
    \includegraphics[width=0.39\linewidth]{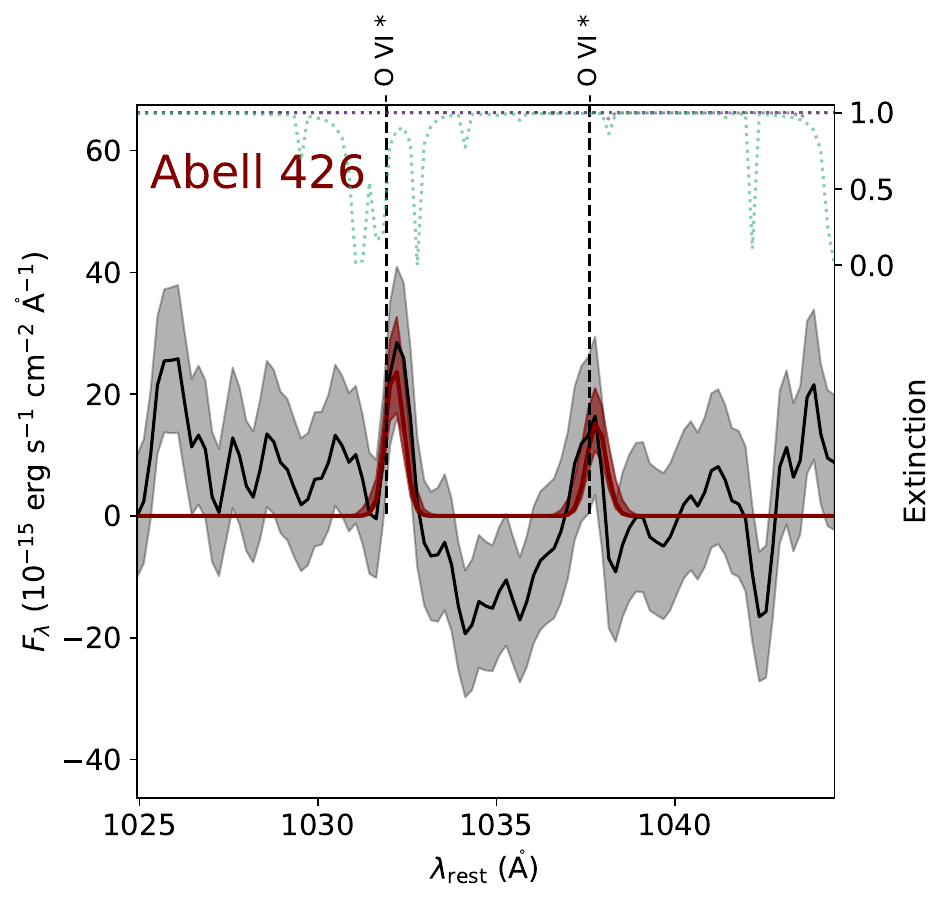}
    \includegraphics[width=0.39\linewidth]{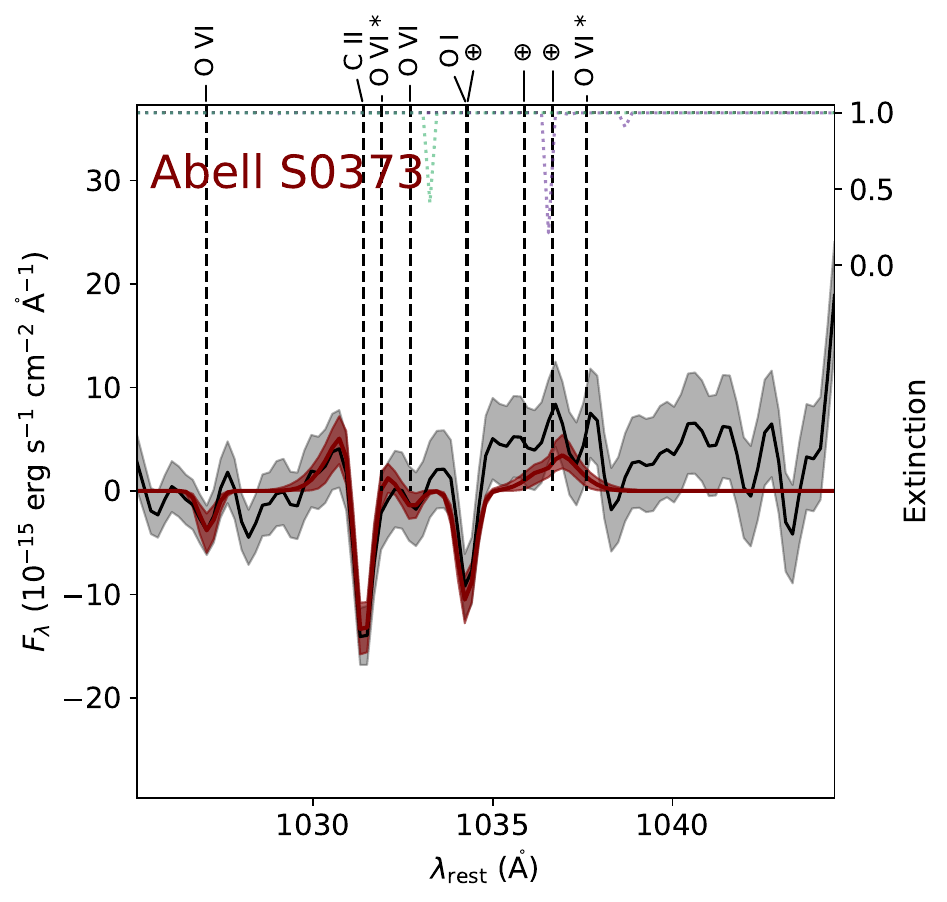}
    \includegraphics[width=0.39\linewidth]{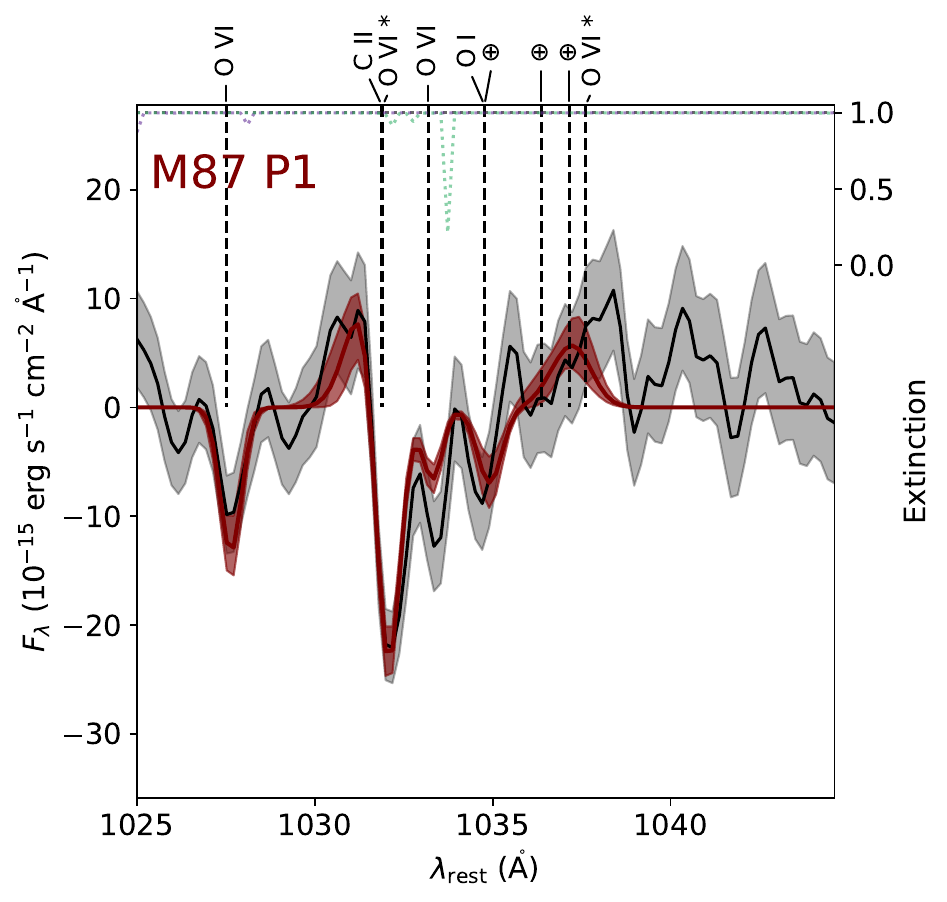}
    \includegraphics[width=0.39\linewidth]{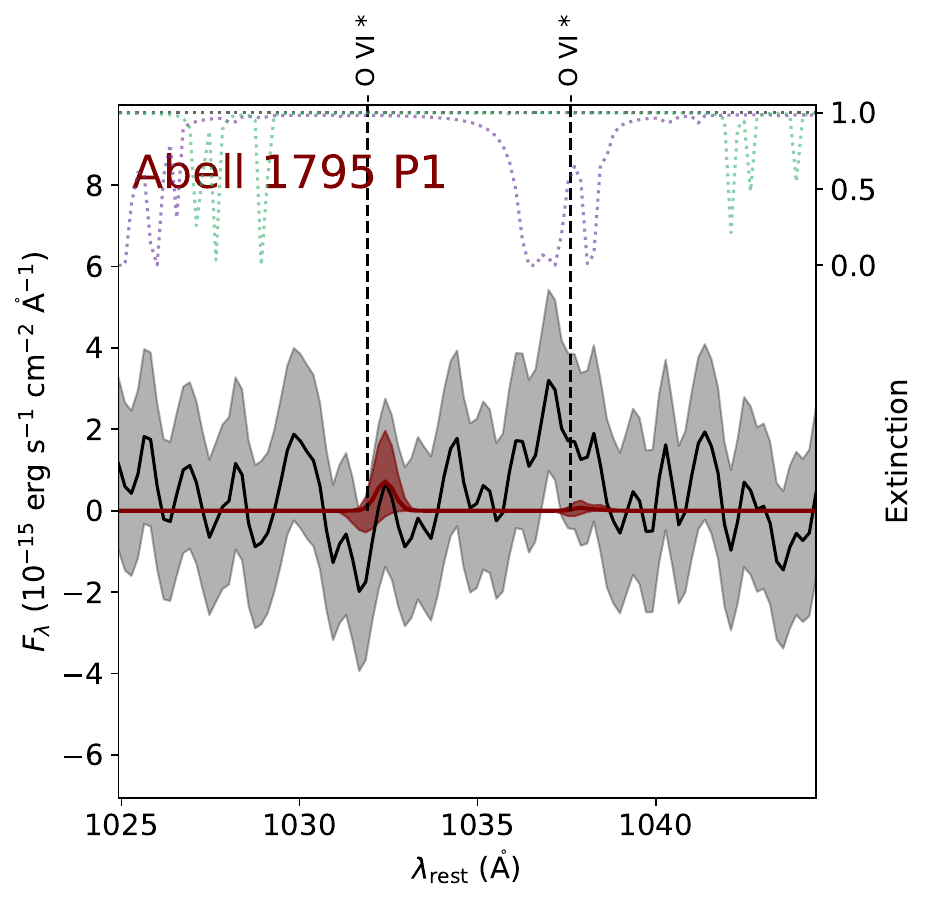}
    \includegraphics[width=0.39\linewidth]{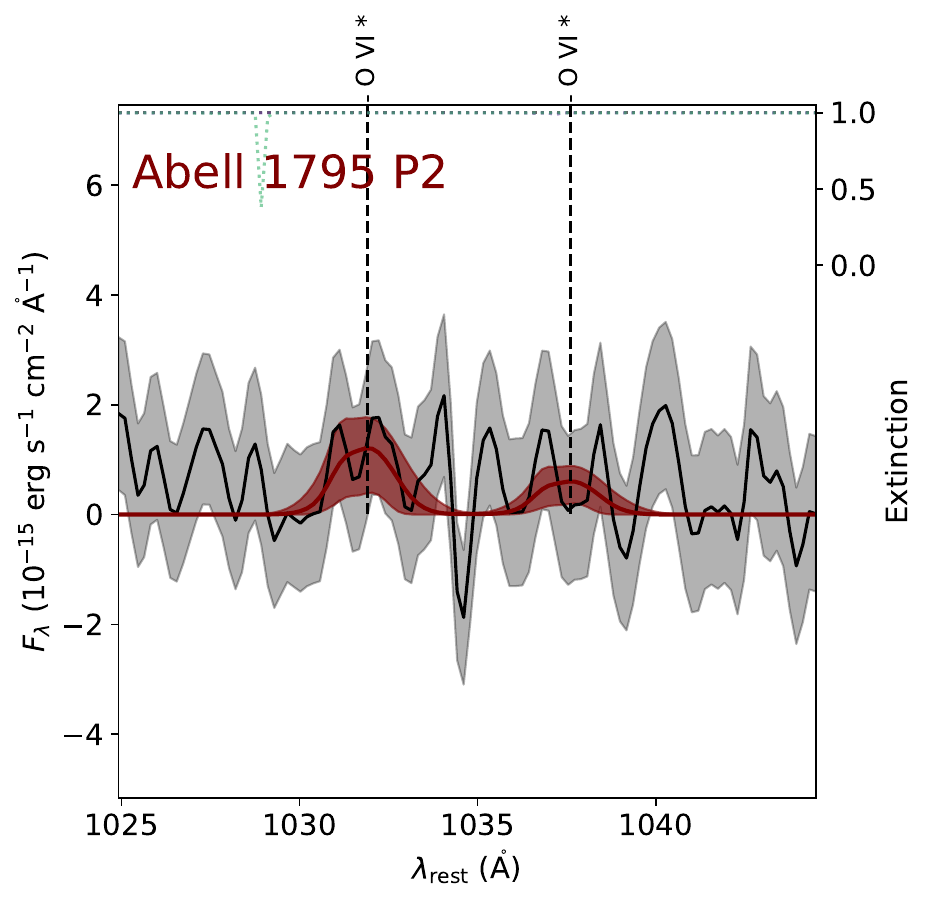}
    \includegraphics[width=0.39\linewidth]{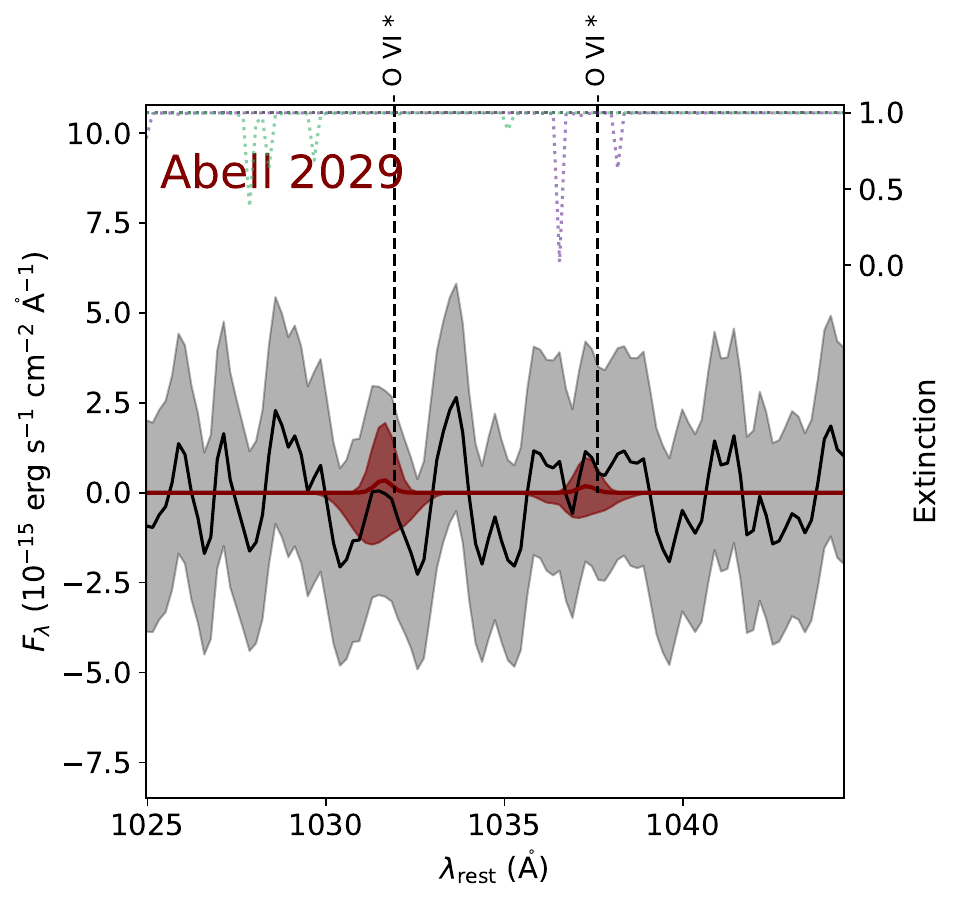}
    \caption{Plots of each system's continuum-subtracted spectrum and model around the \ovi line are given in each panel.  The black line shows the data and the red line shows the median of the bootstrapped models.  The gray and orange shaded regions show the $1\sigma$ uncertainties on the data and the model, respectively.  Emission lines are labeled with vertical dashed lines---asterisks indicate an intrinsic rest-frame feature and no asterisks indicate a Galactic foreground feature.  Telluric features are marked with an $\oplus$.  H$_2$ absorption is shown with the green dotted line (Galactic) and purple dotted line (intrinsic), read from the right axis.  For visual clarity, the data and models have both been smoothed by a Gaussian kernel with $\sigma = 1$ pixel ($0.2~\angstrom$), while the unsmoothed data are used during the fitting process.}
    \label{fig:ovi_fits_1}
\end{figure*}

\begin{figure*}[ht!]
    \centering
    \includegraphics[width=0.39\linewidth]{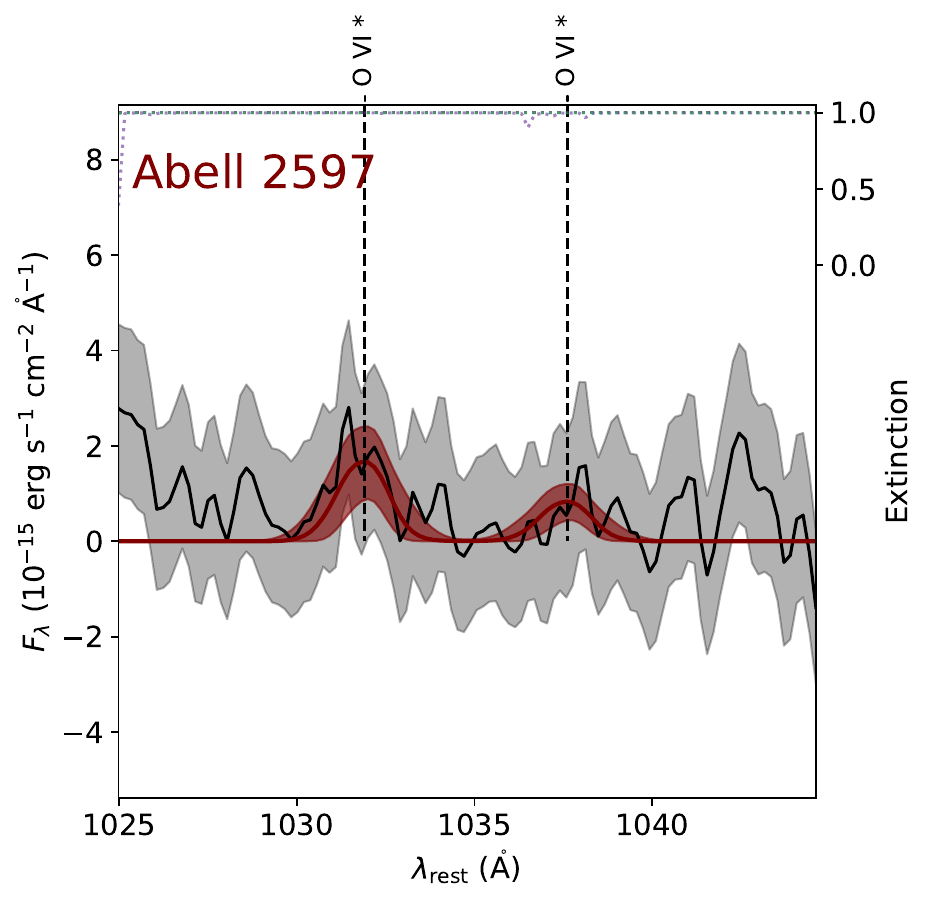}
    \includegraphics[width=0.39\linewidth]{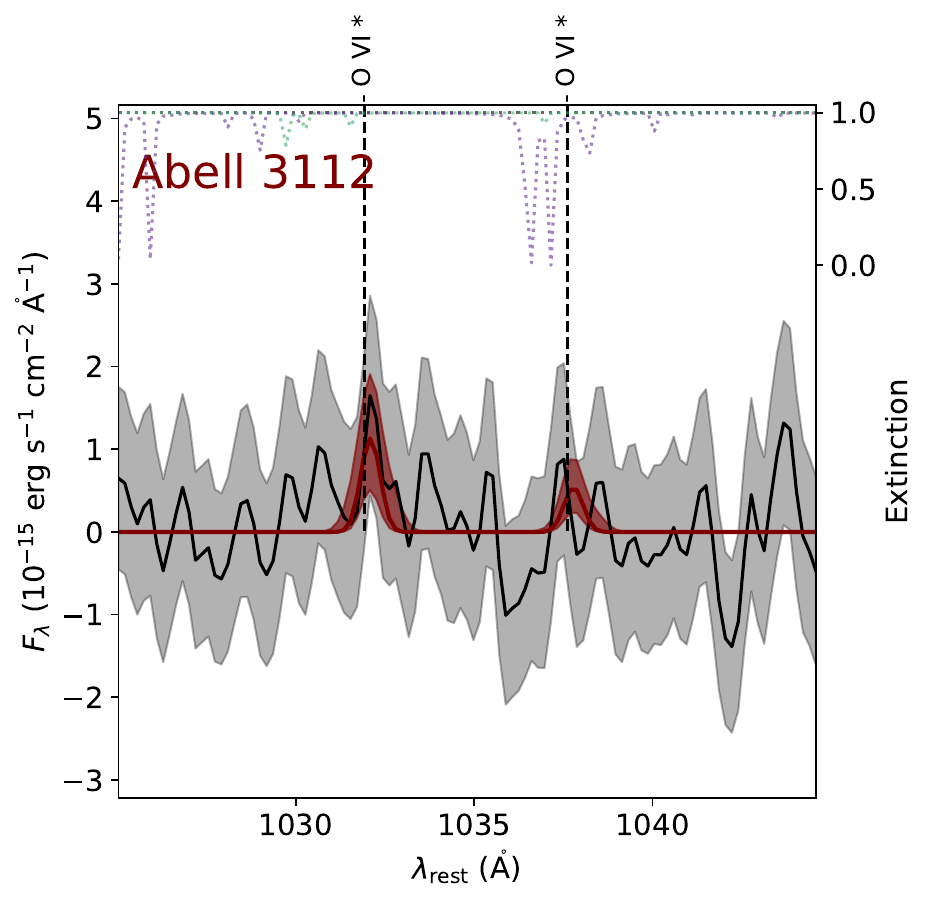}
    \includegraphics[width=0.39\linewidth]{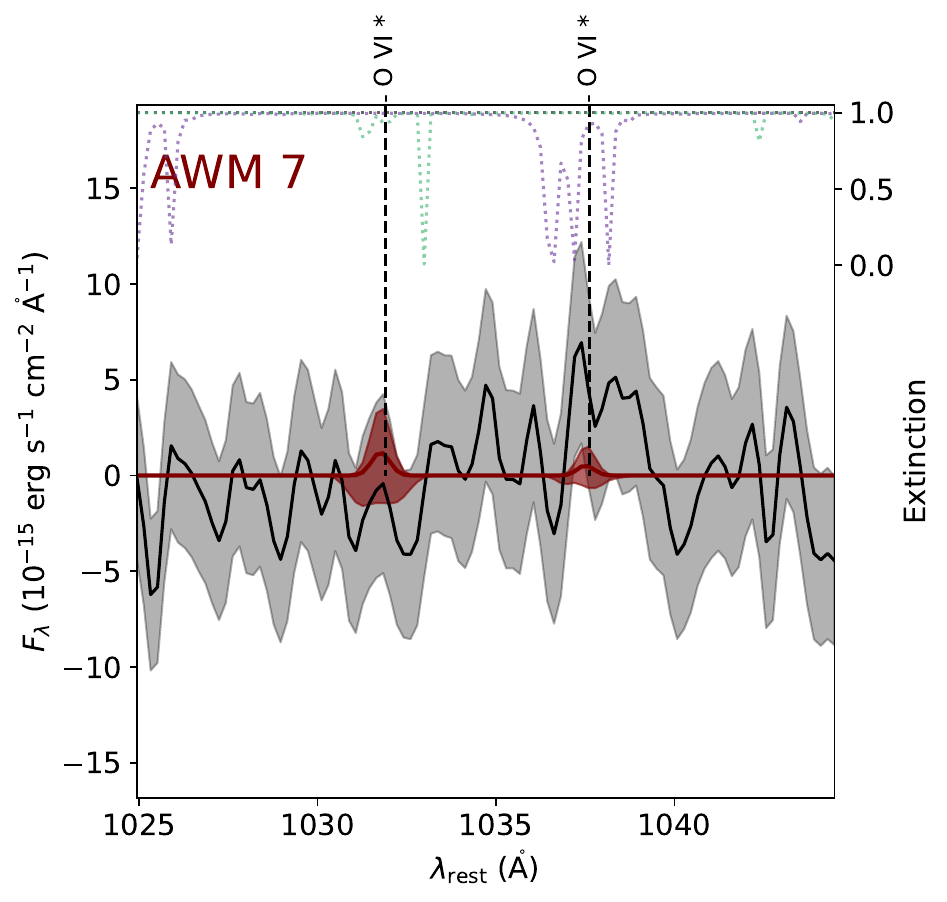}
    \includegraphics[width=0.39\linewidth]{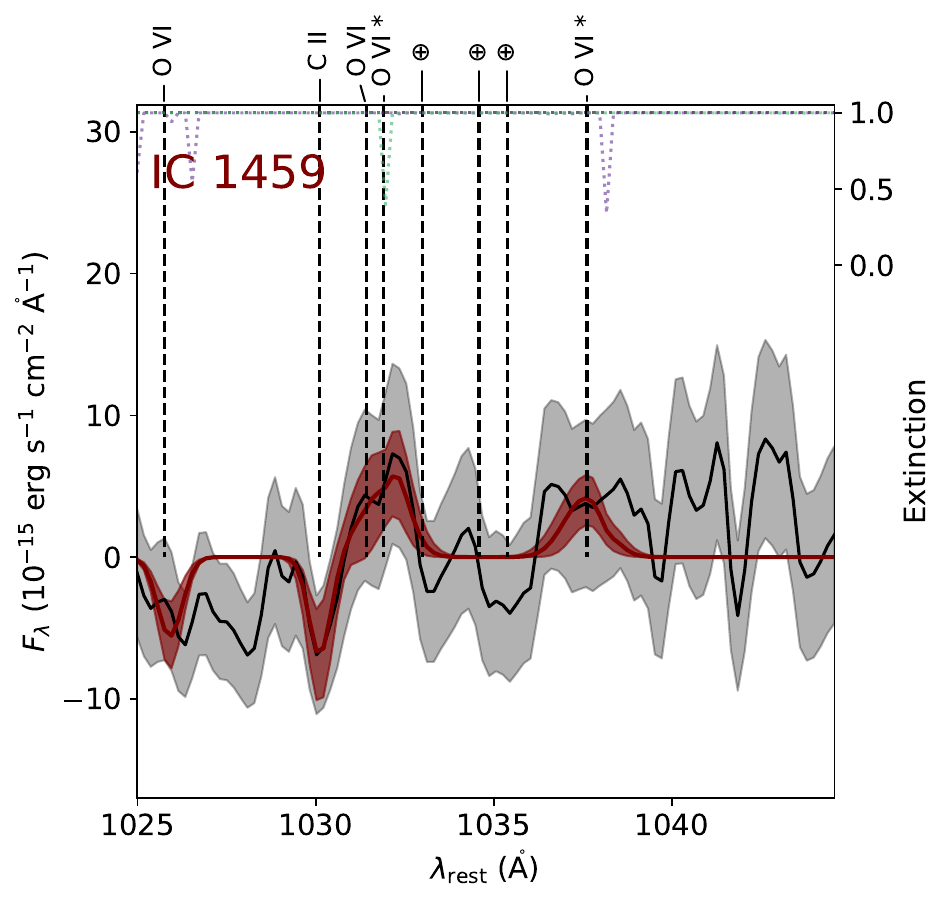}
    \includegraphics[width=0.39\linewidth]{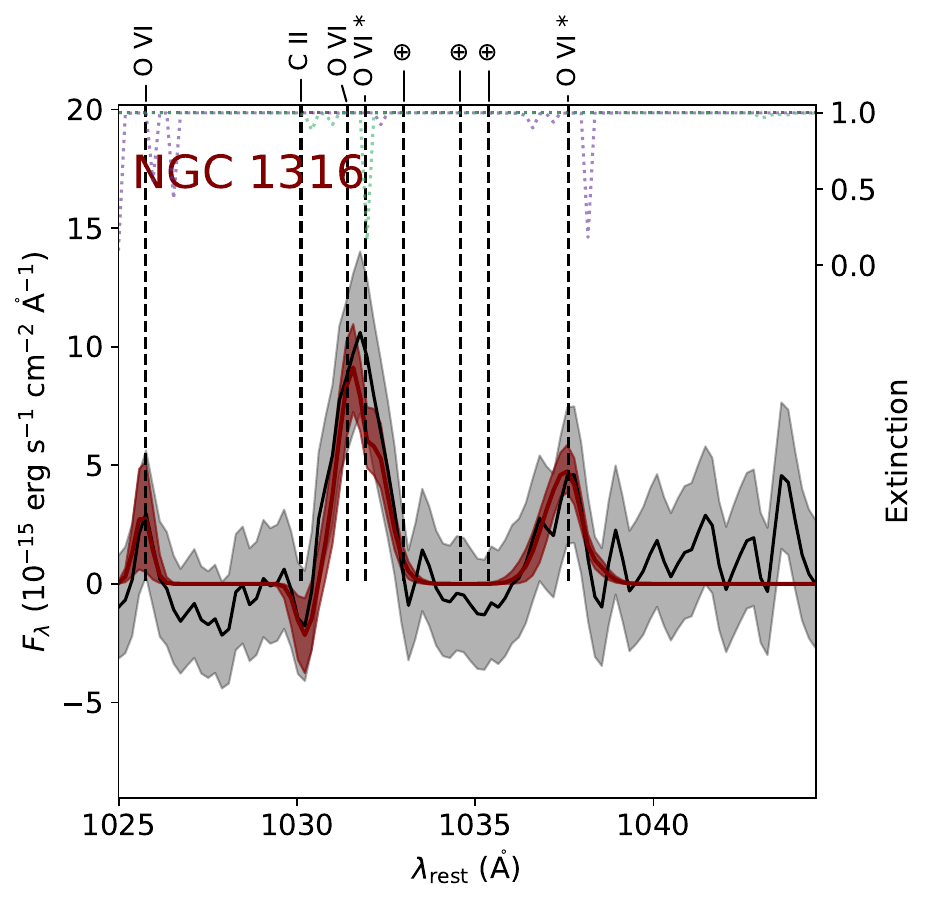}
    \includegraphics[width=0.39\linewidth]{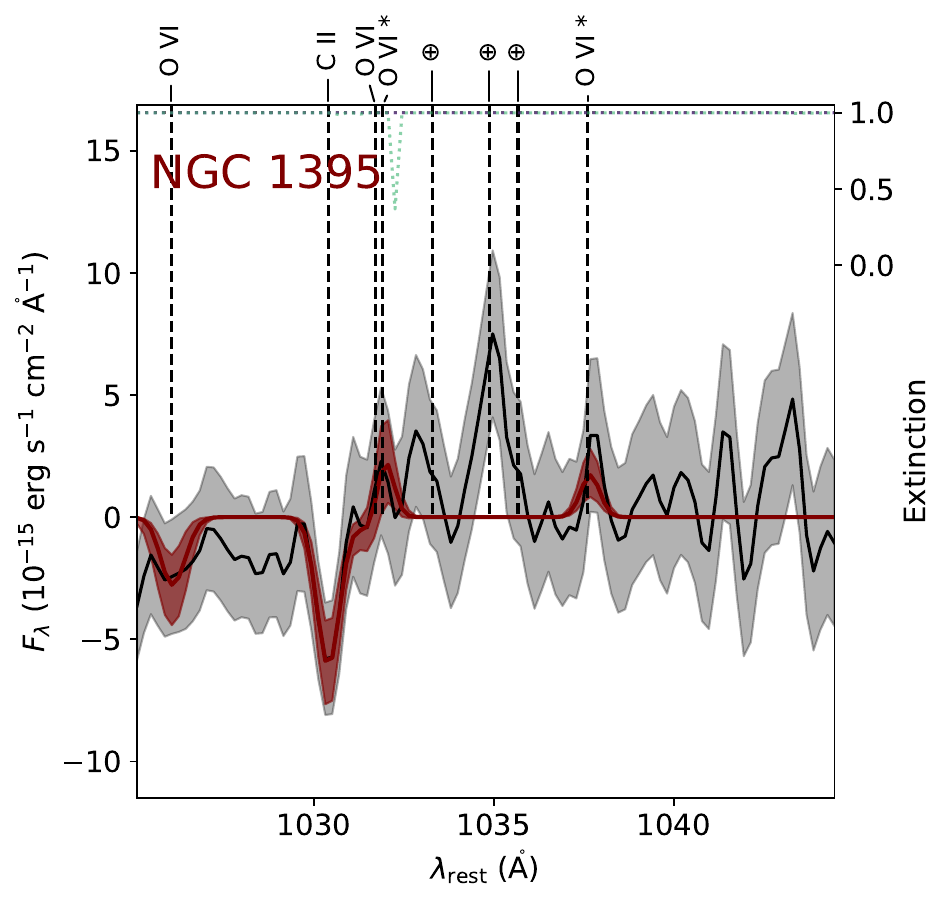}
    \caption{(Continued from Figure \ref{fig:ovi_fits_1})}
    \label{fig:ovi_fits_2}
\end{figure*}

\begin{figure*}[ht!]
    \centering
    \includegraphics[width=0.39\linewidth]{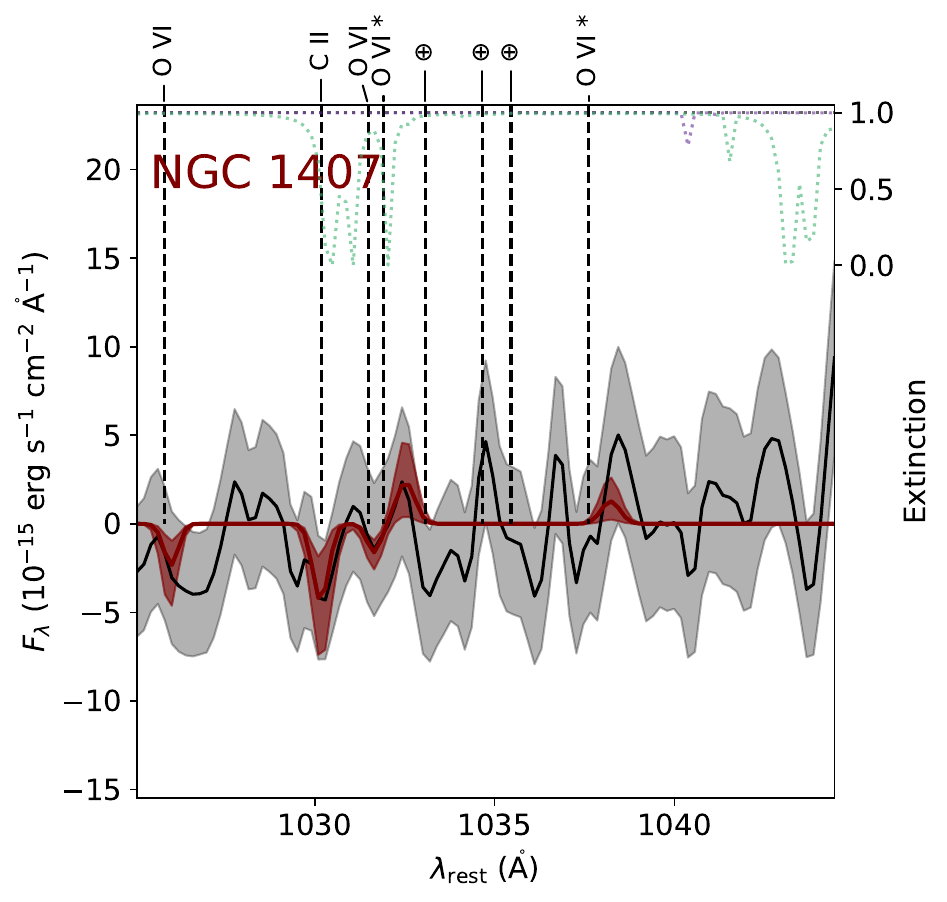}
    \includegraphics[width=0.39\linewidth]{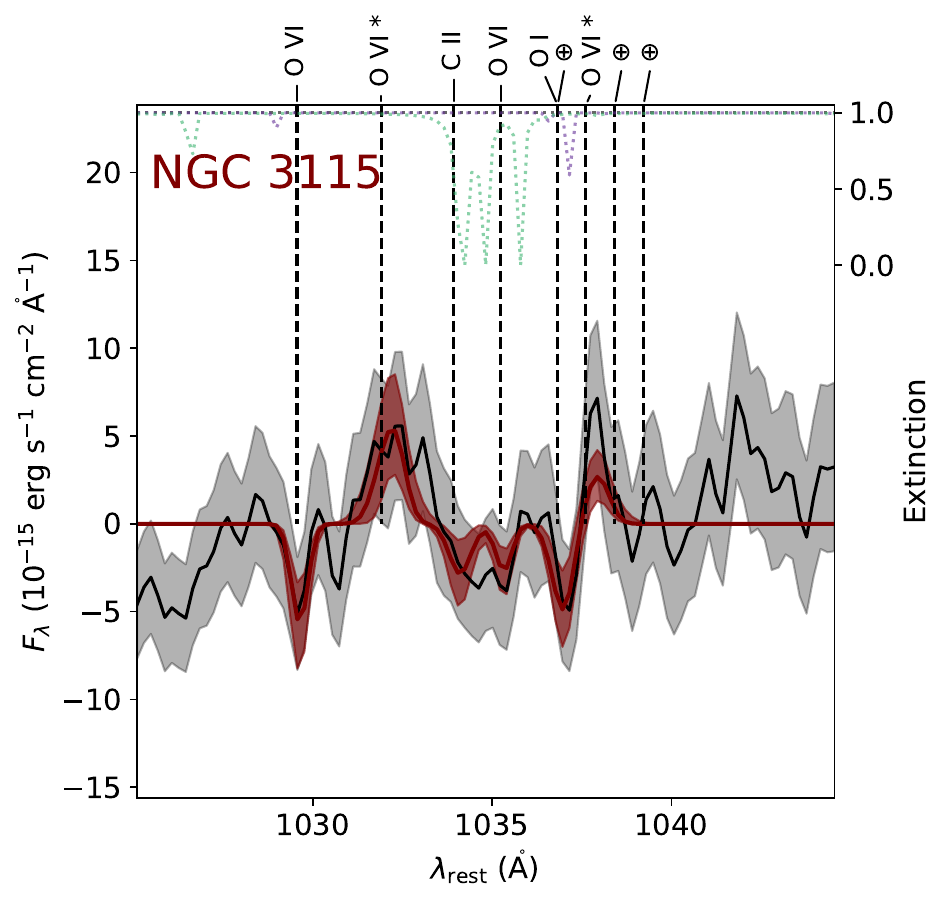}
    \includegraphics[width=0.39\linewidth]{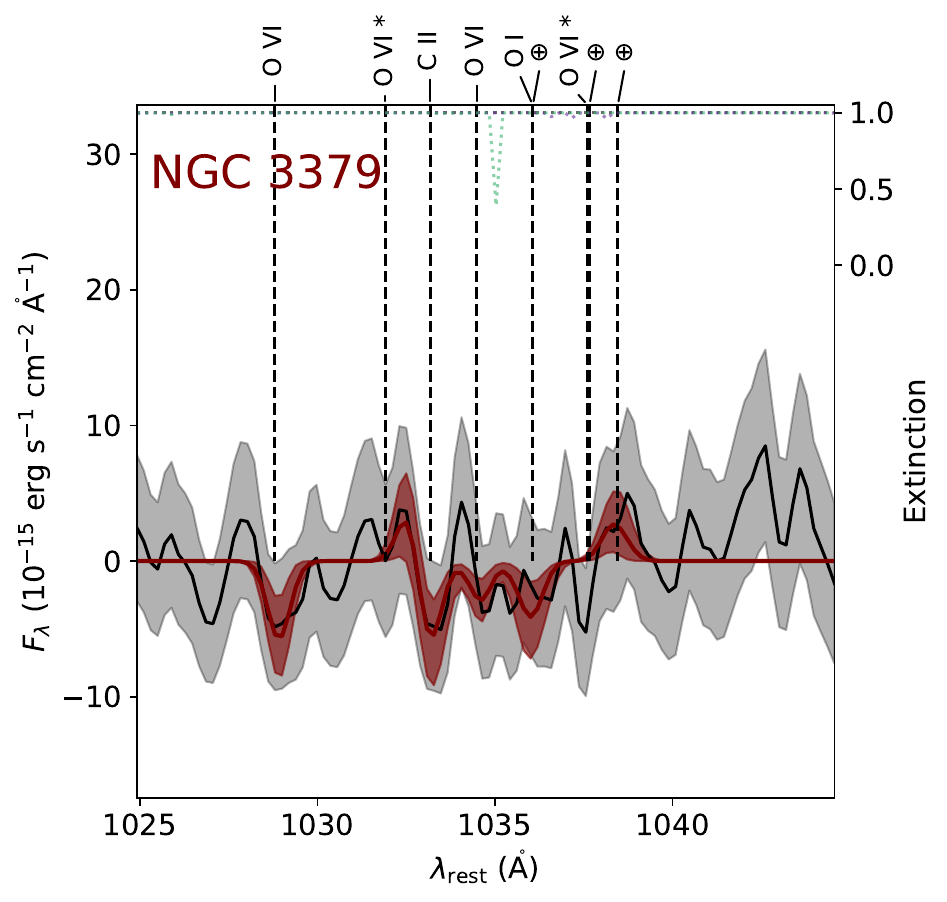}
    \includegraphics[width=0.39\linewidth]{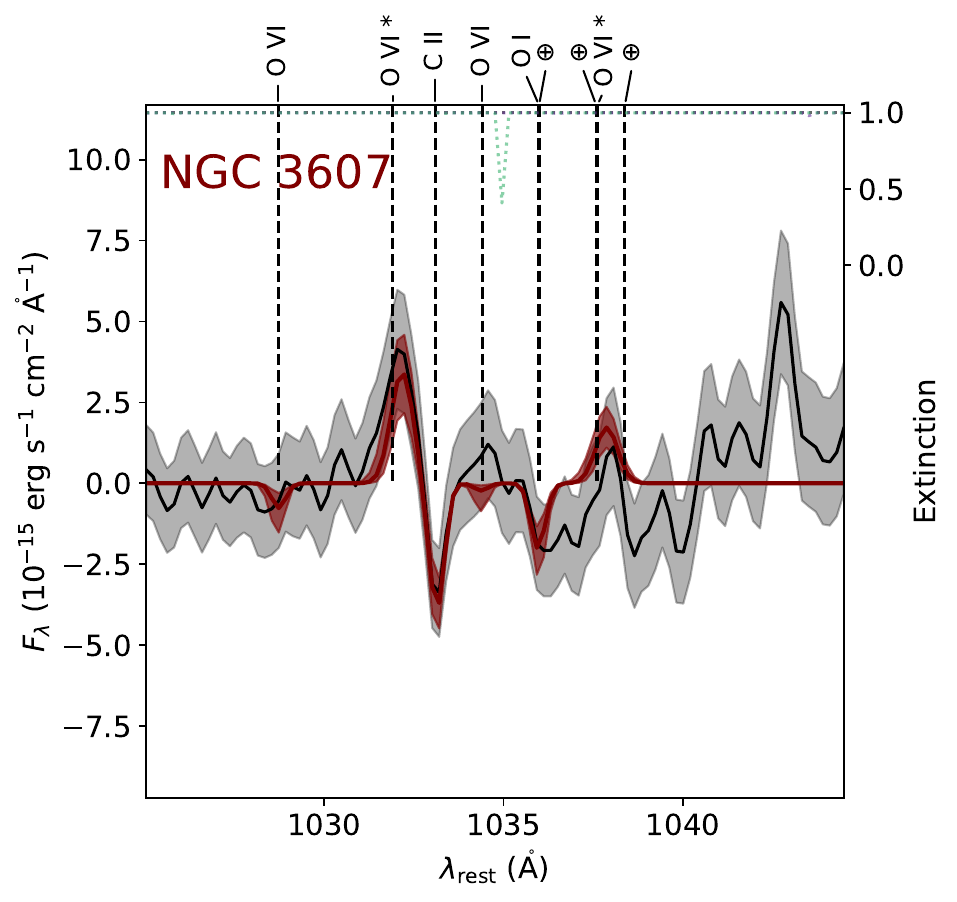}
    \includegraphics[width=0.39\linewidth]{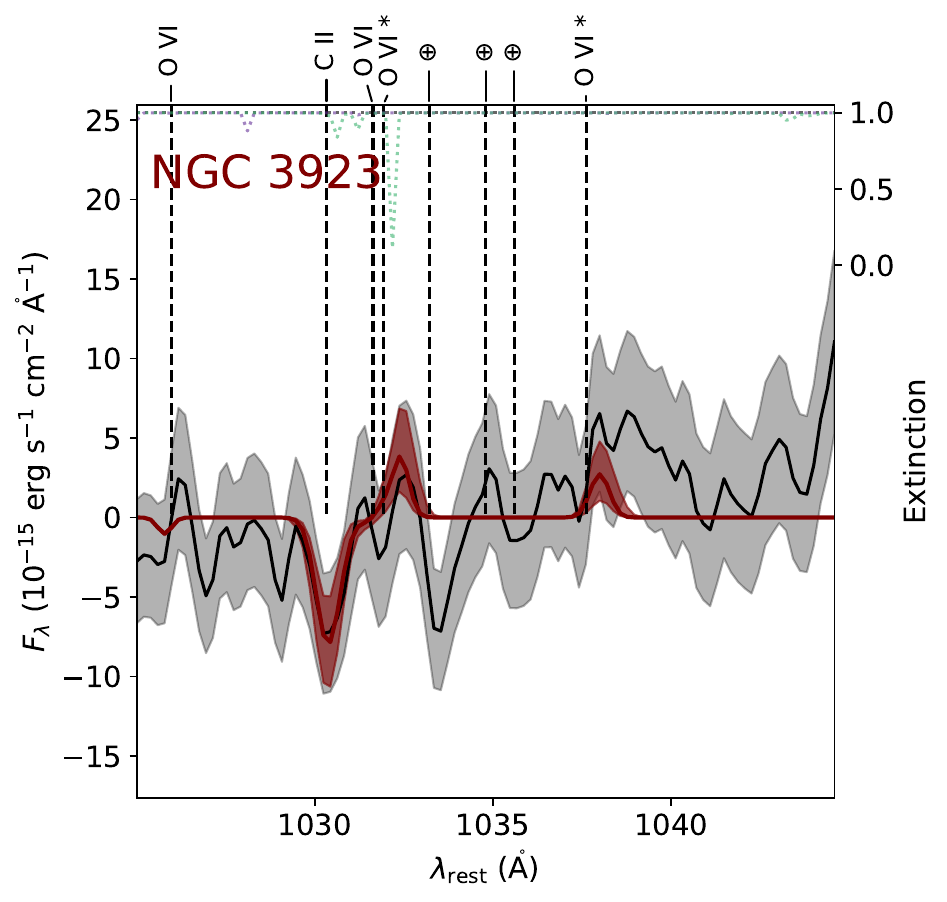}
    \includegraphics[width=0.39\linewidth]{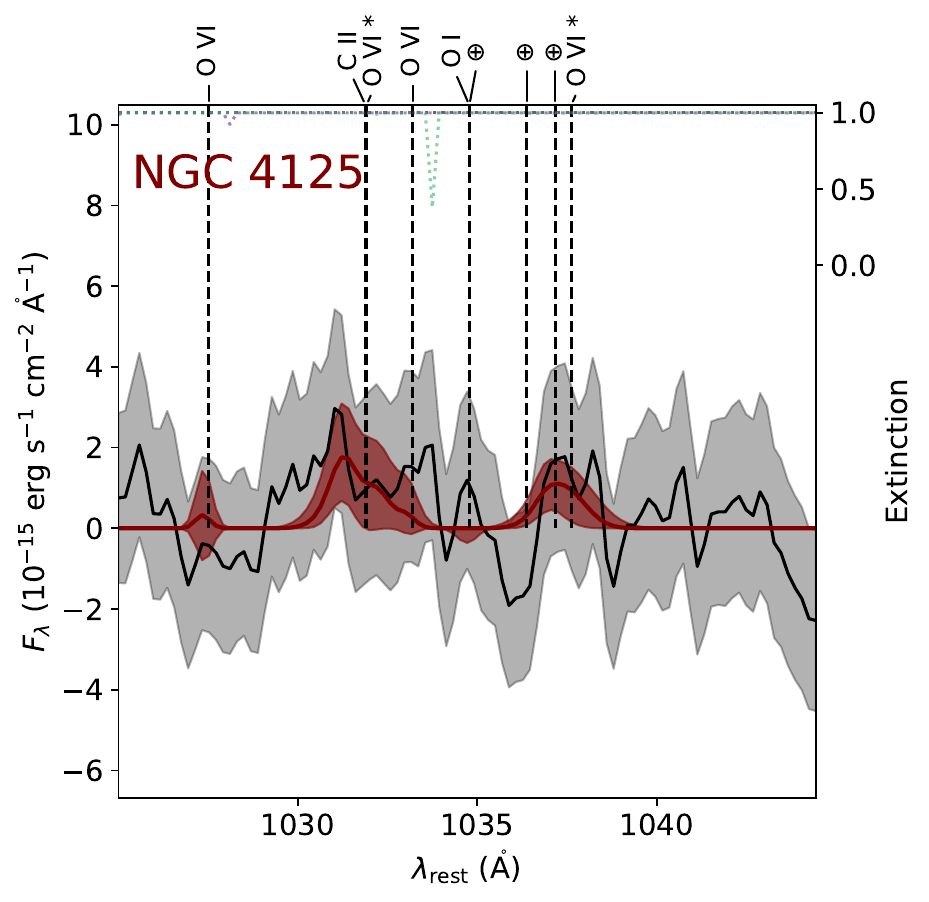}
    \caption{(Continued from Figure \ref{fig:ovi_fits_2})}
    \label{fig:ovi_fits_3}
\end{figure*}

\begin{figure*}[ht!]
    \centering
    \includegraphics[width=0.39\linewidth]{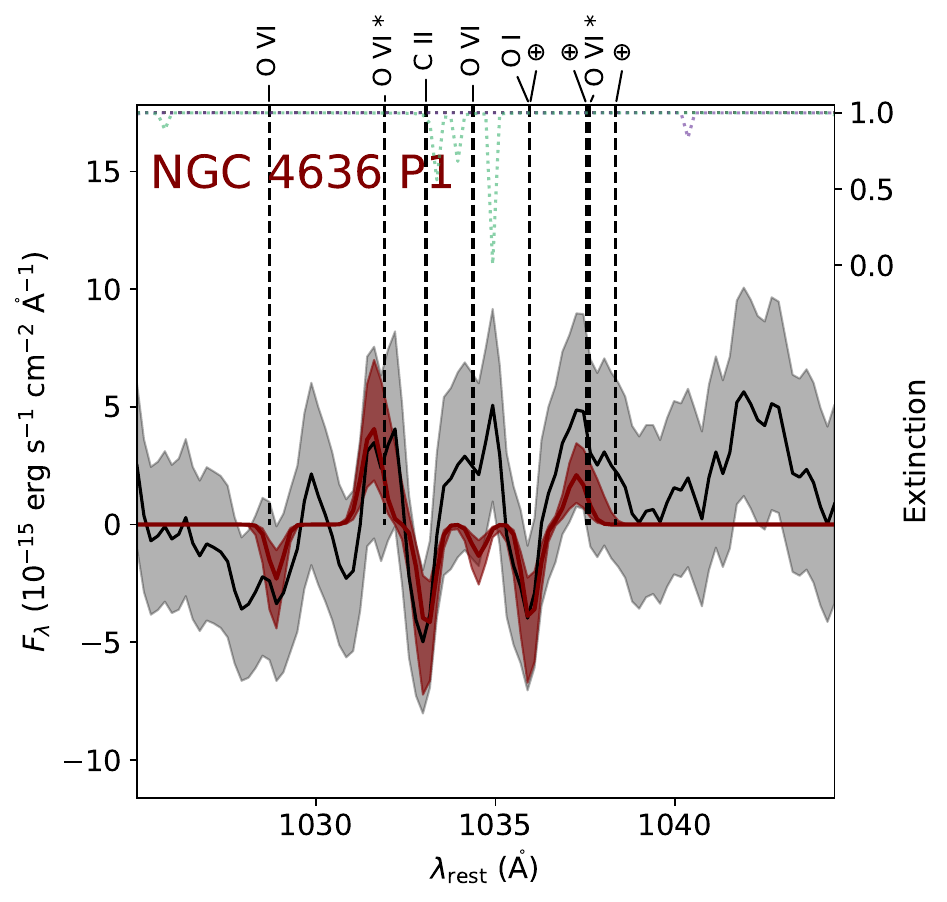}
    \includegraphics[width=0.39\linewidth]{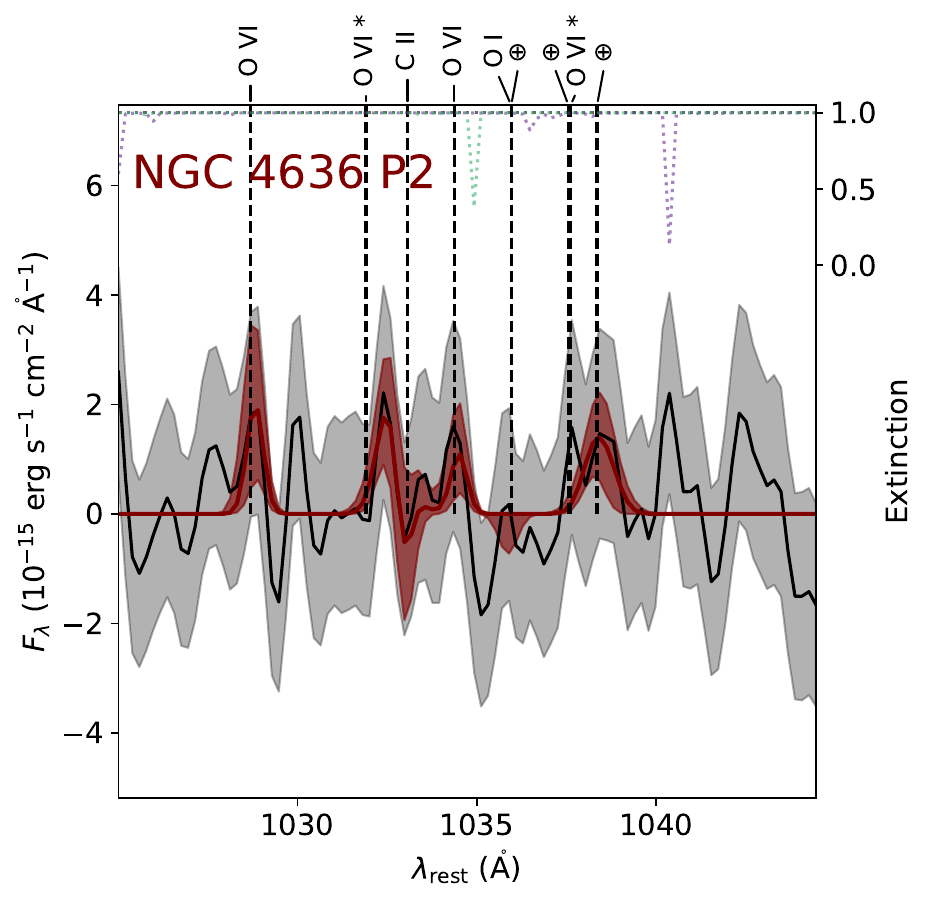}
    \includegraphics[width=0.39\linewidth]{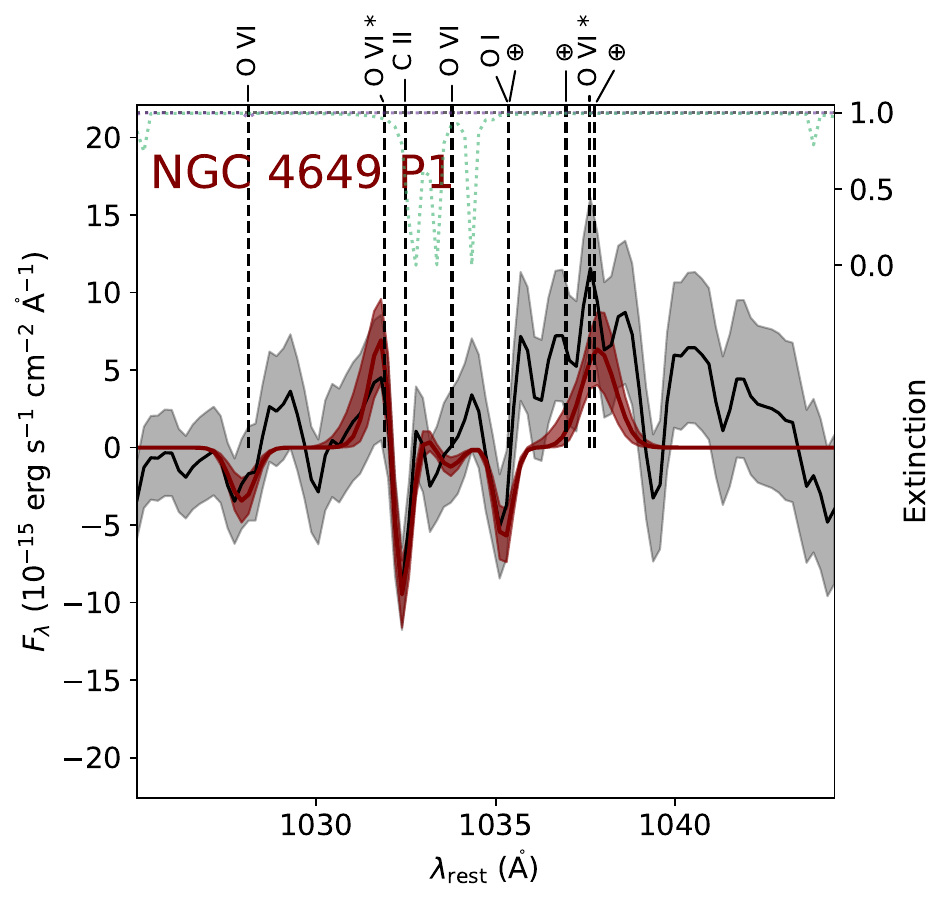}
    \includegraphics[width=0.39\linewidth]{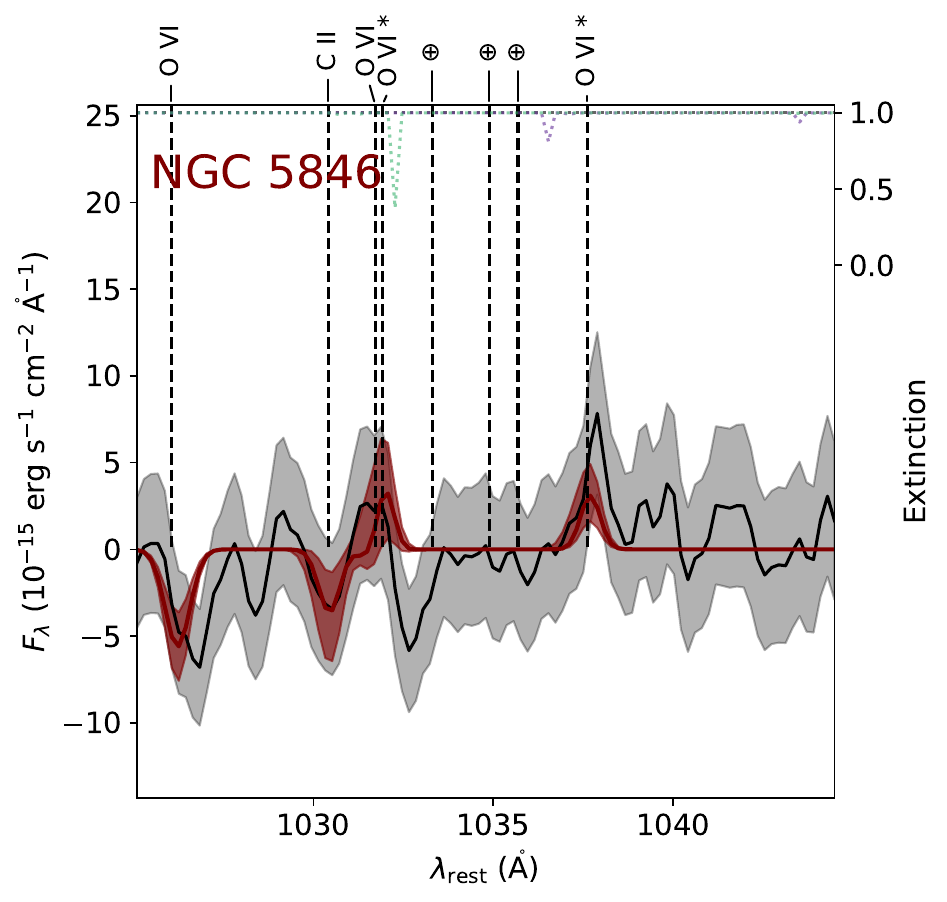}
    \includegraphics[width=0.39\linewidth]{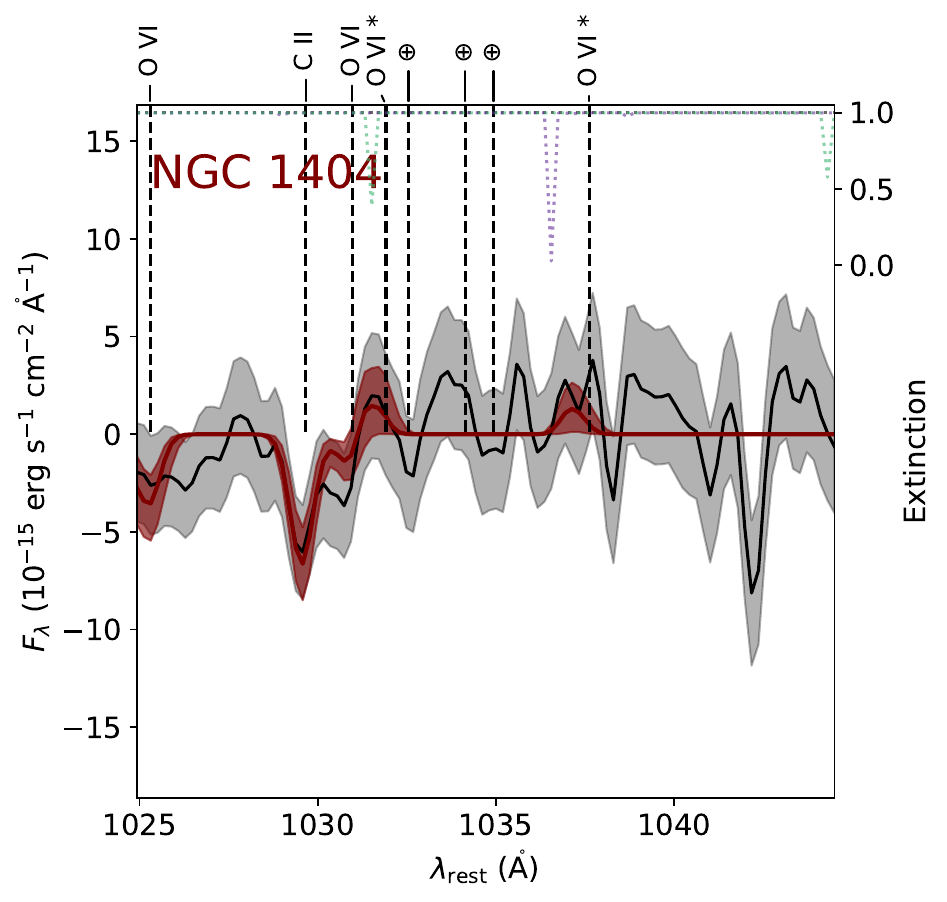}
    \includegraphics[width=0.39\linewidth]{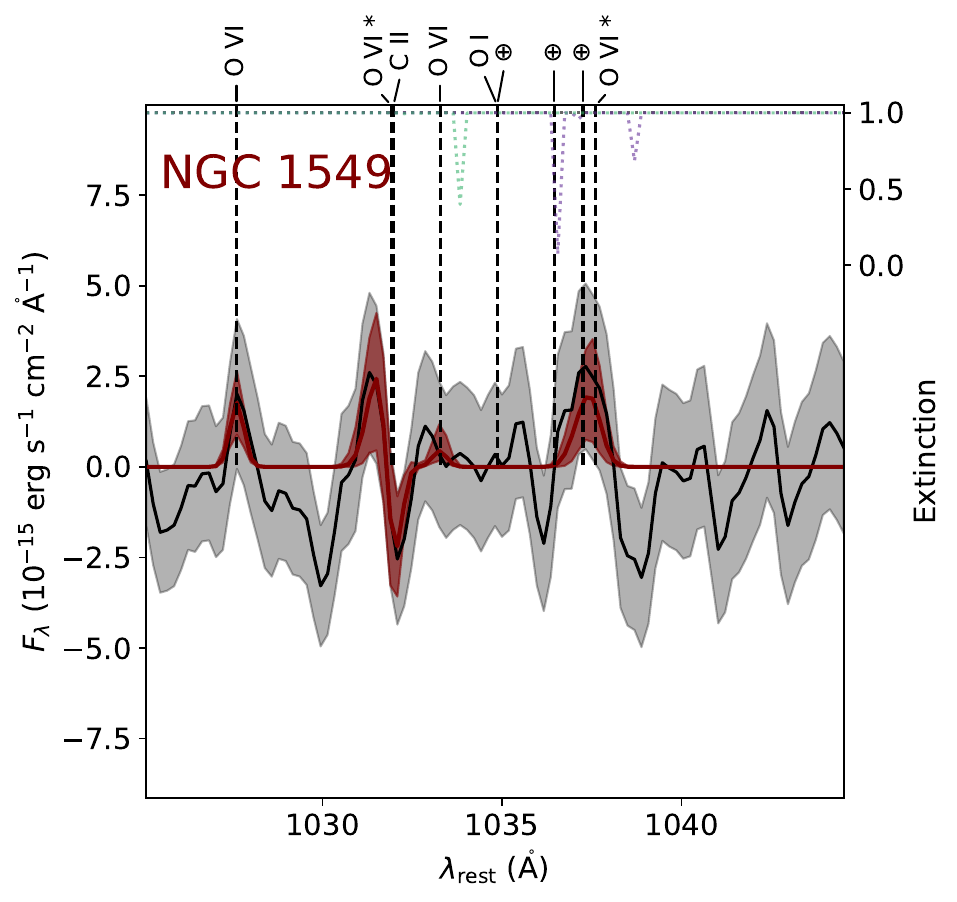}
    \caption{(Continued from Figure \ref{fig:ovi_fits_3})}
    \label{fig:ovi_fits_4}
\end{figure*}

\begin{figure*}[ht!]
    \centering
    \includegraphics[width=0.39\linewidth]{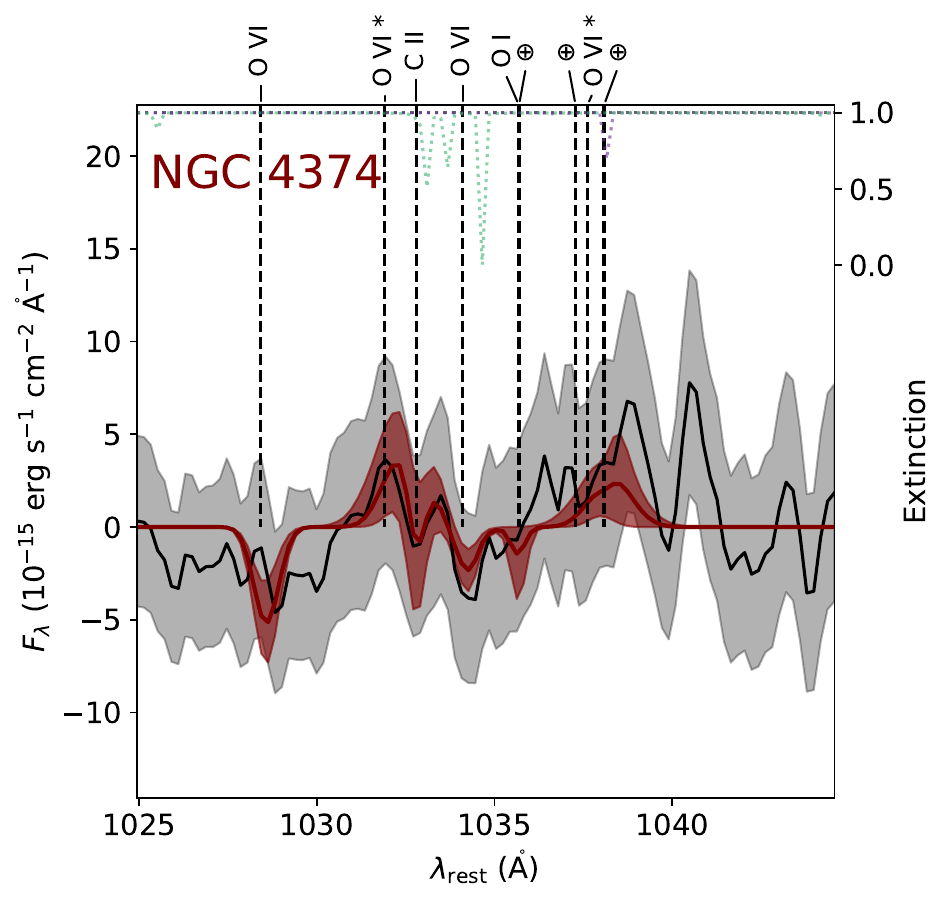}
    \includegraphics[width=0.39\linewidth]{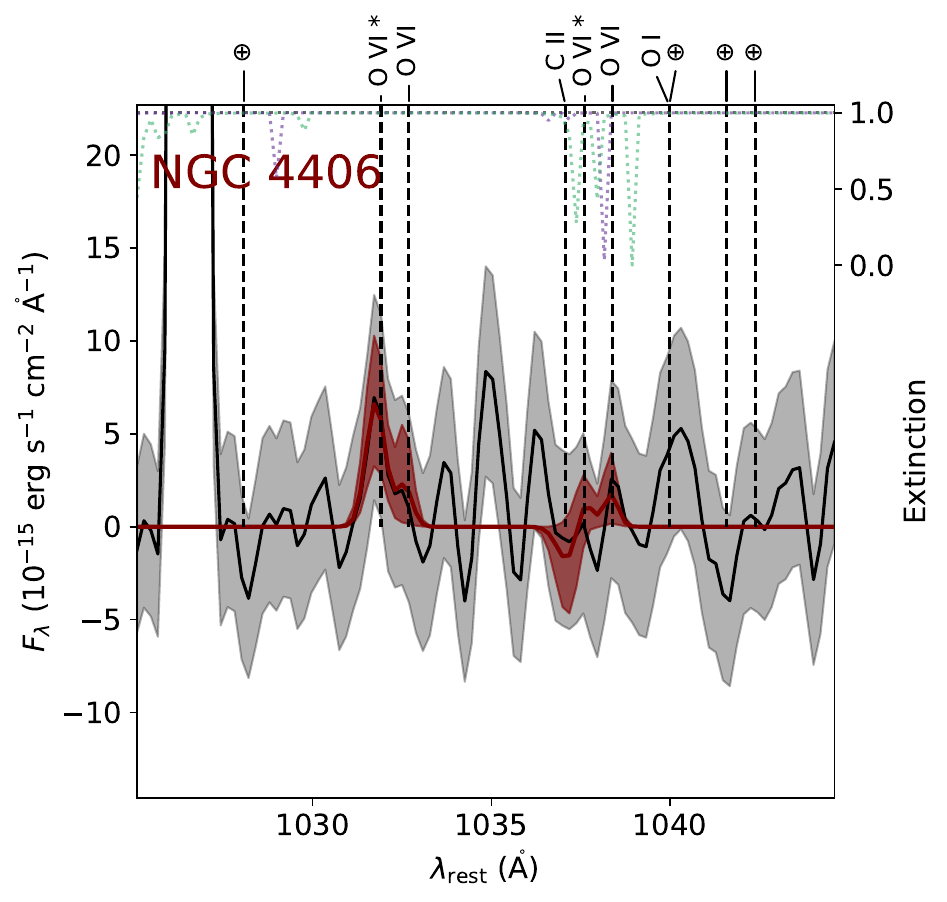}
    \includegraphics[width=0.39\linewidth]{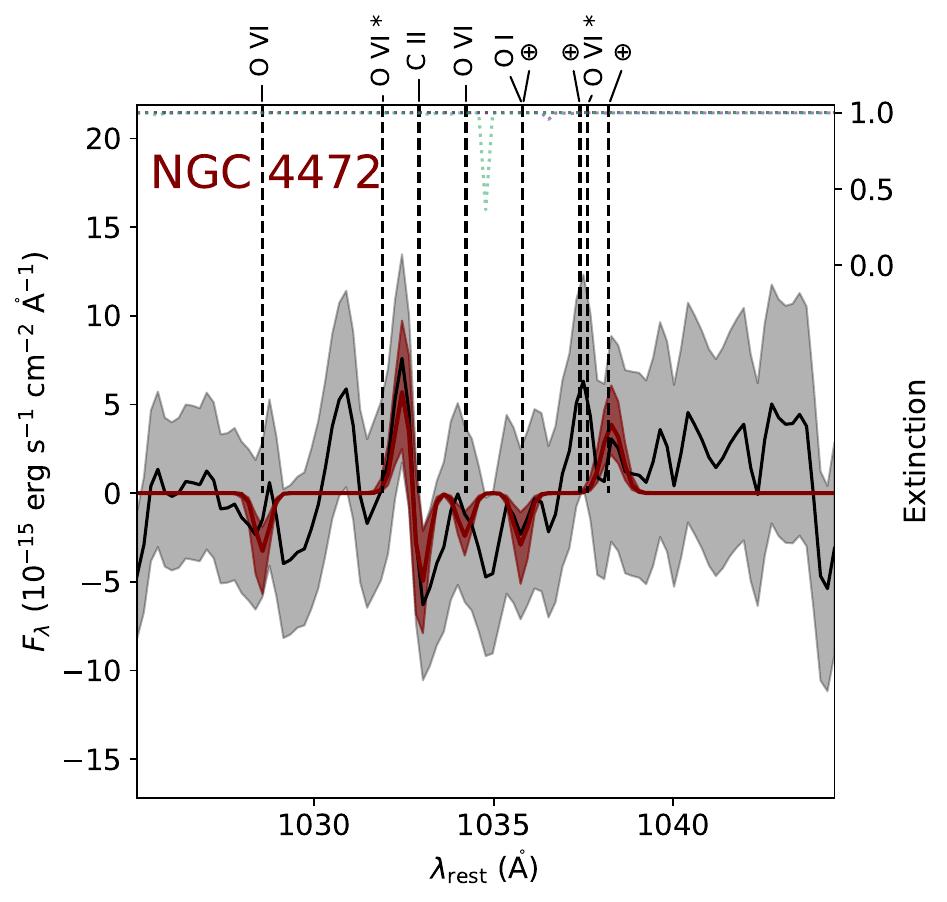}
    \includegraphics[width=0.39\linewidth]{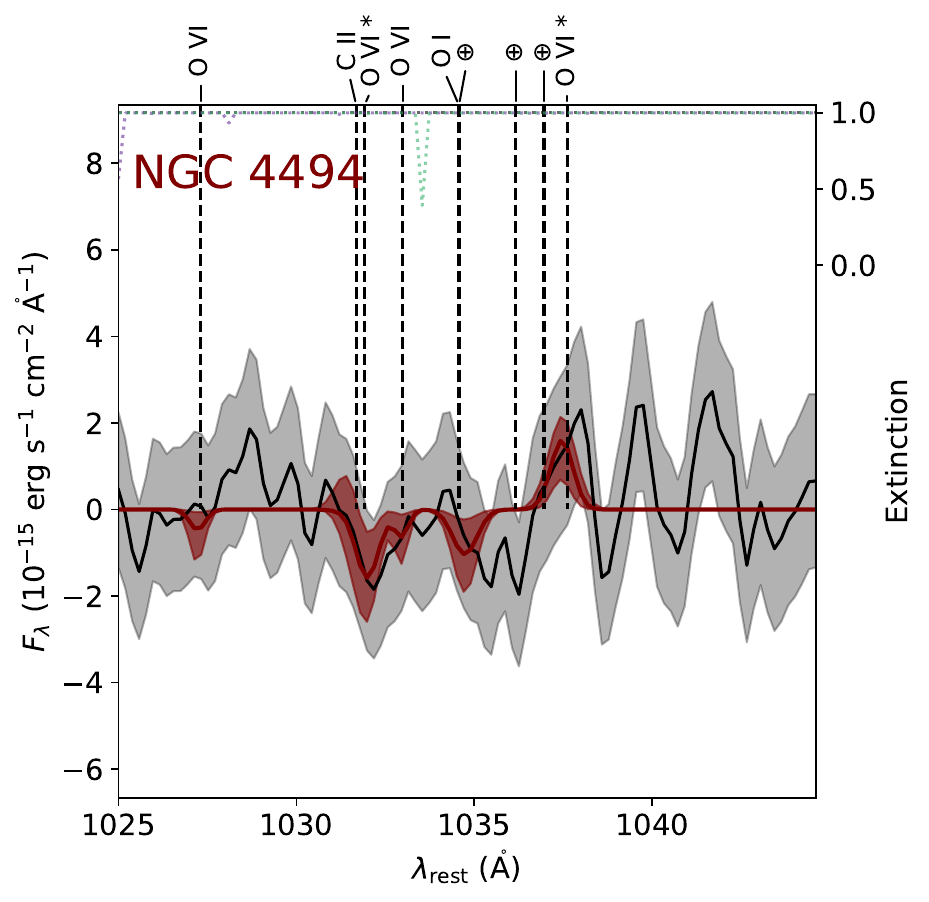}
    \includegraphics[width=0.39\linewidth]{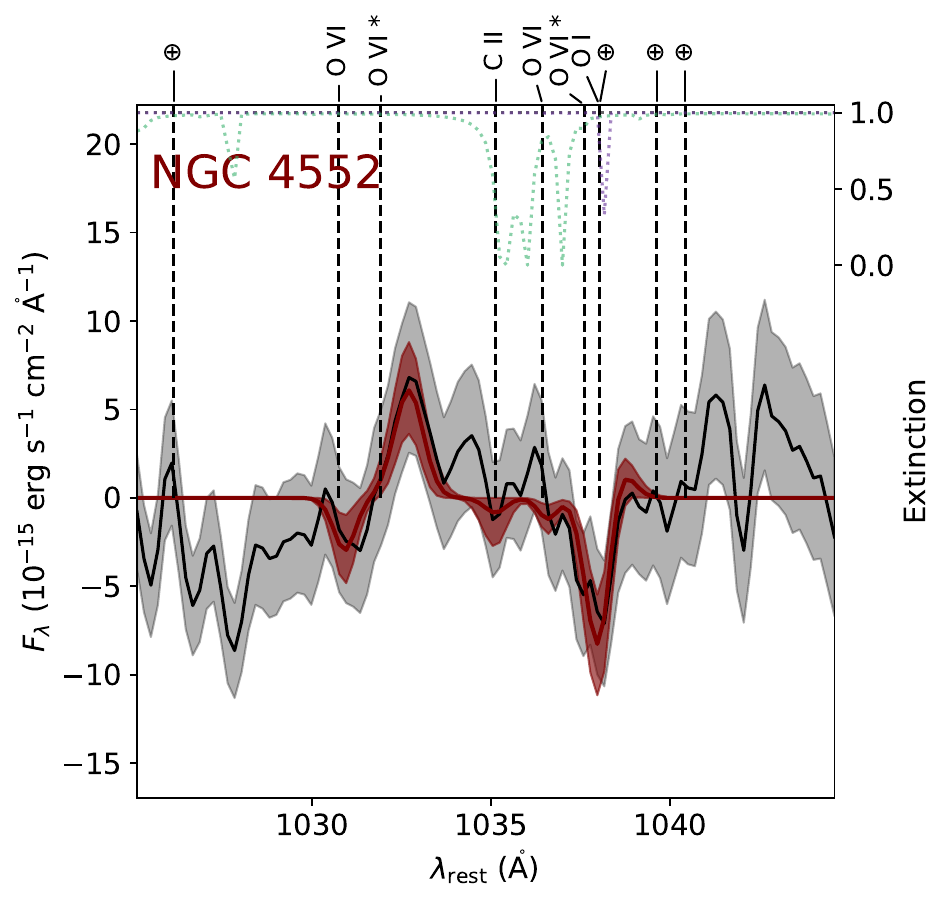}
    \includegraphics[width=0.39\linewidth]{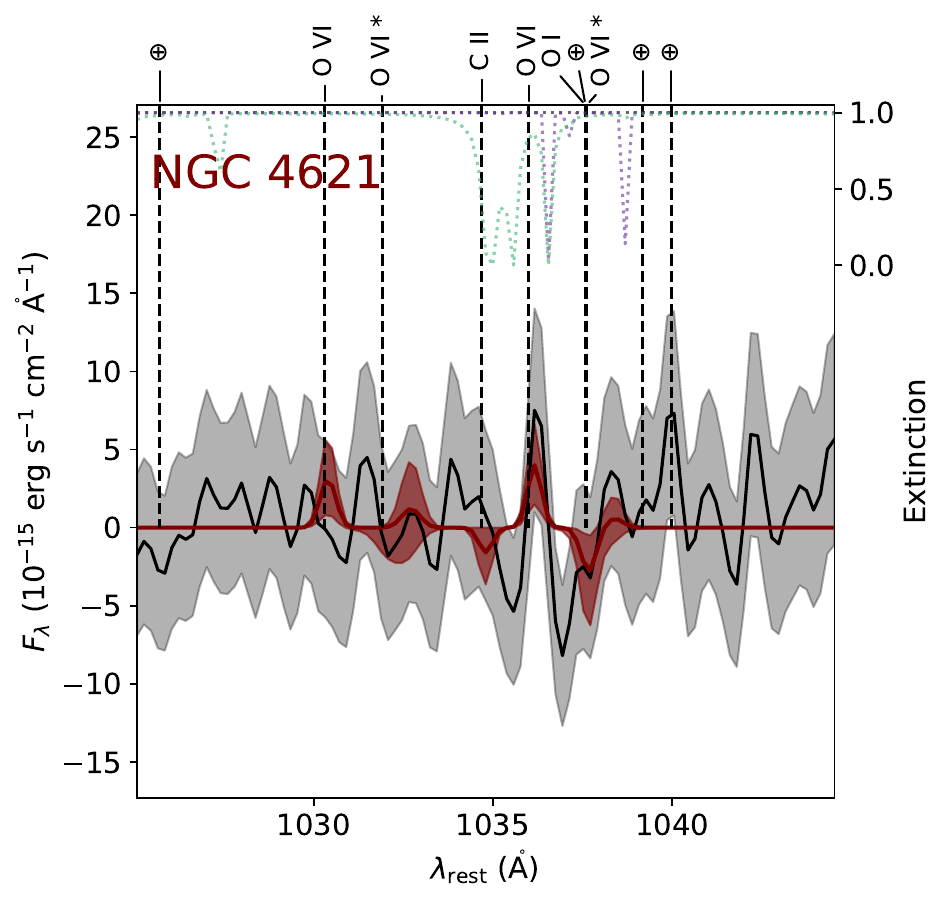}
    \caption{(Continued from Figure \ref{fig:ovi_fits_3})}
    \label{fig:ovi_fits_5}
\end{figure*}

\begin{figure*}[ht!]
    \centering
    \includegraphics[width=0.39\linewidth]{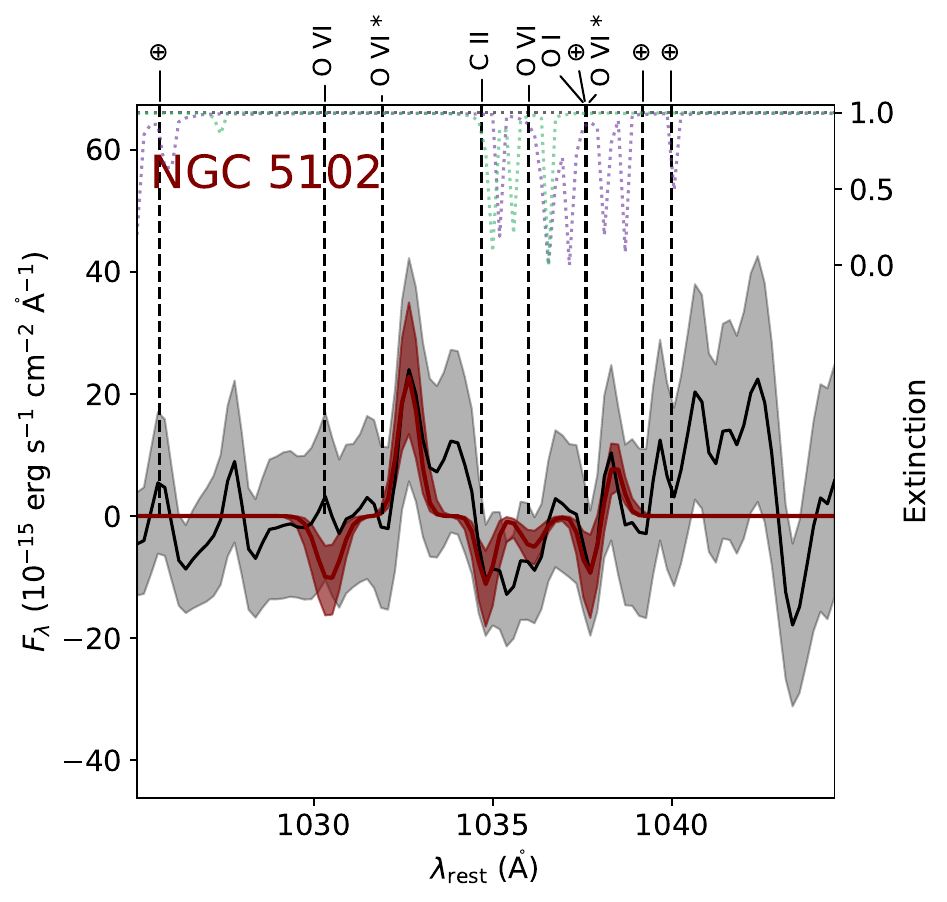}
    \caption{(Continued from Figure \ref{fig:ovi_fits_3})}
    \label{fig:ovi_fits_6}
\end{figure*}

\bibliography{main}{}
\bibliographystyle{aasjournalv7}

\end{document}